\documentclass[fleqn,usenatbib]{mnras}
\usepackage{newtxtext,newtxmath}
\usepackage[T1]{fontenc}
\DeclareRobustCommand{\VAN}[3]{#2}
\let\VANthebibliography\thebibliography
\def\thebibliography{\DeclareRobustCommand{\VAN}[3]{##3}\VANthebibliography}

\usepackage{graphicx}	
\usepackage{amsmath}	
\def    \apjl  		{\rm {ApJL}}
\def    \apj  		{\rm {ApJ}}
\def    \mnras  	{\rm {MNRAS}}
\def    \araa  		{\rm {ARA\& A}}
\def    \apjl  		{\rm {ApJL}}

\def	\cm		{\,{\rm {cm}}}

\def	\mum	{\,{\mu \rm{m}}}

\def	\s		{\,{\rm {s}}}

\def	\B	{\boldsymbol{B}}
\def	\P	{\boldsymbol{P}}

\def	\J	{\boldsymbol{J}}
\def	\ba	{\boldsymbol{a}}

\def \bea {\begin{eqnarray}}
\def \ena {\end{eqnarray}}              

\title[Grain Properties Effects on Synthetic Dust Polarization]{Effects of Grain Magnetic Properties and Grain Growth on Synthetic Dust Polarization of MHD Simulations in Protostellar Environments}

\author[Giang \& Hoang]{Nguyen Chau Giang,$^{1,2}$
Thiem Hoang,$^{1,2}$
\\
$^{1}$ Korea Astronomy and Space Science Institute, Daejeon 34055, Republic of Korea\\
$^{2}$Department of Astronomy and Space Science, University of Science and Technology, 217 Gajeong-ro, Yuseong-gu, Daejeon, 34113, Republic of Korea\\
}

\date{Accepted XXX. Received YYY; in original form ZZZ}

\pubyear{2023}

\begin{document}
\label{firstpage}
\pagerange{\pageref{firstpage}--\pageref{lastpage}}
\maketitle

\begin{abstract}
Thermal dust polarization is a powerful tool to probe magnetic fields ($\B$) and grain properties. However, a systematic study of the dependence of dust polarization on grain properties in protostellar environments is not yet available. In this paper, we post-process a non-ideal MHD simulation of a collapsing protostellar core with our updated POLARIS code to study in detail the effects of iron inclusions and grain growth on thermal dust polarization. We found that superparamagnetic (SPM) grains can produce high polarization degree of $p \sim 10-40\%$ beyond $\sim 500$ au from the protostar because of their efficient alignment by magnetically enhanced Radiative Torque mechanism. The magnetic field turbulence in the envelope causes the decrease in $p$ with increasing emission intensity $I$ as $p\propto I^{\alpha}$ with the slope $\alpha \sim -0.3$. But within 500 au, SPM grains tend to have inefficient internal alignment (IA) and be aligned with $\B$ by RATs only, producing lower $p \sim 1\%$ and a steeper slope of $\alpha \sim -0.6$. For paramagnetic (PM) grains, the alignment loss of grains above $1\mum$ in the inner $\sim 200$ au produces $p << 1\%$ and the polarization hole with $\alpha \sim -0.9$. Grain growth can increase $p$ in the envelope for SPM grains, but cause stronger depolarization for SPM grains in the inner $\sim 500$ au and for PM grains in the entire protostellar core. Finally, we found the increase of polarization angle dispersion function $S$ with iron inclusions and grain growth, implying the dependence of B-field strength measured using the DCF technique on grain alignment and grain properties.
 
\end{abstract}

\begin{keywords}
stars: formation, magnetic fields, low-mass stars, dust extinction, polarization  
\end{keywords}


\section{Introduction}\label{sec:intro}
Magnetic fields ($\B$) are thought to play an essential role in the collapse of dense cores and the formation of protostar, protostellar disks and planetary systems (\citealt{Shu_1987}, \citealt{Mckee_2007}, \citealt{Allen_2003}, \citealt{Yusuke_2022}). The dominant effect of magnetic fields over gravity produces strong magnetic pressure, which prevents the collapse of the core in the early stage (\citealt{Nakano_1978}, \citealt{Krumholz_2019}), but helps to maintain the cloud structure against the destruction from radiation and mechanical feedback from protostars in later period (\citealt{Krumholz_2014}, \citealt{Pabst_2019}). The dominant magnetic energy over turbulence energy also helps to guide the infall gas motion inside the self-collapsing cloud, which shapes the structure of molecular clouds and filaments (\citealt{Hennebelle_2009}, \citealt{Pattle_2018}), protostellar cores and disks, and controls the coherence of magnetic field morphology across different scales in star-forming regions (\citealt{Fiedler_1993}, \citealt{Galli_1993a}, \citealt{Galli_1993b}, \citealt{Allen_2003}, \citealt{Kataoka_2012}, (\citealt{Seifried_2015})).

During the Class 0/I stage of Young Stellar Objects (YSOs), magnetic fields are predicted to help transport the disk angular momentum outward via the magnetic braking effect (\citealt{Mestel_Spitzer_1956}, \citealt{Mouschovias_1979}, \citealt{Mouschovias_1980}, \citealt{Basu_1994},  \citealt{Allen_2003}, \citealt{Melon_2008}). Basically, the coupling between the fast-rotating disk and the poloidal component of magnetic fields produces the magneto-centrifugal force that transports the matter from the disk away along the disk rotation axis, forming a protostellar outflow (\citealt{Pudritz_1983}, \citealt{Bally_2016}). The removal of matter and angular momentum via the outflow saves the life of protostellar disks from being fragmented by the strong centrifugal force and allows the remaining matter to continue feeding the growth of the central protostar (\citealt{Galli_2006}, \citealt{Li_2014}). However, if the magnetic field is strong and aligned with the disk rotation axis, the magnetic braking can become too efficient in removing the disk angular momentum. As a result, the remaining matter in the disk will be quickly accreted onto the protostar that stops the protostellar disk formation, named as the magnetic braking catastrophe (\citealt{Allen_2003}, \citealt{Galli_2006}, \citealt{Mellon_2008}, \citealt{Hennebelle_2008}, \citealt{Machida_2010}). One of the possible scenarios to solve the above problem is to introduce the initial misalignment between the rotational axis and the magnetic field orientation (\citealt{Hennebelle_2009}, \citealt{Joos_2012}, \citealt{Krumholz_2013}). Besides, the effect of non-ideal magnetohydrodynamic (MHD), i.e., ambipolar diffusion (\citealt{Duffin_2009}, \citealt{Mellon_Li_2009}), Ohmic dissipation (\citealt{Dapp_2012}, \citealt{Machida_2014}, \citealt{Tomida_2015}, \citealt{Lam_2019}), the Hall effect (\citealt{Tsukamoto_2015}, \citealt{Wurster_2018}), turbulence-induced reconnection inside the Keplerian disk (\citealt{Santos_2012}, \citealt{Santos_2013}, \citealt{Li_2014}, \citealt{Seifried_2015}), are also suggested to be able to solve the magnetic braking catastrophe. However, different scenarios lead to different outcomes (in terms of disk size, outflow strength, and magnetic field morphology). Consequently, accurate measurements of both the morphology and strength of magnetic fields in all spatial scales of protostellar environments become a key for accurately understanding the role of magnetic fields in regulating the core collapsing and guiding the star, disk, and outflow formation.
 
The most popular technique to measure magnetic fields in protostellar environments is to use polarized thermal dust emission (see, e.g., \citealt{Hull_2019}, \citealt{Pattle_2022} for recent reviews). Given the perpendicular alignment between the grain longest axis and $\B$ that produces polarization vectors $\P \perp \B$ \citep{Lazarian_Hoang_2007, Anderson_2015, Lazarian_2015}, measurements of dust polarization from protostellar cores of 0.1 pc down to protostellar disks of 100 au are widely conducted. This is done using both single-dish telescopes such as the James Clerk Maxwell Telescope (JCMT, \citep{Matthews_2009} and interferometric arrays such as the Jansky Very Large Array (JVLA), Submillimeter Array (SMA, \citep{Ho_2014}, and the Atacama Large Millimeter/submillimeter Array (ALMA). Especially, the operation of ALMA, which can resolve up to $10$ au around the protostar, currently opens a golden era for using dust polarization to study the role of magnetic fields in star and planet formation. Many low/intermediate-mass class 0/I Young Stellar Objects (YSOs) reveal a consistent picture of magnetic fields from the core to envelope scale \citep{Ching_2017, Hull_2014, Davidson_2014, Hull_2017b}, with well-ordered hourglass-shaped $\B$ fields preserved from 1000 au to 100 au \citep{Gigart_1999, Gigart_2006, Goncalves_2008, Rao_2009, Stephens_2013, Maury_2018, Kwon_2019, Sadavoy_2019}. These features match well with theoretical predictions of the magnetically-regulated collapsing core (i.e., \citep{Allen_2003}, revealing the important dynamics of $\B$ in the formation of these objects. Some other sources, in contrast, do not clearly show the correlation of magnetic fields between core, envelope, and disk scales (i.e., Serpens SMMI1 studied by \citealt{Hull_2017a}), revealing the dynamic importance of turbulence and gas kinematics over magnetic energy in guiding core and disk formation (see review by \citep{Hull_2019}.
 
Focusing on observations probed by ALMA, the situation becomes much more complicated since magnetic fields significantly deviate from the inner envelope of $\sim 500$ au to the disk scale within $\sim 100$ au (\citealt{Hull_2014}, \citealt{Cox_2015}, 
\citealt{Cox_2018}, \citealt{Takahashi_2019}, \citealt{Sadavoy_2019}). For example, Class 0 Protobinary System L1448 IRS 2 studied by \cite{Kwon_2019} clearly reveals the change from the hourglass-shaped $\B$-fields in the envelope to the pinched fields along the major disk axis within $\sim 100$ au. Similarly, magnetic fields from Class 0 OMC-3/MMS 6 studied by \cite{Takahashi_2019} show the significant transition from the spiral pattern within $\sim 800$ au to the radial field at $\sim 200$ au, and to the pinched field in the disk scale within $\sim 100$ au. These authors suggest that the pinched or circular field inside the protostellar disk could be induced by the formation of the toroidal $\B$-fields wrapping by the fast rotation disk. But it also could be explained by self-scattering of thermal emission from sub-millimeter grains (\citealt{Yang_2016}, \citealt{Cox_2018}, \citealt{Takahashi_2019}, \citealt{Sadavoy_2018}, \citealt{Sadavoy_2019}, \citealt{Lam_2021}), whose polarization signal does not contain any magnetic field information. Furthermore, observations of dust polarization toward the inner 100 au region of NGC 1333 IRAS4A (\cite{Ko_2020}) and OMC-3/MMS 6 (\citealt{Liu_2021}) even reveal the $90^{\circ}$ flipping of the polarization pattern between millimeter and submillimeter wavelengths, which challenges our understanding about origin of polarization signal obtained by ALMA. The uncertainty of the polarization origin inside the inner $\sim 100$ au region raises the serious question of whether we could use dust polarization to study the role of $\B$ fields in building the star and disk formation in this scale.
 
Besides the complex polarization pattern, interpreting the properties of dust polarization measured toward $100-1000$ au of Class 0/I YSOs by ALMA poses challenges. The most prominent feature detected in this class is the reduction of the polarization degree toward the center, named as the polarization hole, from a rather high $p > 5-40\%$ in the envelope to $p \sim 1\%$ inside the disk (\citealt{Hull_2014}, \citealt{Cox_2015}, \citealt{Cox_2018}, \citealt{Maury_2018}, \citealt{Sadavoy_2018}, \citealt{Kwon_2019}, \citealt{Takahashi_2019}, \citealt{Valentin_2019}, \citealt{Valentin_2023a}). The depolarization effect was first attributed to the low resolution of single-dish telescopes, as they could not resolve the complicated magnetic field in the center. However, the polarization hole remains clearly detectable with the generations of SMA, JVLA, and ALMA at all observed wavelengths. Thus, the distortion of $\B$ fields by turbulence and the weak alignment efficiency of dust grains with magnetic fields (\citealt{Lazarian_2007}, \citealt{Yang_2021}) are usually referenced to explain the low polarization degree in the center. However, the exact contribution of each effect to depolarization and the reasons for the poor grain alignment around the protostar remain unclear. Furthermore, as grains are expected to be inefficiently aligned with $\B$ in protostellar environments due to strong gas randomization, the detection of high $10-40\%$ polarization in the envelope by ALMA, and the subsequent significant reduction of $p(\%)$ from such high values to $1\%$ inside the disk challenge our understanding of grain alignment in these regions. 

The key to answering the above questions is to perform synthetic modeling of polarized dust emission based on a realistic model of grain alignment in Class 0/I YSOs. Grain alignment involves two main stages: 1) internal alignment, which aligns the grain angular momentum $\J$ with the axis of maximum inertia moment ${\ba}_{\rm 1}$ (\citealt{Lazarian_2007}), and 2) external alignment, which results in the alignment between $\J$ and $\B$ (see \citealt{Anderson_2015}, \citealt{Lazarian_2015}, \citealt{Hoang+2022} for a review). For grains containing iron atoms (i.e., silicate grains), which are paramagnetic material (PM grains), they can be magnetized via the Barnett effect (\citealt{Barnett_1915}) and have the internal alignment through Barnett relaxation (\citealt{Purcell_1979}). This mechanism brings grains back to the minimum rotational energy level with $\J \parallel {\ba}_{\rm 1}$ by dissipating the grain rotational energy via the precession of the grain magnetic moment around ${\ba}_{\rm 1}$ within the grain's inertia frame. In addition to Barnett relaxation, nuclear relaxation (\citealt{Lazarian_1999}, \citealt{Lazarian_Hoang_2008}), which shares similar physics to the former ones except the origin of the magnetic moment comes from nuclear spins, also can lead $\J$ to align with ${\ba}_{\rm 1}$. Inelastic relaxation  (\citealt{Purcell_1979}, \citealt{Lazarian_Efroisky_1999}, \citealt{Efroimky_2000}) which dissipates grain rotational energy via the deformation of spinning inelastic grains can also induce grain internal alignment. While Barnett relaxation is found to work effectively for sub-micron and micron-sized grains of $\sim 1\mum$,  nuclear relaxation and inelastic relaxation are efficient for larger grains, from a few microns to very large grains (VLGs) above $> 10\mum$ (\citealt{Lazarian_Hoang_2008}, \citealt{Lazarian_Hoang_2019}). Given the presence of external magnetic fields, grains can couple with $\B$ owning to the Larmor precession formed by the interaction between the grain magnetic moment with the ambient magnetic field. Then, RAdiative Torque (RATs) (\citealt{Dolginov_1976}, \citealt{Draine_Weingartner_1996} \citealt{Lazarian_Hoang_2007}, \citealt{Hoang_Lazarian_2008}), or magnetic relaxation (\citealt{David_1951}) plays a role in driving the alignment between $\J$ and $\B$. Magnetic relaxation leads grains to align with $\B$ by dissipating the grain rotational energy via the rotating induced magnetic moment in magnetized environments (\citealt{David_1951}). In contrast, RAT aligns grains with $\B$ via their alignment torque component produced from the differential scattering and absorption between the ambient anisotropic radiation fields and the left- and right-hand of irregular dust grains (\citealt{Dolginov_1976}, \citealt{Draine_1996}, \citealt{Lazarian_Hoang_2007}, \citealt{Hoang_Lazarian_2008}). However, RAT can lift up grains to suprathermal rotation (\citealt{Hoang_Lazarian_2008}), which stabilizes their orientation in space regardless of gas randomization. Consequently, RAT is the most effective mechanism driving the grain alignment with $\B$ in the diffuse interstellar medium (ISM) and molecular clouds (MCs, \citealt{Reissl_2017}, \citealt{Seifried_2018}, \citealt{Reissl_2020}). Indeed, RAT only can drive a part of grains, which is parameterized by $f_{\rm high-J}$, to have perfect magnetic alignment at their suprathermal rotation (or called high-\textit{J} attractors, \citealt{Hoang_Lazarian_2008}). The value of $f_{\rm high-J}$ for PM grains varies between 0.2 - 0.7 (\citealt{Lazarian_Hoang_2008}, \citealt{Hoang_Lazarian_2016b}, \citealt{Herranen_2021}, \citealt{Lazarian_Hoang_2021}), depending complexly on grain properties (size, grain shape, and composition) and their orientation with radiation fields and magnetic fields.  
 
\citealt{Reissl_2016} first combined three-dimensional (3D) Monte-Carlo Radiative Transfer with the RAT alignment physics into the POLArized RadIation Simulator (POLARIS) code, enabling users to simulate polarized dust emission from aligned dust grains by RATs. Using the POLARIS code, \cite{Brauer_2016} conducted numerical modeling of dust polarization by aligned grains from a Bok globule. They suggested that the existence of aligned VLGs of $\sim 10-100\mum$ inside the protostellar disk can cause depolarization at submillimeter wavelengths, as detected in many Class 0/I YSOs (\citealt{Hull_2014}, \citealt{Cox_2015}, \citealt{Cox_2018}, \citealt{Maury_2018} \citealt{Kwon_2019}, \citealt{Sadavoy_2018}). VLGs also can lead to the $90^{\circ}$ flipping of the polarization pattern as found in the protostellar disk of NGC 1333 IRA4A (\citealt{Ko_2020}) and OMC3 MMS6 (\citealt{Liu_2021}) due to the change in the polarization mechanism from dichroic emission at optically thin wavelengths to dichroic extinction at optically thick wavelengths. Regarding the observed level of dust polarization in Class 0/I YSOs, \cite{Valdivia_2019} and \cite{Valentin_2023b} found that the high polarization of $p \geq 5\%$ observed in the envelope (\citealt{Hull_2014}, \citealt{Cox_2015}, \citealt{Sadavoy_2018}, \citealt{Maury_2018}) only can be explained by the presence of aligned VLGs above $10\mum$ beyond 100 au by RATs. The alignment of VLGs in the envelope is also crucial to validate the usage of dust polarization in probing $\B$ fields on this scale (\citealt{Valdivia_2022}). However, \cite{Valentin_2020} and \cite{Valentin_2023b} recently found the RAT alignment mechanism alone cannot explain the high grain alignment efficiency required in Class 0/I YSOs by ALMA. But the perfect alignment model of VLGs does address this issue. Furthermore, the model proposed by \cite{Brauer_2016} is only effective inside the optically thick disk at submillimeter wavelengths, which contain a high amount of aligned VLGs (\citealt{Zielinski_2021}). The question of whether VLGs can exist and have perfect alignment with magnetic fields by RATs in both the protostellar envelope and disk during Class 0/I stage remains unclear.

Back to the RAT alignment paradigm, \cite{Hoang_Lazarian_2016a} found that for grains with embedded iron clusters (specifically, superparamagnetic (SPM) grains), the enhanced magnetic susceptibility (\citealt{Jones_Spitzer_1967}) can increase the magnetic relaxation over gas randomization. Consequently, grains can quickly align with $\B$ with higher $f_{\rm high-J}$ through the joint action of RATs and superparamagnetic relaxation, named the Magnetically enhanced RAdiative Torque (MRAT) alignment mechanism. Additionally, they found that iron inclusions can also strengthen both the Larmor precession and Barnett relaxation over gas randomization (\citealt{Hoang_Lazarian_2008}). This enhancement enables more large grains in dense environments to be aligned with $\B$ and achieve efficient internal alignment.

Recently, \cite{Hoang_2022} and \cite{Hoang+2022} revisited the grain alignment theory of both PM and SPM grains in protostellar environments. They showed that grains larger than $1\mum$ must be SPM to have the magnetic alignment inside protostellar disks and envelopes owing to enhanced Larmor precession caused by iron clusters. Iron inclusions are also necessary to facilitate fast internal relaxation for micron-sized and VLGs at both high and low-\textit{J} attractors through enhanced Barnett relaxation (\citealt{Lazarian_Hoang_2019}). Furthermore, they pointed out that VLGs containing a high amount of iron clusters may achieve perfect magnetic alignment in protostellar envelopes by MRAT mechanism (\citealt{Valentin_2020}, \citealt{Valentin_2023b}).

Given the crucial importance of iron inclusions on grain alignment and resulting dust polarization, \cite{Giang_2022} upgraded the POLARIS code by incorporating grain magnetic properties and detailed grain alignment physics. The significant improvement in this updated version enables us to model both the internal and external alignment of grains consistently based on the level of embedded iron inclusions, instead of fixing the RAT alignment efficiency as in previous studies using the original POLARIS version (\citealt{Brauer_2016}, \citealt{Valdivia_2019}, \citealt{Valentin_2020}, \citealt{Valentin_2023b}). Our recent synthetic modeling of polarized dust emission for the Bok globule in \cite{Giang_2022} using this updated POLARIS code confirmed the numerical predictions from \cite{Hoang+2022}. We found that SPM grains of size $100\mum$ with $10^{3} - 10^{4}$ iron atoms per cluster can be perfectly aligned with $\B$ in the envelope by MRAT mechanism, as implied from \citealt{Valentin_2020} and \citealt{Valentin_2023b} and predicted in \cite{Hoang+2022}. These grains also can produce high $p \sim 10-40\%$, which is consistent with observations of many YSOs by ALMA (\citealt{Cox_2015}, \citealt{Sadavoy_2018}, \citealt{Maury_2018}, \citealt{Kwon_2019}, \citealt{Takahashi_2019}). . On the other hand, we found that inside the disk, PM grains are not aligned with $\B$, while SPM grains above $1\mum$ always have slow internal relaxation at both high and low-\textit{J} attractors. Furthermore, these SPM grains are only aligned with $\B$ by RATs (in contrast to MRAT alignment in the envelope) regardless of the help of iron inclusions. Consequently, we propose that the misalignment of PM grains, or the reduction of the internal and external alignment degree of SPM grains is the major origin causing depolarization in protostellar cores.

However, our previous study \citep{Giang_2022} used the idealized Bok globule model with uniform $\B$ fields and the Bonnor-Ebert gas density distribution. These assumptions may overestimate the impact of iron inclusions, and eliminate the turbulence and geometrical effect of magnetic fields on polarized dust emission. In addition, we found that VLGs tend to have inefficient internal and external alignment due to strong gas randomization. Therefore, in contrast to the positive correlation between grain growth activities and the observed degree of polarization (i.e., larger $a_{\rm max}$ induces larger $p$), found in \cite{Valdivia_2019} and \cite{Valentin_2023b}, we predict grain growth will suppress polarized dust emission. However, this effect was not addressed in \cite{Giang_2022}. Therefore, our following paper aims to comprehensively understand the effects of grain magnetic properties and grain growth on polarized thermal dust emission in the realistic magnetohydrodynamic (MHD) simulation of the protostellar core. In this study, we focus on the protostellar envelope of $\sim 1000$ au and the protostellar disk of $\sim 100$ au. A study of grain alignment and polarized dust emission inside the bipolar outflow will be presented in our upcoming paper (Giang et al. in prep). In addition, since this paper focuses on the relationship between iron inclusions and polarized dust emission, we consider only Barnett relaxation as the major internal alignment mechanism of dust grains. The effects of nuclear relaxation and inelastic relaxation on the net degree of grain alignment (\citealt{Lazarian_Hoang_2019}, \citealt{Hoang+2022}) will be discussed and studied in detail in future work. The structure of our paper is organized as follows. We first describe the MHD datacube in Section \ref{sec:MHD}, our model setups, and how to post-process the MHD simulation with POLARIS in Section \ref{sec:POLARIS}. Results for grain alignment is presented in Section \ref{sec:MCMC+Alignment_results}. The effects of iron inclusions and maximum grain size on the polarization map, the variation of $p(\%)$, polarization angle dispersion function $S$, and the alignment efficiency characterized by quantity $p\times S$ with intensity are shown in Sections \ref{sec:iron_pol_map}, \ref{sec:p_I} and \ref{sec:iron_S_pS}, respectively. The influence of iron inclusions on dichroic extinction at submillimeter wavelengths is studied in Section \ref{sec:extinction}. Further discussions and conclusions of our study are provided in Sections \ref{sec:discuss} and \ref{sec:summary}, respectively.

\begin{figure*}
\centering
    \includegraphics[width=\textwidth,height=\textheight,keepaspectratio]{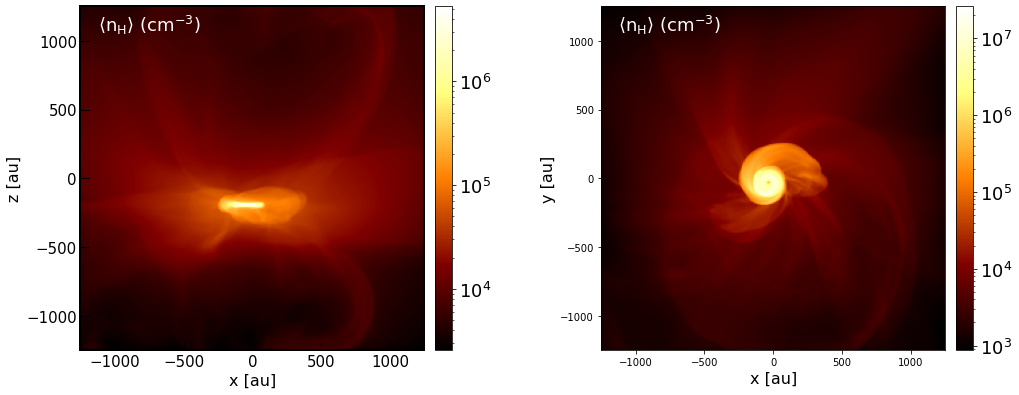}
    \caption{2D map of the mean gas density obtained on the x$-$z plane (left panel) and x$-$y plane (right panel). The material in the envelope is accumulated into the equatorial plane under the contraction from gravity, forming the protostellar disk of radius $\sim 100$ au around the protostar with the mean gas density of $n_{\rm H} \sim 10^{8} - 5\times 10^{9}\cm^{-3}$. The gas density in the envelope is low of $n_{\rm H} \sim 10^{5}-10^{6}\cm^{-3}$.} 
     \label{fig:gas_density}
\end{figure*}

\section{MHD Simulation of Collapsing Cores}\label{sec:MHD}
\subsection{Simulation setup and gas density distribution}

We use a datacube that shows the clearest formation of a protostellar disk from the series of non-ideal MHD simulations of the self-collapsing low-mass core from \cite{Lam_2019} (Model M1.0A10.0, Table 1). The simulation starts with a cubic box of 5000 au, containing a spherical core of total gas mass $M = 0.5M_{\odot}$ and a radius of 2000 au. The core's gas density follows the Bonor-Ebert density profile, and it is assumed to rotate around the z direction with an angular velocity of $\Omega = 6\times10^{-14} \s^{-1}$. Initial magnetic fields are assumed to be uniform along the z direction. The core is supercritical with the dimensionless mass-to-flux ratio of $\lambda \sim 2.6$, and turbulence is transonic with the sonic Mach number $M_{s} = 1$. The ambipolar diffusion coefficient of the simulation is 10 times higher than the standard values of $Q_{\rm A,0} = 95.2\rm g^{1/2}\cm^{-3/2}$, assuming the iron-neutral drag coefficient of $\gamma = 3.5\times10^{13}\cm^{3} \rm g^{-1} \rm s^{-1}$ and the cosmic ray ionization rate of $10^{-17}\rm s^{-1}$ (\citealt{Shu_1992}). The core is assumed to collapse isothermally with the gas temperature of $T_{\rm g} = 10$ K. 

\cite{Lam_2019} used the ATHENA code and the uniform cubic grids of $512^{3}$, which corresponds to the spatial resolution of $\Delta x = 20$ au, to simulate the collapse of the dense core and the formation of the protostellar disk. After a sink particle forms, they rebin the cubic box of $256^{3}$ around the sink particle to the new size of $512^{3}$ to better capture the disk formation with a higher resolution of $\Delta x = 5$ au. 

To perform synthetic observations of dust polarization in both the protostellar core and disk, we take the same snapshot used in \cite{Lam_2021} when the cloud evolves 0.158 Myrs. In this snapshot, the protostellar disk with a radius of $\sim 100$ au is clearly formed around the sink particle whose mass is about $M_{\rm star} = 0.22M_{\odot}$. The mean gas density distribution on the x$-$z and x$-$y plane taken from the snapshot is shown in Figure \ref{fig:gas_density}. One can see the existence of the disk on the equatorial plane (at $z\sim 0$) when looking along the edge-on direction (left panel) and the clear accretion of material into the disk scale of 100 au in the anti-clockwise direction when looking along the face-on direction (right panel).  The gas density inside the disk is about $n_{\rm H} \sim 10^{8}-10^{9}\cm^{-3}$ and it decreases outward to $n_{\rm H} \sim 10^{6}-10^{4}\cm^{-3}$ at $500$ au and $\sim 1000$ au, respectively.

  \subsection{Mean magnetic field orientation}\label{sec:B_fields_MHD}
\begin{figure*}
\centering
    \includegraphics[width=\textwidth,height=\textheight,keepaspectratio]{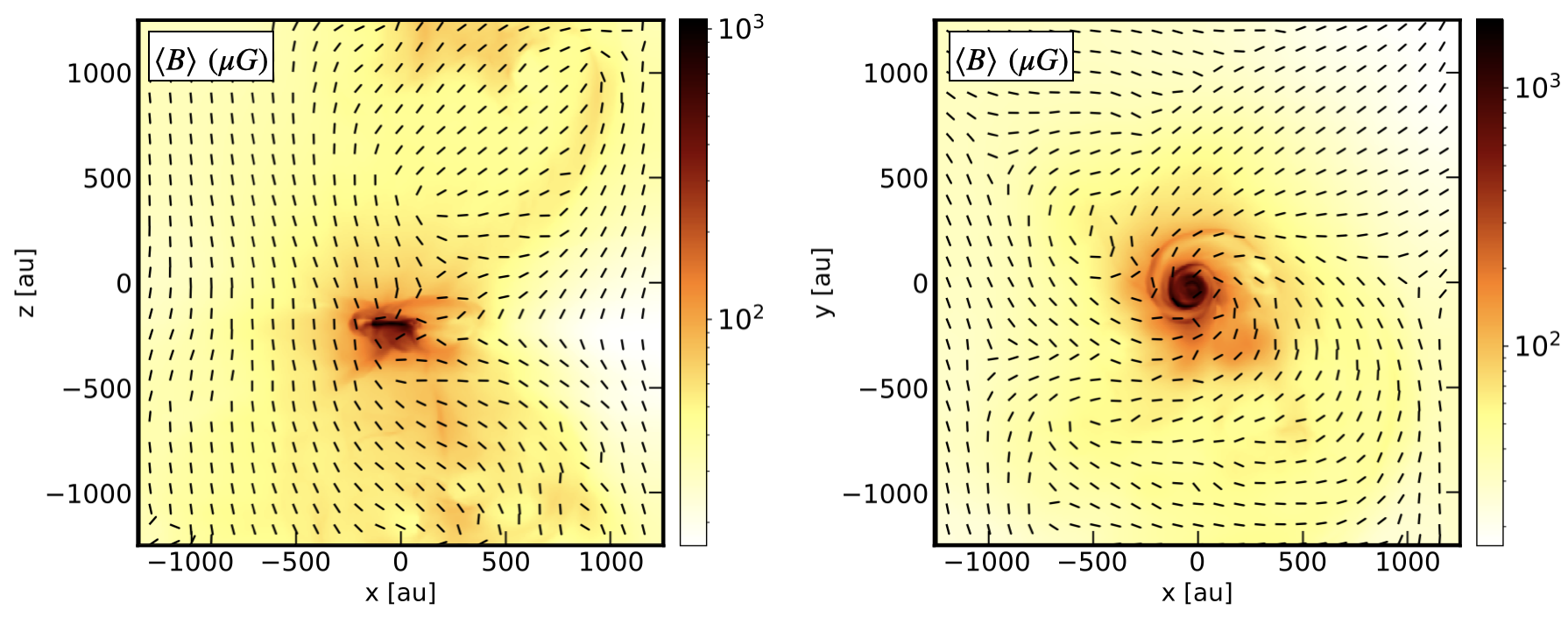}
    \caption{The mean orientation of magnetic fields on the x$-$z (left panel) plane and x$-$y (right panel) (black segments) overplotted over the density-weighted magnetic field strength (color code) integrated along the y and z-direction, respectively. In the x$-$z plane, the magnetic field changes from the hourglass-shaped field along the z direction to the pinch field perpendicular to the outflow in the center induced by the development of the toroidal field from the fast rotation of the disk. On the x$-$y plane, the magnetic field follows the spiral pattern driven by the infall and accretion motion of material from the envelope into the central protostar. The magnetic field strength increases continuously from $B \sim 100\mu G$ in the envelope to $B \sim 5\times 10^{3}\mu G$ in the disk due to the magnetic flux freezing.} 
     \label{fig:magnetic_field}
\end{figure*}

Figure \ref{fig:magnetic_field} shows the mean orientation of magnetic fields integrated along the y- and z-direction overplotted with the density-weighed magnetic field strength $\langle B \rangle$ on x$-$z and x$-$y plane, respectively. The values of $\langle B \rangle$ is given by:
\bea 
\langle B \rangle = \frac{\int_{\rm los} B n_{\rm H} dl}{\int_{\rm los} n_{\rm H} dl},
\label{eq:density-weighted}
\ena 
where $B$ and $n_{\rm H}$ are the magnetic field strength and the gas density in each cell along the observed direction. We weigh magnetic field strength (and later other quantities, Section \ref{sec:MCMC+Alignment_results}) with gas density to better visualize the amplification of $\B$-fields in dense regions due to the magnetic flux freezing effect. For the mean orientation of magnetic fields on the plane of the sky (POS), we use the same approach proposed by \cite{Valdivia_2022} to emphasize the magnetic field structure in dense regions and to later compare with the inferred magnetic fields from dust polarization in Section \ref{sec:recovery_rate}. We first compute the density-weighted sinus and cosines of $2\phi_{\rm B}$, $\langle \sin(2\phi_{\rm B}) \rangle$ and $\langle \cos(2\phi_{\rm B}) \rangle$, along the y- and z-direction by using Equation (\ref{eq:density-weighted}), where $\phi_{\rm B}$ is the angle of projected magnetic field on the POS to the North direction in each cell. The mean angle of $\B$ to the North direction from MHD simulation $\langle \phi_{\rm B} \rangle$ is then given by:
\bea 
\langle \phi_{\rm B} \rangle = \frac{1}{2} \arctan \frac{\langle \sin(2\phi_{\rm B}) \rangle}{\langle \cos(2\phi_{\rm B}) \rangle}.
\ena 

Due to the magnetic flux freezing, the initial uniform magnetic fields along the z-direction are dragged by infall gas toward the center, forming the hourglass-shaped magnetic fields seen in the x$-$z plane (left panel) and the spiral pattern seen in the x$-$y plane (right panel). The magnetic field along the outflow (left panel) is slanted toward the right, and the large-scale hourglass shape magnetic field is bent to be pinched along the disk within $\sim 200$ au due to the formation of the toroidal $\B$ component wrapped by the fast rotation of the disk. The interaction between the outflow and infalling gas develops the small-scale toroidal pattern seen on the x$-$y plane.

\section{Synthetic Polarization Observations of MHD simulations with the updated POLARIS}\label{sec:POLARIS}
We perform synthetic multi-wavelength observations of polarized dust emission from the protostellar core and disk by post-processing the above MHD simulation with the updated POLARIS code introduced by \cite{Giang_2022}. Here, we first introduce our setup for the radiation source and the dust model in Section \ref{sec:source+dust_model}, then we briefly explain our working flow in POLARIS from Sections \ref{sec:MCMC} to \ref{sec:Pol_RT}.

\subsection{Radiation source and dust model}\label{sec:source+dust_model}
We consider the stellar radiation from the sink particle to be the unique heating source in our modeling. The contribution from the interstellar radiation field is neglected due to their inefficient effect on heating dust grains at $\sim 2500$ au scale of the protostellar core. For the stellar radiation source, we assume the sink particle is a black body with the radius of $R_{\rm star} = 2.5R_{\odot}$ and temperature of $T_{\rm star} = 9707$ K, which corresponds to the stellar luminosity $L_{\rm star} = 50L_{\odot}$ (or low-mass protostar). The choice of $L_{\rm star}$ is motivated by the study of \cite{Valentin_2023b} who found that the high $p \sim 20\%$ in the outflow cavity of Class 0/I low-mass YSOs only can be reproduced if the protostar has $L_{\rm star} \sim 50-100L_{\odot}$. We consider the wide radiation spectrum with the lower cutoff at $\lambda_{\rm min} = 0.1\mum$ at which UV photons are mainly absorbed by Hydrogen atoms, and the upper cutoff at $\lambda_{\rm max} = 2$ mm in which the dust-radiation interactions become insignificant. We assume the source emits $N_{\rm photon} = 10^{7}$ photons per each wavelength. 

For the dust model, we assume the uniform distribution of dust grains in the entire protostellar core with the typical gas-to-dust mass ratio $\eta = 0.01$ found in ISM. Dust grains are assumed to be in the composite form with $67.5\%$ of silicate and $32.5\%$ of graphite, and follow the standard Mathis Rumpl Nordsieck (MRN) distribution of $dn/da = n_{\rm H} C a^{-3.5}$ (\citealt{Mathis_1977}) with $C$ the normalization constant determined from the value of $\eta = 0.01$ (see \citealt{Giang_2022} for detail). We assume the composite dust model because sub-micron grains dense regions such as in protostellar environments have higher probability to collide and stick together, forming larger grains (\citealt{Okuzumi},\citealt{Kataoka_2013}). Under this assumption, composite grains containing a mixture of silicate with iron inclusions and carbonaceous material become magnetic and are aligned by the similar alignment physics as silicate grains. Thus, both silicate and carbonaceous material within composite dust grains will produce polarized dust emission. We consider the wide grain size distribution from the minimum size of $a_{\rm min} = 3.5$ \AA $~$ to the maximum size varying from $a_{\rm max} = 1\mum$ to $a_{\rm max} = 100\mum$. The choice of VLGs within the $2500$ au scale stems from the detection of large grains via the SED fitting (see, e.g., \citealt{Kwon_2009,Miotello_2014,Galametz_2019,Liu_2021}) as well as the synthetic observations of dust polarization (see, e.g., \citealt{Valdivia_2019}, \citealt{Valentin_2023b}).

\subsection{Monte-Carlo Radiative Transfer and Dust temperature calculation}\label{sec:MCMC}
The first step of our post-processing is to perform the radiative transfer of photons emitting from the sink particle using the Monte-Carlo technique introduced by \cite{Lucy_1999}. POLARIS simulates the scattering, absorption, and dust emission processes for all dust grains in the MHD simulation box. For each absorption event, they manage the energy conservation between the stellar absorption and thermal dust emission by immediately correcting the grain temperature and sending one lower energy photon into the grid space. The energy density distribution inside each cell is updated continuously when photons enter, interact with dust grains, and leave the cell until the end of the simulation. POLARIS stores the direction of each photon inside the cell to calculate the anisotropic degree $\gamma_{\rm rad}$ and the radiative torques required to model the grain alignment by RATs in the latter step.
 
After the radiative transfer simulation, we calculate the absorption rate for each grain size using the energy density distribution stored in each cell. The grain temperature is determined by solving the energy conservation between radiative heating and cooling of dust grains. The average dust temperature $T_{\rm d}$ inside the cell is calculated by integrating the grain temperature over the grain size distribution, and the gas temperature is given by multiplying $T_{\rm d}$ by a correlation factor. We consider $T_{\rm g} = T_{\rm d}$ in our simulation because gas-grain collision is the main heating source of gas in protostellar environments.

\subsection{Grain alignment physics in the updated POLARIS}\label{sec:Align_POLARIS}
Taking into account the complexity of grain alignment physics in protostellar environments (\citealt{Hoang_2022}, \citealt{Hoang+2022}), we use the updated version of POLARIS introduced by \cite{Giang_2022} to model in detail the alignment of grains with $\B$ in the MHD model of the protostellar core and disk. To understand the role of grain magnetic properties, we consider three types of grains: paramagnetic (PM) grains with the iron fraction $f_{\rm p} = 0.1$, superparamagnetic (SPM) grains with a moderate value of $N_{\rm cl} = 100$ iron atoms/cluster, and SPM grains with the high value of $N_{\rm cl} = 10^{4}$ iron atoms/cluster. We assume the same volume filling factor of iron clusters $\phi_{\rm sp} = 0.1$ for two types of SPM grains corresponding to $\sim 30\%$ of iron abundance locked inside dust grains in the form of iron clusters.
 
Following the RAT theory (\citealt{Lazarian_Hoang_2007}, \citealt{Hoang_Lazarian_2008}),  we first calculate the radiative torques acting on each grain size based on the radiation field obtained from the Monte-Carlo radiative transfer simulation (Section \ref{sec:MCMC}, \citealt{Reissl_2016}, \citealt{Giang_2022}). Then we calculate the maximum rotational rate that grains gained by RATs (\citealt{Hoang_Lazarian_2014}) and select the grain size which rotates at least three times faster than their thermal rotation as the minimum alignment size $a_{\rm align}$ (\citealt{Hoang_Lazarian_2008}). The maximum alignment size $a_{\rm max,JB}^{\rm Lar}$ is determined based on the competition between the Larmor precession and the gas randomization. In detail, grains rotating suprathermally by RATs only be considered to receive $\B$ as their alignment direction if their Larmor precession timescale is at least ten times shorter than the gas damping timescale (\citealt{Yang_2021}, \citealt{Giang_2022}). For grains within the alignment range, their internal alignment is considered to have fast internal relaxation if their Barnett relaxation timescale is shorter than the gas damping timescale, constrained by the range $a_{\rm min,aJ} - a_{\rm max,aJ}$ (\citealt{Hoang_2022}, \citealt{Hoang+2022}). Grains beyond this limit will have inefficient IA by slow internal relaxation. Taking into account the dependence of the Barnett relaxation with the grain rotational rate, we determine in detail the internal alignment of each grain size in their alignment state at both high-\textit{J} and low-\textit{J} attractors (see \citealt{Giang_2022} for detailed calculations).

\begin{figure}
    \includegraphics[width=0.45\textwidth,height=0.45\textheight,keepaspectratio]{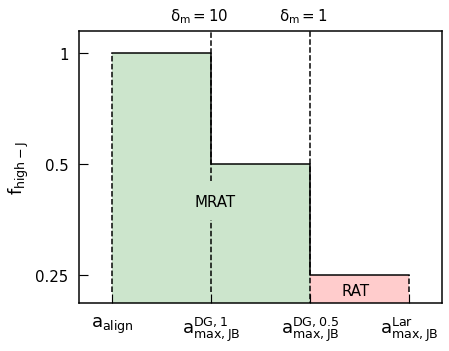}
    \caption{Illustration of the variation of $f_{\rm high-J}$ with grain sizes within the alignment range $[a_{\rm align} - a_{\rm max,JB}^{\rm Lar}]$. The green area determines the size of grains being aligned with $\B$ by the MRAT mechanism. We consider $f_{\rm high-J} = 1$ for grains having high magnetic relaxation ratio $\delta_{\rm m}\geq 10$, and $f_{\rm high-J} = 0.5$ for grains having $1 \geq \delta_{\rm m} \leq 10$. The red area determines the size range for RAT alignment with  $f_{\rm high-J} = 0.25$. The maximum size for grains having $f_{\rm high-J} = 1$ and $f_{\rm high-J} = 0.5$ by MRAT alignment is denoted by $a_{\rm max,JB}^{\rm DG,1}$ and $a_{\rm max,JB}^{\rm DG,0.5}$, respectively.} 
     \label{fig:fhighJ}
\end{figure}

Following the RAT paradigm, grains can have the external alignment with $\B$ by RATs or MRAT alignment depending on their magnetic properties (\citealt{Hoang_Lazarian_2016a}, \citealt{Hoang+2022}). \cite{Giang_2022} determine the external alignment mechanism for each grain size based on the magnetic relaxation parameter $\delta_{\rm m}$, which characterizes the magnetic relaxation efficiency over the gas randomization. As shown in Figure \ref{fig:fhighJ}, MRAT will be the major alignment mechanism for grains having $\delta_{\rm m} > 1$, and RATs will play the main role in driving the alignment of small grains having inefficient magnetic relaxation, i.e., $\delta_{\rm m} < 1$. For RAT alignment, we assume about $\sim 25\%$ of aligned dust grains to be efficiently aligned with $\B$ at high-\textit{J} attractors (or $f_{\rm high-J} = 0.25$). For MRAT alignment, we consider higher $f_{\rm high-J} = 0.5$ if grains have $1 < \delta_{\rm m} < 10$, and $f_{\rm high-J} = 1$ for grains having very efficient magnetic relaxation with $\delta_{\rm max} \geq 10$.

POLARIS determines the overall alignment degree for each grain size with magnetic fields (\citealt{Hoang_Lazarian_2014}) by using the Rayleigh reduction factor $R$ (\citealt{Greenberg_1968}) given by:

\bea 
R = f_{\rm high-J} Q_{\rm X}^{\rm high-J} Q_{\rm J}^{\rm high-J} + (1-f_{\rm high-J})Q_{\rm x}^{\rm low-J}Q_{\rm J}^{\rm low-J}, 
\label{eq:R}
\ena 

where $Q_{\rm X}$ and $Q_{\rm J}$  characterize the internal and external alignment degree. 

For grains having fast internal relaxation, i.e., grains within $a_{\rm min,aJ}-a_{\rm max,aJ}$, their internal alignment is perfect if grains aligning with $\B$ at high-\textit{J} attractors, but will be imperfect for grains at low-\textit{J} attractors due to their internal thermal fluctuation (\citealt{Purcell_1979})). We consider $Q_{\rm X}^{\rm high-J} = 1$ for the former case and describe $Q_{\rm X}^{\rm low-J}$ for the latter case by the local thermal equilibrium (TE) Boltzmann distribution (\citealt{Lazarian_Roberge_1997}). 

For grains having slow internal relaxation, i.e., grains beyond the range $a_{\rm min,aJ}-a_{\rm max,aJ}$, they still can have right IA if they are aligned with $\B$ at high-\textit{J} attractors, but grains can have right IA or wrong IA (i.e., grains rotate around their longest axis and thus the grain longest axis will be aligned parallel with the magnetic field direction) if they have the magnetic alignment at low-\textit{J} attractors (\citealt{Hoang_Lazarian_2009}). We characterize the right IA by using the positive value of $Q_{\rm X}^{\rm high-J}$ for grains at high-\textit{J} and positive $Q_{\rm X}^{\rm low-J}$ for grains at low-\textit{J}, and characterize the wrong IA by using the negative $Q_{\rm X}^{\rm low-J}$. 

The value of $Q_{\rm X}$ for grains having slow internal relaxation is controlled by hand in POLARIS due to the difficulty of studying the dynamic of grains without internal relaxation (\citealt{Hoang_Lazarian_2009}). However, we expect this value should be close to zero due to their weak internal alignment degree. We use the similar setup in \cite{Giang_2022} with $Q_{\rm X}^{\rm high-J} =0.15$ for grains at high-\textit{J}, and $Q_{\rm X}^{\rm low-J} = 0.05$ for the case of right IA and $Q_{\rm X}^{\rm high-J} = -0.1$ for the case of wrong IA at low-\textit{J} attractors to study the effect of grains with slow internal relaxation on the observed polarization signal, named as model rIA and wIA.

For the external alignment between $\J$ and $\B$, we consider perfect alignment for grains at both high and low-\textit{J} attractors, or $Q_{\rm J}^{\rm high-J} = Q_{\rm J}^{\rm low-J} = 1$. The adoption of $Q_{\rm J}^{\rm high-J} = 1$ for high-\textit{J} state is reasonable due to the grain suprathermal rotation. In terms of $Q_{\rm J}^{\rm low-J} = 1$, this adoption may overestimate the realistic external alignment efficiency of grains at low-\textit{J}. However, since the net polarization signal is dominated by the emission from grains at high-\textit{J}, the overestimated external alignment efficiency of grains at low-\textit{J} may not significantly affect our final results. Therefore, it is acceptable to adopt $Q_{\rm J}^{\rm low-J} = 1$. 

Finally, knowing the internal and external alignment degrees of all grain sizes within the alignment range, we thus can determine the overall alignment degree of grains with $\B$ using Equation (\ref{eq:R}). The Rayleigh reduction factor $R$ will be zero for grains beyond the range of alignment. The summary of parameters used in our model is given in Table \ref{tab:parameter}.

Besides two models of grain alignment rIA and wIA described in the above paragraph, we also consider the ideal case that all grains above $a > a_{\rm align}$ have perfect alignment with $\B$, named as model PA. By comparing model PA with models rIA and wIA, one thus can separate the effect of turbulence and magnetic field geometry from grain alignment efficiency, which helps to clarify the contribution of each factor on the dust polarization. The summary of our model names and their setups is shown in Table \ref{tab:model_table}.

\subsection{Polarized Radiative Transfer of Stokes Parameters}\label{sec:Pol_RT}
Given the alignment degree and alignment direction of all grain sizes with the ambient magnetic field, we finally perform the synthetic observation of dust polarization by solving the polarized radiative transfer of Stokes parameters (\citealt{Reissl_2016}). We put the plane detector with $256\times256$ pixels at 100 pc to the source to observe the envelope of $2500$ au and then the disk of $500$ au scale. Each pixel on the detection will cover the area with a sidelength of $\Delta x = 9.4$ au when observing the full map of 2500au scale and $\Delta x = 1.95$ au when zooming onto the protostellar disk. We perform the multi-wavelength observations with $\lambda = 2$ mm, $870\mum$, $450\mum$, and $250\mum$ with different inclination angles from the face-on to the edge-on direction to understand behaviors of dust polarization at different wavelengths with different light of sight (LOS). The polarization degree in terms of percentage $p(\%)$ in each pixel is given by:
\bea 
p(\%) = \frac{\sqrt{Q^{2} + U^{2}}}{I}\times 100\%,
\ena
and the polarization angle $\psi$ in the unit of radian is:
\bea 
\psi = \frac{1}{2}\arctan\frac{U}{Q}
\ena
where $I$ is the first Stokes parameter describing the thermal intensity, and $Q$ and $U$ describe the linear polarization states of dust emission.

We calculate the polarization angle dispersion function $S$ to study the effect of B-field tangling on the observed polarization map. For each cell at position $x$, we calculate the difference in $\psi$ between this cell and their neighborhoods within the sphere of lag $\delta$ given by \citep{PLanck_2015}: 

\bea 
S(x,\delta) = \sqrt{\frac{1}{N} \sum_{\rm i = 1}^{\rm N} [\psi(x+\delta_{\rm i}) -\psi(x)]^{2} },
\ena 

where $N$ is the total number of cells inside the spherical window. We choose $\delta = 4 \Delta_{\rm x}$ with $\Delta_{\rm x}$ the region covered by one pixel inside the detector plane to understand the effect of B-field tangling in cases of very high resolution.

   \begin{figure*}
\centering
    \includegraphics[width=0.32\textwidth,height=0.45\textheight,keepaspectratio]{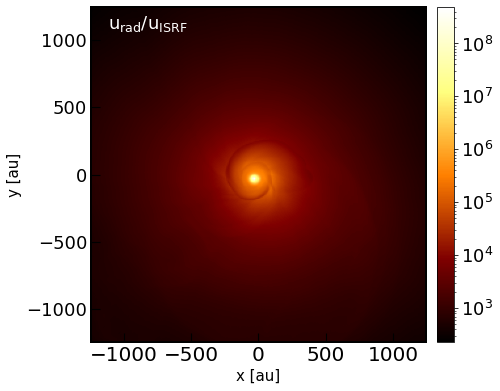}
     \includegraphics[width=0.32\textwidth,height=0.45\textheight,keepaspectratio]{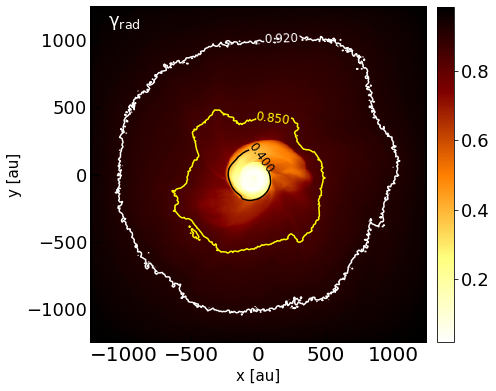}
     \includegraphics[width=0.32\textwidth,height=0.45\textheight,keepaspectratio]{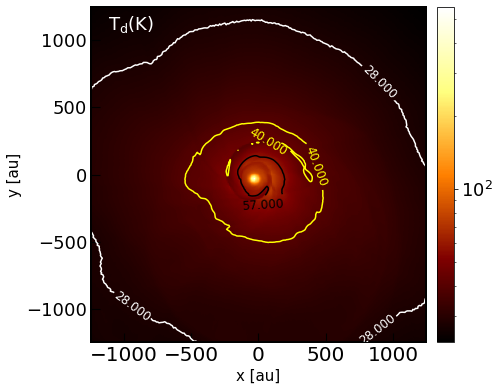}
     
    \caption{Spatial distribution of the density-weighted radiation field strength $U$ (left panel), the mean anisotropic degree $\gamma_{\rm rad}$ (right panel), and the mean dust temperature $T_{\rm d}$ (right panel) on the x$-$y plane. The radiation field strength decreases continuously outward due to the extinction of surrounding dust grains, inducing the decrease of dust temperature from $T_{\rm d} \geq 100 $K around the protostar to $T_{\rm d} \sim 20$ K at 2500 au. The radiation field is highly isotropic with $\gamma_{\rm rad} \sim 0.1 - 0.4$ inside the disk resulting from the strong interaction between dust grains and stellar radiation fields, but it becomes more anisotropic when moving outward, i.e., $\gamma_{\rm rad} \sim 0.9$, due the weak interaction between thermal dust emission at infrared wavelengths with cold dust grains there.} 
     \label{fig:U_gamma_Td}
\end{figure*}

\begin{table}
\centering
\caption{Setup for the radiation field and dust model in POLARIS}
\begin{tabular}{lll}
\hline
Quantity & Symbol & Value \\   
\hline
\multicolumn{3}{c}{\textbf{ Radiation sources}}\\
\hline 
Stellar radius              & $R_{\rm star}$       & $2.5R_{\odot}$ \\ 
Effective temperature       & $T_{\rm star}$       & 9707 K  \\
Stellar luminosity          & $L_{\rm star}$       & $50 L_{\odot}$ \\
 
\hline
\multicolumn{3}{c}{\textbf{Dust model}}\\
\hline
Grain axial ratio                & $s$                    & 0.5\\
Dust-to-gas mass ratio      & $\eta$               & 0.01\\
Grain size distribution           & $\rm dn/da$              & C$ a^{-3.5}$\\
Minimum grain size          & $a_{\rm min}$        & 3.5\AA  \\ 
Maximum grain size          & $a_{\rm max}$        & $1, 5, 10, 20, 50, 100\mu m$  \\
Fraction of silicate        &                      & $67.5\%$  \\
Fraction of graphite        &                      & $32.5\%$  \\

\hline
\multicolumn{3}{c}{\textbf{Grain magnetic properties}}\\
\hline
Iron fraction & $f_{\rm p}$ & 0.1 \\
Iron atom/cluster & $N_{\rm cl}$ & 100 $\&$ $10^{4}$ \\
Volume filling factor  & $\phi_{\rm sp}$ & 0.1 \\
of iron clusters & & \\

 \hline
\multicolumn{3}{c}{\textbf{Internal alignment degree }}\\
\hline
\multicolumn{3}{c}{Grains with fast internal relaxation}\\
\hline
High-$J$ attractors & $Q_{\rm X}^{\rm high-J}$ & 1 \\
Low-$J$ attractors & $Q_{\rm X}^{\rm low-J}$ & TE Boltzmann distribution\\

 \hline
\multicolumn{3}{c}{\textbf{External alignment degree}}\\
\hline
High-$J$ attractors & $Q_{\rm J}^{\rm high-J}$ &  1 \\
Low-$J$ attractors & $Q_{\rm J}^{\rm low-J}$ & 1 \\

  \hline
    \label{tab:parameter}
    \end{tabular}   
\end{table}

 \begin{table*}
  \centering
         \caption{Parameters of the grain alignment model}
  \begin{tabular} {cccccc}
  \hline 
\textbf{Model name} & \textbf{Alignment range} & \textbf{Slow internal relaxation} & $Q_{\rm X}^{\rm high-J}$ & $Q_{\rm X}^{\rm low-J}$ & $f_{\rm high-J}$ \\
  \hline 
PA & $a_{\rm align} - a_{\rm max}$ & No & -- & -- & 1 \\
 
rIA   & $a_{\rm align} - a_{\rm max,JB}^{\rm Lar}$ & Yes & 0.15 & 0.05 & Depend on grain magnetic properties \\ 
  
wIA   & $a_{\rm align} - a_{\rm max,JB}^{\rm Lar}$ & Yes & 0.15 & -0.1 & Depend on grain magnetic properties \\ 
  \hline 

    \label{tab:model_table}
    \end{tabular}\\

    \footnotesize{\textbf{Note}: Value of $Q_{\rm X}^{\rm low-J}$ and $Q_{\rm X}^{\rm high-J}$ here are for the internal alignment degree of grains with slow internal relaxation.}\\
\end{table*}

\section{Numerical Results for Grain Alignment}\label{sec:MCMC+Alignment_results}
In this section, we first show the distribution of the radiation field strength and dust temperature on the x$-$y plane. We then show results about grain alignment with $a_{\rm max} = 100\mum$ for PM grains and SPM grains from Section \ref{sec:align_range} to Section \ref{sec:amax_JB_DG}. Results of grain alignment with different maximum grain sizes from $a_{\rm max} = 1\mum$ to $a_{\rm max} = 100\mum$ are shown in Appendix \ref{sec:grain_alignment_amax}.

\subsection{Radiation field and dust temperature}\label{sec:urad_gamma_Td}
Figure \ref{fig:U_gamma_Td} shows the spatial distribution of the radiation field strength $U$ (left panel), the anisotropic degree $\gamma_{\rm rad}$ (central panel), and the dust temperature $T_{\rm d}$ (right panel) on the x$-$y plane, in which each quantity is weighted with gas density along the z-direction. The radiation field strength is strongest in the center where the protostar forms and decreases outward due to the increasing extinction from surrounding dust grains. The radiation field is highly isotropic, $\gamma_{\rm rad} \leq 0.2$, within the inner 100 au due to the strong scattering between stellar radiation and dust grains inside the dense disk. Then it will become highly anisotropic when moving outward, i.e., $\gamma_{\rm rad} \sim 0.8$, due to weak interactions of thermal emission from warm grains in the center with $ T_{\rm d} \geq 100$ K with colder grains in the envelope with $T_{\rm d}  \sim 40$ K. 
 \begin{figure*}
\centering
    \includegraphics[width=\textwidth,height=\textheight,keepaspectratio]{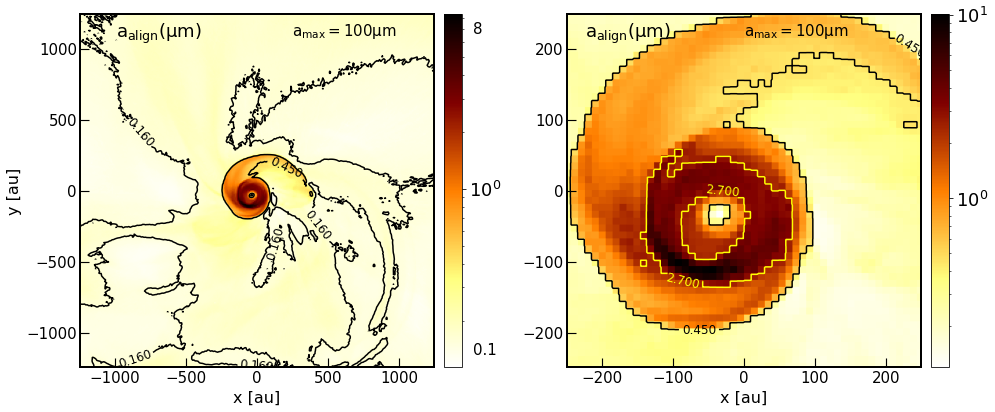}
     
    \caption{The density-weighted minimum alignment size $a_{\rm align}$ on the x$-$y plane obtained in the full 2500au scale (left panel) and in the inner 500 au region (right panel), assuming $a_{\rm max} = 100\mum$. The contour line shows the boundary of $a_{\rm align} = 0.16\mum$ in the envelope, $0.45\mum$ in the inner part of the envelope, and $2.7\mum$ found in the disk. Sub-micron size of $a_{\rm align} \sim 0.2\mum$ around the protostar can be aligned with $\B$ due to the efficient RATs in the strong stellar radiation field. However, this value increases quickly to $a_{\rm align} \sim 8\mum$ inside the disk due to the inefficient RATs induced by the high gas damping, then it decreases again to $a_{\rm align} \sim 0.7\mum$ in the envelope because of the reduced gas density there. } 
     \label{fig:align}
\end{figure*}

\subsection{Grain Alignment Size Range}\label{sec:align_range}
 
 \begin{figure*}
\centering
    \includegraphics[width=\textwidth,height=\textheight,keepaspectratio]{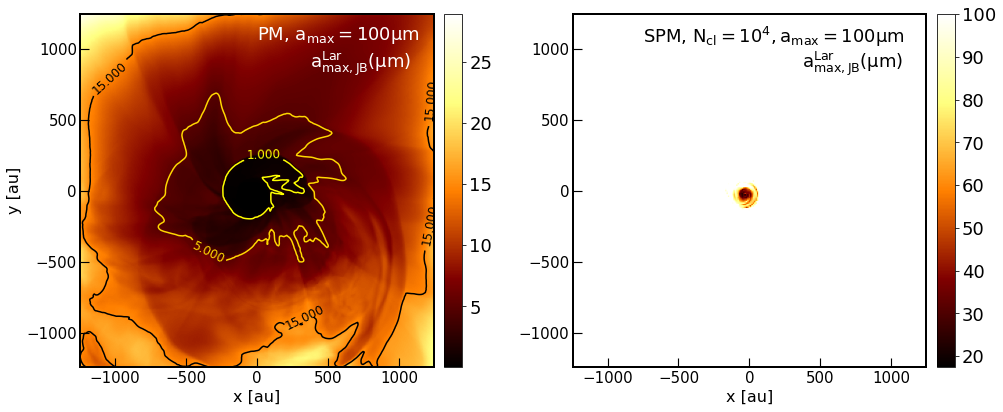}

    \caption{Distribution of the maximum alignment size $a_{\rm max,JB}^{\rm Lar}$ weighed with gas density on the x$-$y plane for PM grains (left panel) and SPM grains with $N_{\rm cl} = 10^{4}$ (right panel). The maximum alignment size generally reduces toward the center due to the significant increase in gas randomization. Large PM grains of $a > 10\mum$ basically cannot be aligned with $\B$ in the entire protostellar core due to the weak Larmor precession, but this problem can come over if grains are SPM with a high amount of iron inclusions. } 
     \label{fig:amaxJB}
\end{figure*}

Figure \ref{fig:align} shows the map of the weight-density minimum alignment size $a_{\rm align}$ on the x$-$y plane obtained in the full 2500 au scale (left panel) and in the inner 500 au region (right panel). Around the protostar, sub-micron grains of $a_{\rm align} \sim 0.2\mum$ (right panel) can be aligned with $\B$ due to the efficient RATs in the strong stellar radiation field. The alignment size then increases quickly to $a_{\rm align} \sim 1 - 8\mum$ inside the disk (right panel) and decreases again to $a_{\rm align} \sim 0.1-0.8\mum$ in the envelope scale (left panel). The reduced RAT efficiency in the disk is caused by the attenuation of the radiation field strength, the increase in the field isotropic degree, and the enhanced gas damping by the high gas density. In the protostellar envelope, more sub-micron grains can rotate suprathermally and be able to have the magnetic alignment by RATs due to the increased anisotropic degree of the radiation field and decreased gas damping. 

Figure \ref{fig:amaxJB} shows the density-weight maximum alignment size $a_{\rm max,JB}^{\rm Lar}$ calculated on the x$-$y plane for PM grains (left panel) and SPM grains with $N_{\rm cl} = 10^{4}$ (right panel). For both PM and SPM grains, one can clearly see that large grains inside the disk are hard to be aligned with $\B$ than the other due to the strong gas randomization on the Larmor precession. However, grains with higher embedded iron inclusions have higher possibilities to align with $\B$ due to the enhanced Larmor precession by their larger magnetic susceptibility. For example, PM grains are not aligned with $\B$ in the disk, i.e., $a_{\rm max,JB} < a_{\rm align}$ (Figure \ref{fig:align}), and only grains below $10\mum$ can have the magnetic alignment in the envelope. In contrast, all SPM grains with $N_{\rm cl} = 10^{4}$ up to $a_{\rm max} = 100\mum$ beyond the disk can be aligned with $\B$. Inside the inner 100 au, grains up to $30\mum$ can have the magnetic alignment, but larger grains have random orientation because of their slow Larmor precession driven by their higher inertia moment.

\subsection{Grain sizes with fast internal relaxation} \label{sec:aaJ}
 \begin{figure*}
\centering
    \includegraphics[width=\textwidth,height=\textheight,keepaspectratio]{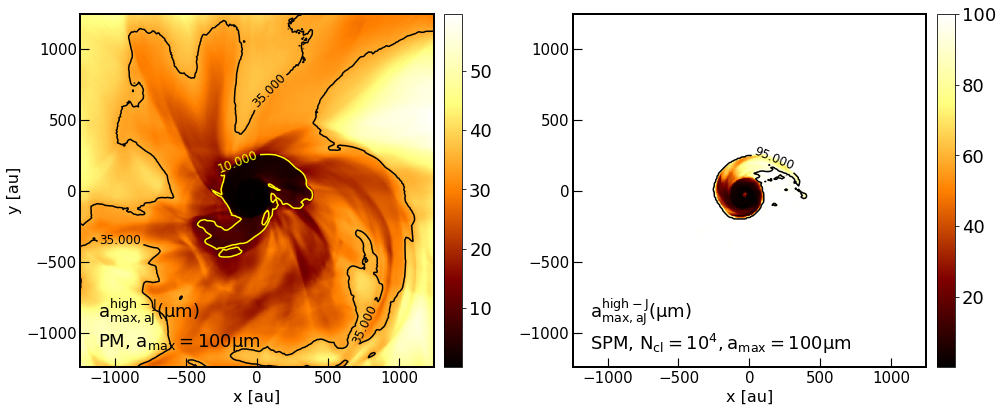}
     
    \caption{Spatial distribution of the density-weighed maximum size that grains have fast internal relaxation at high-\textit{J} attractors $a_{\rm max,aJ}^{\rm high-J}$ for PM grains (left panel) and SPM grains with $N_{\rm cl} = 10^{4}$ (right panel) on x$-$y plane. Generally, large grains inside the disk tend to have slow internal relaxation, i.e., smaller $a_{\rm max,aJ}^{\rm high-J}$, due to the strong gas randomization. The size range of grains having fast internal relaxation is extended to larger sizes if grains have higher magnetic susceptibility. However, it is insufficient to cause large SPM grains above $20\mum$ inside the disk to have fast internal relaxation. } 
     \label{fig:aaj_highJ}
 
\centering
    \includegraphics[width=\textwidth,height=\textheight,keepaspectratio]{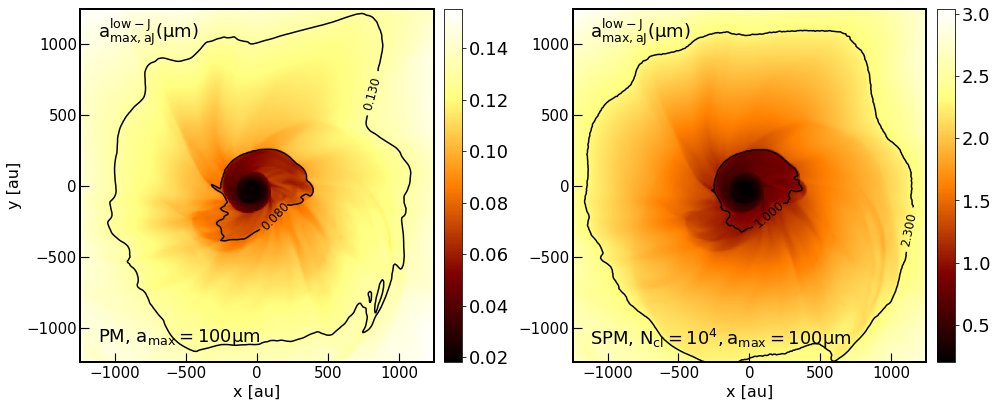}
 
    \caption{ Similar results as Figure \ref{fig:aaj_highJ} but for grains aligning with $\B$ at low-\textit{J} attractors $a_{\rm max,aJ}^{\rm low-J}$. The size range of grains having fast internal relaxation reduces toward the center region and can extend to larger sizes with increasing the amount of iron clusters locked inside dust grains. However, all micron-sized grains above $1\mum$ tend to have slow internal relaxation at low-\textit{J} attractors in the entire protostellar core regardless of the grain magnetic properties due to their slow rotation in space. } 
     \label{fig:aaj_lowJ}
\end{figure*}

Figures \ref{fig:aaj_highJ} and \ref{fig:aaj_lowJ} show the density-weighted maximum size that grains have fast internal relaxation for PM grains (left panel) and SPM grains (right panel). The first figure is for grains at high-\textit{J} attractors $a_{\rm max,aJ}^{\rm high-J}$ and the second is for grains at low-\textit{J} attractors $a_{\rm max,aJ}^{\rm low-J}$. One can see that in both high and low-\textit{J} attractors, large grains in the protostellar disk are hard to have fast internal relaxation compared to grains in the envelope due to the stronger gas randomization during their internal alignment stage. The maximum size for fast internal relaxation increases for grains having higher embedded iron inclusions due to the enhanced Barnett relaxation by higher magnetic susceptibility. Furthermore, this size significantly increases if grains are aligned with $\B$ at high-\textit{J} attractors. For example, beyond $\sim 200$ au for SPM grains with $N_{\rm cl} = 10^{4}$, only small grains below $1-3\mum$ can have fast internal relaxation at low-\textit{J} attractors (Figure \ref{fig:aaj_lowJ}, right panel), while this value can excess to $100\mum$ for grains at high-\textit{J} attractors (Figure \ref{fig:aaj_highJ}, right panel). The huge difference in the value of $a_{\rm max,aJ}$ between high and low-\textit{J} is caused by the difference in the rotation rate. In particular, grains rotating suprathermally experience faster Barnett relaxation and thus have higher possibilities of having fast internal relaxation in dense environments. That also explains why we see the slight increase of $a_{\rm max,aJ}^{\rm high-J}$ toward the strong stellar radiation source in the case of high-\textit{J} attractors (Figure \ref{fig:aaj_highJ}).

 \begin{figure*}
\centering
    \includegraphics[width=\textwidth,height=\textheight,keepaspectratio]{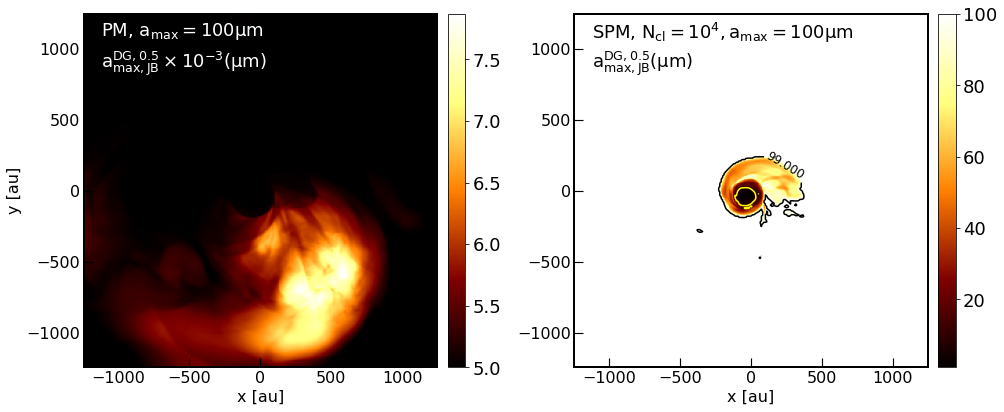}

    \caption{Spatial distribution of the maximum size that $50\%$ of grains can be aligned with $\B$ by MRAT alignment $a_{\rm max,JB}^{\rm DG,0.5}$ for PM grains (left panel) and SPM grains with $N_{\rm cl} =10^{4}$ (right panel). Visually, PM grains are aligned with $\B$ by RATs due to the weak magnetic relaxation, i.e., $a_{\rm max,JB}^{\rm DG,0.5} \sim 0.005\mum$, but SPM grains can be aligned via the MRAT alignment due to the enhanced magnetic relaxation by iron inclusions. However, the perfect alignment of VLGs cannot happen in the disk even if grains contain a high amount of iron clusters due to the high gas randomization there.} 
     \label{fig:adg_1}
\end{figure*}
 
\begin{figure*}
\centering
    \includegraphics[width=\textwidth,height=\textheight,keepaspectratio]{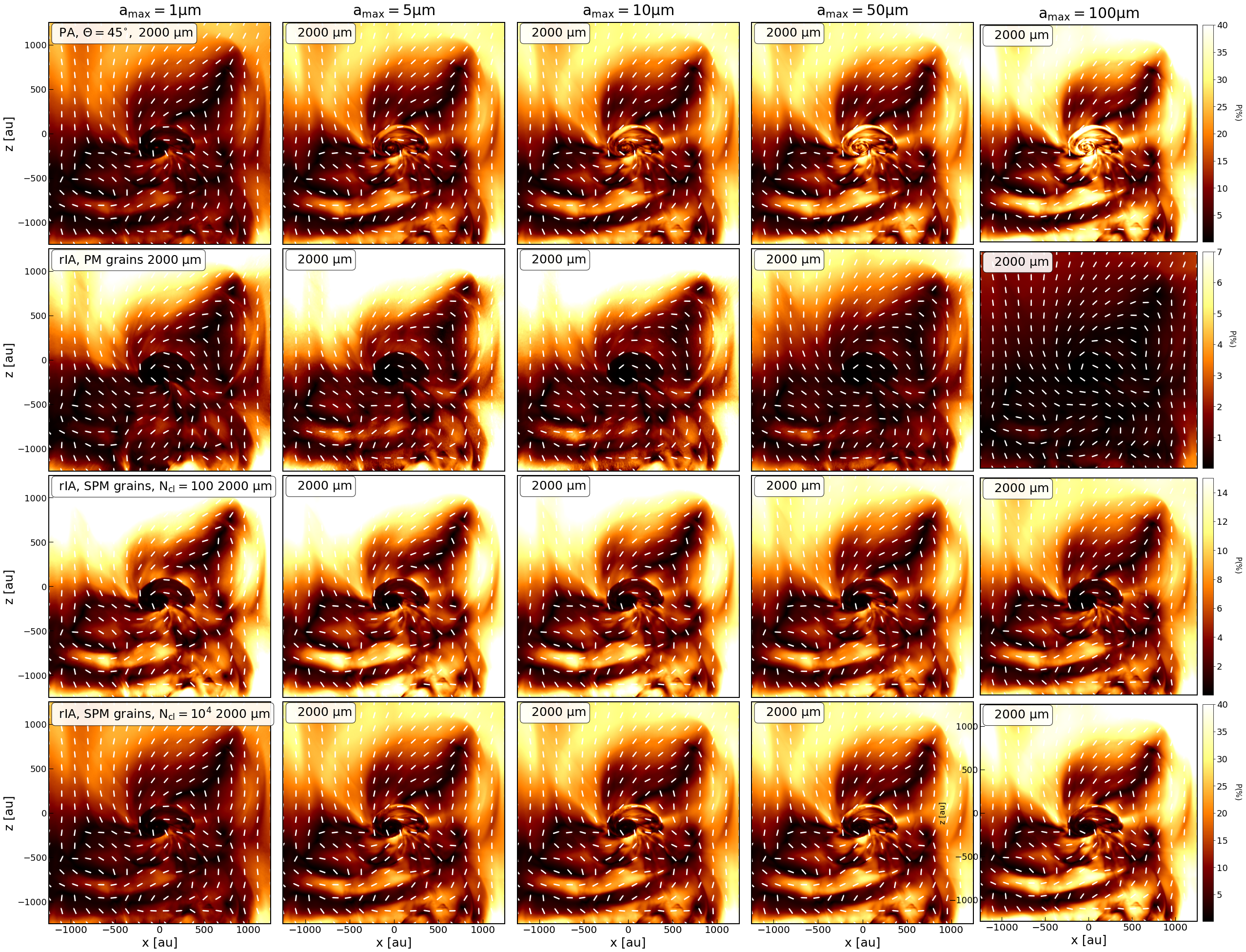}
    \caption{The inferred magnetic field map obtained in the entire protostellar core from dust polarization at 2mm with the inclination angle of $45^{\circ}$. The color code shows the polarization degree $p(\%)$ and white segments show magnetic field orientation obtained by rotating the polarization vector $\P$ by $90^{\circ}$. The first row shows results for model PA, while the second to fourth rows show results for model rIA of PM grains and SPM grains with $N_{\rm cl} = 100$ and $N_{\rm cl} = 10^{4}$, respectively. With each dust model, we show results with different maximum grain sizes from $a_{\rm max} = 1\mum$ in the left to $a_{\rm max} = 100\mum$ in the right. Dust polarization from both model PA and rIA reveals the spiral magnetic field pattern driven by the gas infalling motion inside the rotational-collapsing core. However, in terms of polarization degree, model rIA for PM grains reveals much lower $p(\%)$ compared with model PA and model rIA for SPM grains due to their weak alignment with $\B$. Besides, for grains with low iron inclusions, their polarization degree will decrease with increasing $a_{\rm max}$ owing to the increased amount of grains with inefficient alignment. But for grains are SPM with $N_{\rm cl} = 10^{4}$, $p(\%)$ can increase with $a_{\rm max}$ as model PA due to the extension of efficient alignment range to larger sizes.} 
     \label{fig:pol_map_2mm_rIA}
\end{figure*}

\subsection{Effective grain sizes of MRAT alignment}\label{sec:amax_JB_DG}
Figure \ref{fig:adg_1} shows the density-weighed maximum size that $50\%$ of grains will be aligned with $\B$ at high-\textit{J} attractors by MRAT alignment, $a_{\rm max,JB}^{\rm DG,0.5}$. The left panel shows the result obtained for PM grains and the right panel if for SPM grains with $N_{\rm cl} = 10^{4}$. Obviously, RAT is the major mechanism driving the alignment of PM grains in protostellar environments, i.e., $a_{\rm max, JB}^{\rm DG, 0.5} << a_{\rm align}$ (Figure \ref{fig:align}), owing to their weak magnetic relaxation strength. On the contrary, all SPM grains with $N_{\rm cl}=10^{4}$ can be efficiently aligned with $\B$ by MRAT alignment in the envelope as a result of the enhanced magnetic relaxation by iron inclusions. However, within the inner $\sim 200$ au region,  MRAT alignment is only the major alignment mechanism for micron-sized grains below $\sim 20\mum$. For larger grains, RATs play a main role in driving the magnetic alignment due to the reduced magnetic relaxation driven by the higher inertia moment of VLGs.

\section{Effects of Iron Inclusions and Grain Growth on Synthetic Polarization Map}\label{sec:iron_pol_map}
We next move to analyze the effect of iron inclusions and maximum grain size on the inferred magnetic field map from dust polarization. We place the detector at the inclination angle $\Theta = 45^{\circ}$ (to the North direction) and observe the object at an optically thin wavelength $\lambda = 2$ mm. We first analyze results from the full-scale observation of 2500 au in Section \ref{sec:pol_map_envelope}, then from the zoom-in scale of 500 au around the protostar in Section \ref{sec:pol_map_disk}.

\begin{figure*}
\centering
    \includegraphics[width=\textwidth,height=\textheight,keepaspectratio]{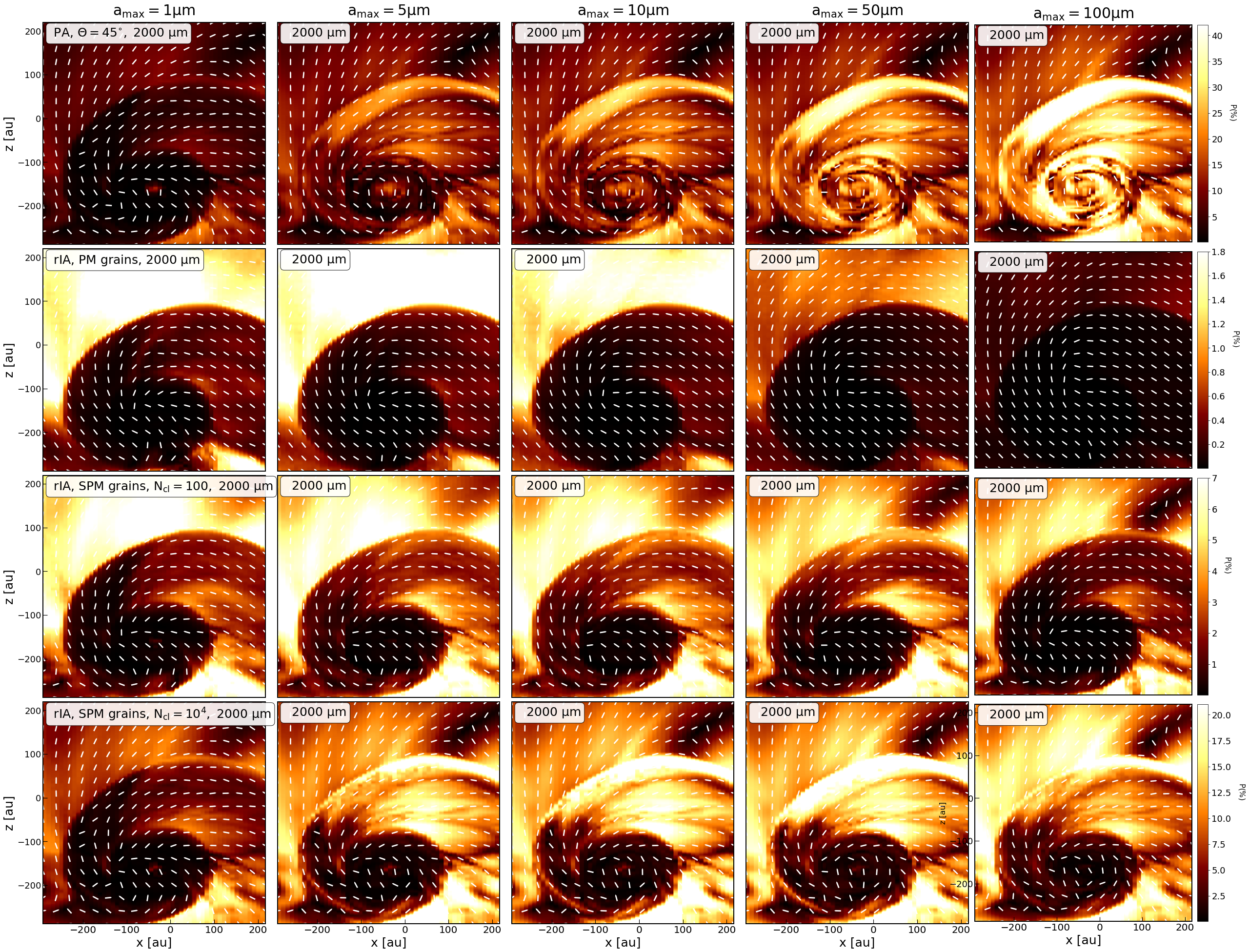}
    \caption{Effect of maximum grain size on the dependence of the polarization degree map and inferred magnetic field orientation from dust polarization obtained in the inner 500 au region from model PA (upper panels) and rIA for PM and SPM grains (second to fourth rows). In model PA, for $a_{\rm max} \geq 5\mum$, dust polarization reveals the complex magnetic field map with $\B$ vectors change from the spiral pattern to be concentric along the disk minor axis due to the projection effect of $\B$-fields on the POS. In model rIA, only SPM grains with $N_{\rm cl} = 10^{4}$ can reveal a similar magnetic field map as model PA because only they can be aligned with $\B$ in the disk. Grains with lower levels of iron inclusions do not reveal the magnetic field information within 200 au around the protostar because they cannot be aligned with $\B$ there. } 
     \label{fig:pol_map_2mm_disk_rIA}
\end{figure*}
\begin{figure*}
\centering
    \includegraphics[width=\textwidth,height=\textheight,keepaspectratio]{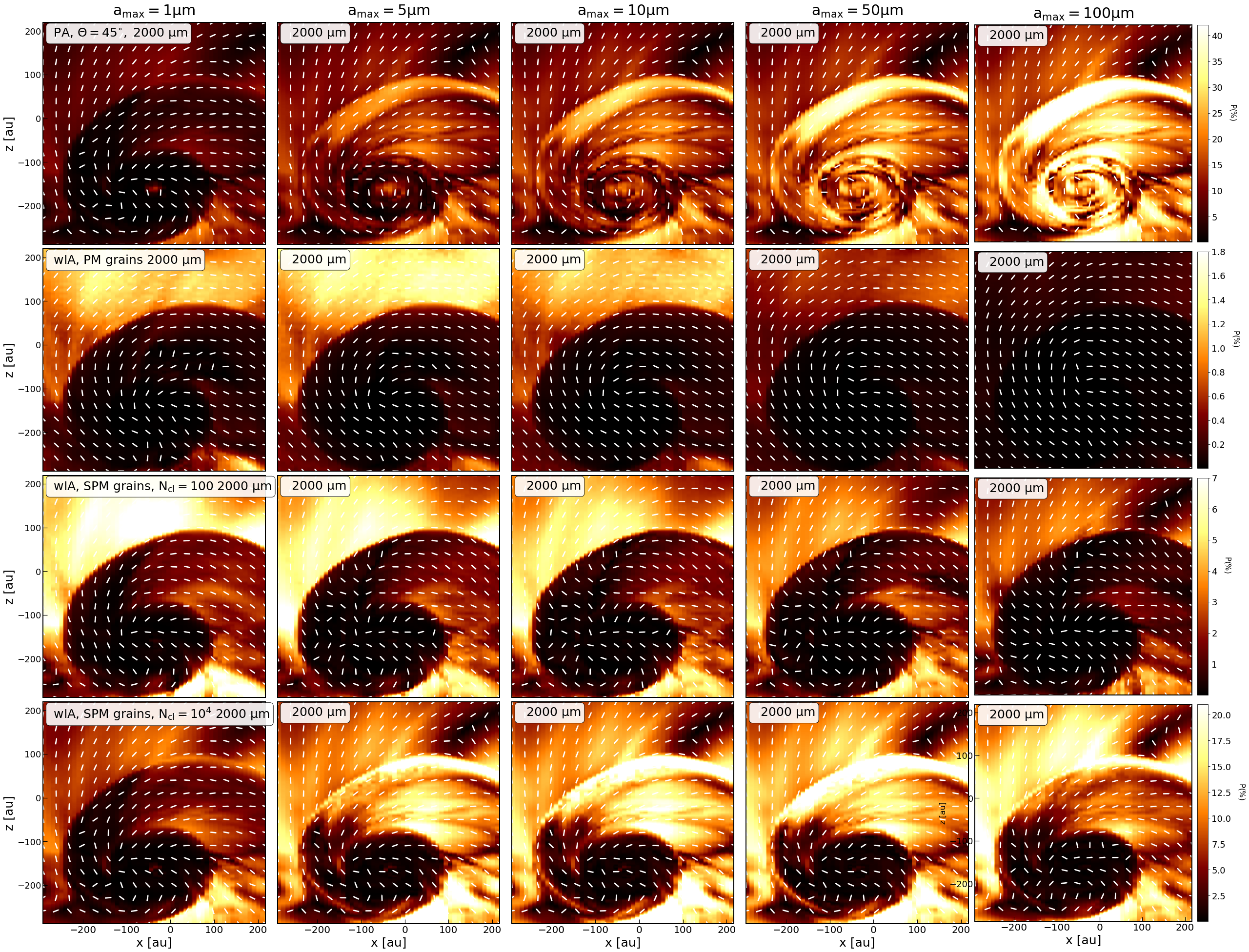}
    \caption{Same as Figure \ref{fig:pol_map_2mm_disk_rIA} but for model wIA. Similar to model rIA, only SPM grains with $N_{\rm cl} = 10^{4}$ can reveal the complex orientation of magnetic fields driven by the gas dynamic in the central 500 au region. However, their inferred magnetic field within 200 au is pinched along the disk major axis, which is $90^{\circ}$ difference from one obtained in model rIA and model PA (Figure \ref{fig:pol_map_2mm_disk_rIA}). The wrong inferred magnetic fields in the disk from model wIA is caused by the wrong interpretation of polarized dust emission emitting from wrong aligned dust grains wholes polarization vectors already tell the magnetic field direction, i.e., $\P \| \B$. }
     \label{fig:pol_map_2mm_disk_wIA}
\end{figure*}

\subsection{Protostellar envelope}\label{sec:pol_map_envelope}
Figure \ref{fig:pol_map_2mm_rIA} shows the effects of iron inclusions and maximum grain size on the polarization degree and the inferred magnetic field direction obtained in the entire protostellar core. In each column, we fix the maximum grain size and show the comparison between model PA (upper row) and model rIA for PM grains (second row), SPM grains with $N_{\rm cl} = 100$ (third row), and SPM grains with $N_{\rm cl} =10^{4}$ (fourth column). For each dust model, we show results for different maximum grain sizes of $a_{\rm max} = 1\mum$, $5\mum$, $10\mum$, $50\mum$, and $a_{\rm max} = 100\mum$, from left to right, respectively. The color code shows the polarization degree in the unit of percentage $p(\%)$, and white segments show the inferred magnetic field direction obtained by rotating the polarization vector $\P$ by $90^{\circ}$. 

 \subsubsection{Effects of grain magnetic properties}
For $a_{\rm max} = 1\mum$ (first column), dust polarization from model PA reveals the spiral magnetic field pattern driven by the gas infall inside the rotating collapsing core. The map of polarization degree also shows the spiral structure with $p \sim 5\%$ inside the arm along the South-West and North-East direction and $p \sim 10-15\%$ in the remaining parts. The complex map of polarization degree in the protostellar envelope is a result of the projection effect of hourglass-shaped magnetic fields on the POS with the inclination angle $\Theta = 45^{\circ}$.

In model rIA (second to fourth panels, first column), polarized dust emission from both PM and SPM grains generally can infer again the spiral $\B$-fields pattern because all grains up to $1\mum$ can be aligned with $\B$ in the envelope scale (Figure \ref{fig:alar_amax}, upper left panel). However, PM grains produce the uniform low $p \sim 1-5\%$ in the entire envelope because most of them have inefficient magnetic alignment, i.e., aligned dust grains above $0.5\mum$ have inefficient IA by slow internal relaxation and only $25\%$ of them will be aligned with $\B$ at high-\textit{J} attractors by RATs (Figures \ref{fig:aaj_highJ_amax}, \ref{fig:aaj_lowJ_amax}, \ref{fig:adg_0.5_amax}, upper left panel). In contrast, SPM grains can produce higher $p \geq 10\%$ and reflect well the spiral pattern in the polarization degree map as model PA owing to their efficient magnetic alignment by MRAT alignment (Figures \ref{fig:aaj_highJ_amax}, \ref{fig:aaj_lowJ_amax}, and \ref{fig:adg_0.5_amax}, center and lower left panels).

 \subsubsection{Effects of maximum grain size}
In model PA (first row), one can see the clear increase of the polarization fraction from $p \sim 10-15\%$ to $p \sim 15-40\%$ when the maximum grain size increases from $a_{\rm max} = 1\mum$ to $a_{\rm max} =100\mum$. The positive correlation between $p$ and $a_{\rm max}$ for model PA is owed to the extension of the alignment range to larger sizes. However, when we take into account the realistic alignment of PM grains (second row), the rise of $p$ with $a_{\rm max}$ only happens when grains grow from $a_{\rm max} = 1\mum$ to $a_{\rm max} = 10\mum$. The further growth in grain size, in contrast, weakens the observed polarization fraction due to the misalignment of VLGs above $10\mum$ with magnetic fields in the envelope (Figure \ref{fig:alar_amax}, upper panels). For SPM grains with low $N_{\rm cl} = 100$, $p$ is higher than the case of PM grains because VLGs can be aligned with $\B$ and achieve efficient IA at high-\textit{J} attractors (Figures \ref{fig:alar_amax} and \ref{fig:aaj_highJ_amax}, central row), but the dependence of $p$ on $a_{\rm max}$ is still the same as the case of PM grains because VLGs cannot achieve efficient external alignment by MRAT alignment as smaller grains (Figures \ref{fig:adg_0.5_amax} and \ref{fig:adg_1_amax}, central row). For SPM grains with high $N_{\rm cl} = 10^{4}$, one can obtain the continuous increase of $p$ with $a_{\rm max}$ as model PA because all grains above $a_{\rm align}$ can achieve perfect magnetic alignment by efficient MRAT alignment there (Figure \ref{fig:aaj_highJ_amax}, \ref{fig:adg_0.5_amax}, \ref{fig:adg_1_amax}, lower panels).

The effect of iron inclusions and maximum grain size on the inferred magnetic fields within thousands au scale from model wIA are similar to model rIA (see the map in Appendix \ref{sec:pol_envelope_wIA}). It is because polarized dust emission from wrong-aligned dust grains in the envelope is subdominant to polarized dust emission from grains with the right IA there.

\begin{figure*}
\centering
    \includegraphics[width=\textwidth,height=\textheight,keepaspectratio]{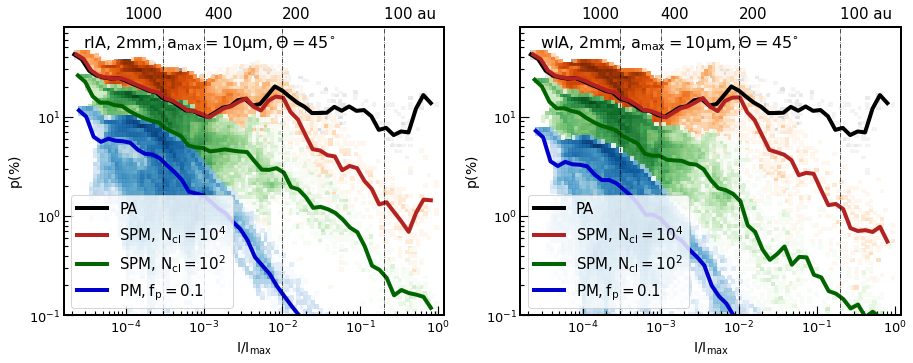}
 
    \caption{Left panel: variation of the mean polarization degree $p(\%)$ with normalized intensity $I/I_{\rm max}$ at 2mm for model PA (black line) and model rIA with different magnetic properties of grains, assuming $a_{\rm max} = 10\mum$ and $\Theta = 45^{\circ}$. Right panel: similar results as the left panel but for results from model wIA. The rough position of $1000$, $400$, $200$, and $100$ au to the protostar is marked on the upper x-axis. In general, the polarization degree obtained from all models tends to decrease with increasing intensity toward the central region. However, for model PA, the depolarization effect is only caused by the narrower of alignment range toward the disk (Figure \ref{fig:align}). For model rIA and wIA, the reduction of $p$ with $I/I_{\rm max}$ is faster than model PA due to the reduced grain alignment efficiency in the central region. The value of $p(\%)$ is smaller and the depolarization will appear more prominent with decreasing amount of iron locked inside dust grains. In addition, model wIA produces slightly lower $p(\%)$ compared with model rIA due to the additional suppression of polarized dust emission arising from the coexistence of grains with the right and wrong IA.}
     \label{fig:P_I_Ncl}
 
\centering
    \includegraphics[width=\textwidth,height=\textheight,keepaspectratio]{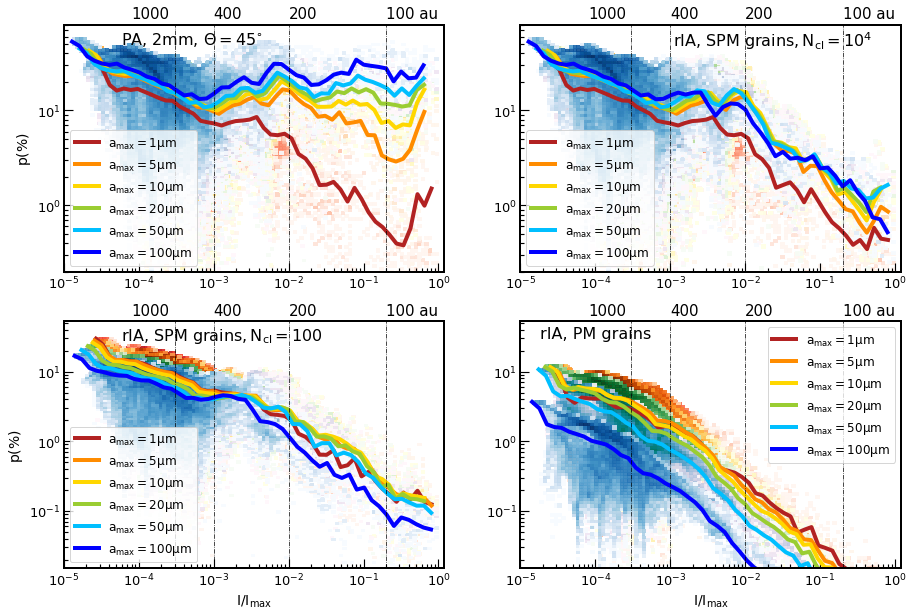}
    \caption{Effect of maximum grain size on the variation of $p(2\rm mm)(\%)$ with $I/I_{\rm max}$, assuming $\Theta = 45^{\circ}$. The upper left panel shows results for model PA, the upper right panel is for model rIA with SPM grains having high $N_{\rm cl} = 10^{4}$, the lower left panel is for SPM grains with low $N_{\rm cl} = 100$, and the lower right panel is for PM grains. In model PA, $p$ increases with increasing $a_{\rm max}$ due to the extension of the alignment range. However, in model rIA, the grain growth in dense environments suppresses the observed degree of polarization due to the extension of the amount of large grains with inefficient magnetic alignment. The decrease of $p$ with increasing $a_{\rm max}$ appears clearer for grains containing lower levels of iron inclusions. } 
     \label{fig:P_I_amax}
\end{figure*}

\subsection{Protostellar disk} \label{sec:pol_map_disk}
\subsubsection{Model rIA}
 Figure \ref{fig:pol_map_2mm_disk_rIA} shows similar results as Figure \ref{fig:pol_map_2mm_rIA}, but zooming in the central region of 500 au, which reveals dust polarization in the protostellar disk. In model PA with $a_{\rm max} = 1\mum$, the inferred magnetic field from dust polarization continues to follow the spiral pattern driven by the accretion of material from the inner part of the envelope onto the disk. The obtained polarization fraction is quite low of $p < 5\%$ because of the narrow alignment range ($a_{\rm align} \sim 0.95\mum$, Figure \ref{fig:align_amax}, right panel). When grains grow to above $a_{\rm max} \geq 5\mum$, $p$ increases due to the extension of the alignment range to large sizes, from $p \sim 15\%$ for $a_{\rm max} = 5\mum$ to $p \sim 30-35\%$ for $a_{\rm max} = 100\mum$. The inferred $\B$-fields are similar to the case of $a_{\rm max} = 1\mum$. But moving toward the disk scale, $\B$ vectors change to be pinched along the disk minor axis, which could be explained by the projection effect of $\B$-fields on the POS. The reason why we do not obtain this feature in the case of $a_{\rm max} = 1\mum$ is because we do not have enough tracers inside the disk need to reflect well the change of $\B$-fields around the protostar.
 
Taking into account the realistic model of grain alignment, one can see that for PM (second row) and SPM grains with low $N_{\rm cl} =100$ (third row), the dust polarization from all models of $a_{\rm max}$ only can reflect the spiral $\B$-field pattern beyond $\sim 200$ au. The deviation of $\B$-fields inside the disk is not revealed because of the alignment loss of micron-sized grains within $\sim 200$ au around the protostar (see Figure \ref{fig:align_amax}, right panel and Figure \ref{fig:alar_amax}, upper and central rows). The polarization degree obtained in $\sim 500$ au region is low of $p < 3\%$ and $p$ clearly decreases with increasing maximum grain sizes as a result of increasing amount of misaligned dust grains (Figure \ref{fig:alar_amax}) and grains with inefficient IA by slow internal relaxation (Figures \ref{fig:aaj_highJ_amax}, \ref{fig:aaj_lowJ_amax}, and \ref{fig:adg_0.5_amax} ).

For SPM grains with high $N_{\rm cl} = 10^{4}$, polarized dust emission from all models of $a_{\rm max}$ clearly reveals the change from the spiral $\B$-fields pattern beyond $\sim 200$ au to the pinched field along the disk minor axis as found in model PA. It is because micron-sized grains having high amounts of iron inclusions can have magnetic alignment in the disk scale by their fast Larmor precession (Figure \ref{fig:alar_amax}, third row). However, their polarization degree is lower than model PA, i.e., $p \sim 2.5 - 20\%$, because dust grains are not able to have perfect alignment with $\B$ inside the disk owning to the high gas randomization here (see Appendix \ref{sec:grain_alignment_amax}, Figures \ref{fig:adg_0.5_amax} and \ref{fig:adg_1_amax}). In addition, the increase in maximum grain size from $1\mum$ to $10\mum$ can help to increase $p$ due to the extension of the alignment range. However, beyond $a_{\rm max} = 10\mum$, the further growing of grains will suppress polarized dust emission due to the alignment loss of VLGs and the presence of micron-sized grains with inefficient internal and external alignment with magnetic fields.
  
 \subsubsection{Model wIA} 
Figure \ref{fig:pol_map_2mm_disk_wIA} shows the effect of iron inclusions (from top to bottom), and maximum grain size (from left to right) on the inferred magnetic field morphology in the inner 500 au region for model wIA. Similar to model rIA, only SPM grains with high $N_{\rm cl} = 10^{4}$ can produce the detected polarization degree above $1\%$ in this area and are able to trace the change of magnetic fields inside the protostellar disk as model PA. The grain growth activities inside the inner 500 au region  help to increase $p$ for grains having high amount of iron inclusions but decrease $p$ for grains with low magnetic susceptibility.

However, one can see that for model wIA, the inferred magnetic field from SPM grains with low $N_{\rm cl} = 100$ is more distorted than results found in model rIA, with some areas showing $\B$ vectors being perpendicular to the large-scale spiral pattern, i.e., the edge of the disk in the West direction (third row). Furthermore, magnetic fields within 100 au inferred from dust polarization of SPM grains with high $N_{\rm cl} = 10^{4}$ are pinched along the disk major axis (fourth row), which is $90^{\circ}$ difference from one obtained from model PA and rIA. The change in the inferred magnetic fields in the above areas results from the wrong interpretation of the polarization signal originating from grains with the wrong IA, whose polarization vectors $\P$ already show the magnetic field direction. That explains why rotating $\P$ in these regions by $90^{\circ}$ shows $90^{\circ}$ difference in $\B$ compared with results from model PA and model rIA. The effect of wrong aligned dust grains on dust polarization becomes prominent when grains just grow to above $5\mum$ because micron-sized grains always tend to have slow internal relaxation and be aligned with $\B$ at low-\textit{J} attractors inside the disk owning to the strong gas randomization around the protostar.

\section{Effects of Iron Inclusions and Grain Growth on Synthetic Dust Polarization degree}\label{sec:p_I}
We next move to analyze the variation of the mean polarization degree $p(\%)$ with intensity. The first section is on the effect of iron inclusions on the $p-I/I_{\rm max}$ with $I_{\rm max}$ the maximum intensity of thermal dust emission, the second section is about the effect of maximum grain size on $p-I/I_{\rm max}$ , and the last section is the effect of iron inclusions and grain growth on the slope of $p-I/I_{\rm max}$.

\subsection{Effects of grain magnetic properties} \label{sec:p_I_magnetoc}
The left panel of Figure \ref{fig:P_I_Ncl} shows the variation of the mean polarization degree $p(\%)$ (color lines) with $I/I_{\rm max}$ obtained in the entire protostellar core at $\lambda = 2$ mm for model PA (black line) and model rIA, assuming different grain magnetic properties and $a_{\rm max} = 10\mum$. The boundary of the inner 100 au, 200 au, 400 au, and 1000 au from the protostar are roughly marked in the upper x-axis to distinguish the variation of $p$ with $I/I_{\rm max}$ in the envelope (beyond 400au) and in the disk scale. In model PA, the polarization degree beyond 400 au slightly decreases with increasing intensity, from $\sim 30\%$ to $\sim 10\%$, due to the effect of turbulence and the disorganization of $\B$-fields along the LOS. Moving toward the inner region, $p$ continues to reduce to $p\sim 8\%$ at $\sim 100$ au then increases again to $\sim 20\%$ owing to the narrower of the alignment range in the dense protostellar disk and the enhanced alignment of small grains by efficient RATs near the protostar, respectively (Figures \ref{fig:align}). 

Taking the realistic model of grain alignment (model rIA) into account, one obtains a similar reduction of $p$ with increasing $I/I_{\rm max}$, but the slope of $p-I$ is much steeper compared with model PA due to the significant reduction of the grain alignment efficiency toward the center. The obtained polarization degree depends on the magnetic susceptibility of dust grains. For instance, PM grains (red line) show $p \sim 4\%$ in the envelope because they are only aligned with $\B$ by RATs and almost all of them have inefficient IA by slow internal relaxation. Moving toward the inner region, their polarization degree decreases significantly to negligible values of $p << 0.1\%$ owing to the alignment loss around the protostar (Figures \ref{fig:align_amax}, \ref{fig:alar_amax} and \ref{fig:aaj_highJ_amax}, upper left panels). In contrast, SPM grains with low $N_{\rm cl} = 100$ can produce higher $p \sim 20\%$ in the envelope because of their enhanced internal and external alignment by iron inclusions. But they still produce $p \sim 0.1\%$ in the inner 100 au due to the misalignment of grains above $1\mum$ there. For SPM with higher $N_{\rm cl} = 10^{4}$, MRAT can drive all grains of $a \geq a_{\rm align}$ to be perfectly aligned with $\B$ in the envelope, producing the similar high $p\sim 10-40\%$ as model PA. But within 400 au, $p$ quickly decreases from $\sim 10\%$ to $\sim 1\%$ because of the change of the alignment mechanism from MRAT to RATs and the presence of VLGs with slow internal relaxation in this area.

The right panel of Figure \ref{fig:P_I_Ncl} shows similar results as the left panel but for model wIA. In general, the polarization degree tends to decline from the envelope to the disk scale following the significant reduction of grain alignment efficiency with increasing gas density. $p$ produced from grains with higher embedded iron inclusions is higher owing to the enhanced magnetic alignment of dust grains. However, comparing to model rIA, one can see that reduction of $p(\%)$ versus $I/I_{\rm max}$ in model wIA is slightly stronger, which is caused by the self-suppression of dust polarization signal radiating from grains with right and wrong IA in the protostellar core (Figure \ref{fig:pol_map_2mm_disk_wIA}).

 \begin{figure*}
\centering
    \includegraphics[width=\textwidth,height=\textheight,keepaspectratio]{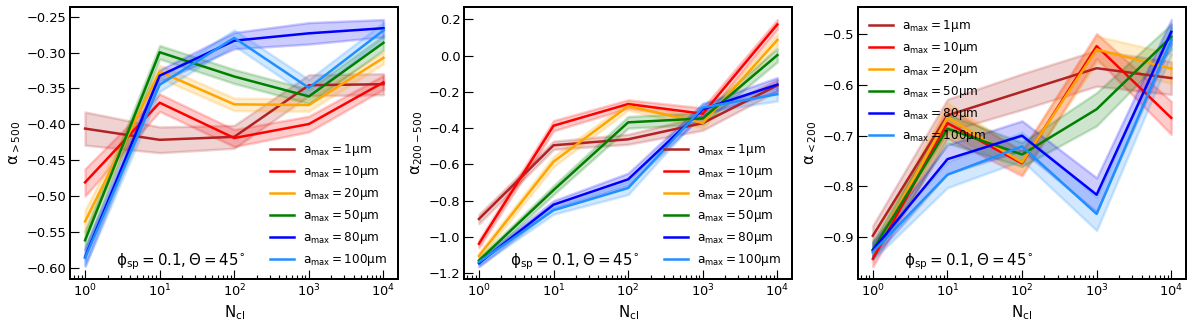}
    \caption{Variation of the slope of $p-I$ relation obtained from model rIA at 2mm: $\alpha_{\geq 500}$ (left panel), $\alpha_{\rm 200-500}$ (central panel), and $\alpha_{\leq 200}$ (right panel) as the function of $a_{\rm max}$ for different grain magnetic properties, assuming $\Theta = 45^{\circ}$. In general, $\alpha_{\geq 500}$, $\alpha_{\rm 200-500}$, and $\alpha_{\leq 200}$ increases with increasing amount of iron inclusions inside grains but tend to decreases
    with increasing maximum sizes. The reduction of $p-I$ is steeper in the central region due to the higher significant effect of gas randomization on both internal and external alignment of dust grains in the innermost region of the protostellar core. } 
     \label{fig:alpha_amax}
\end{figure*}

\subsection{Effects of grain growth}\label{sec:p_I_amax}
The upper left panel of Figure \ref{fig:P_I_amax} shows the effect of maximum grain size on the variation of $p-I/I_{\rm max}$ for model PA. As the maximum grain size increases, the overall polarization degree obtained in thousands au scale around the protostar increases due to the extension of the alignment range toward larger sizes. Besides, the grain growth eliminates the effect of alignment range on the values of $p$ obtained in the protostellar disk. In detail, in a model with $a_{\rm max} = 1\mum$, the increase of alignment size toward the disk by increasing gas density (Figure \ref{fig:align_amax}, right panel) induces the decrease of $p$ from $\sim 8\%$ at $\sim 400$ au to $p \sim 0.7\%$ at $\sim 100$ au. Then, the decrease of $a_{\rm align}$ toward the protostar by increasing RATs efficiency (Figure \ref{fig:align_amax}) induces the increase of $p\sim 0.7\%$ at $\sim 100$ au to $p\sim 1\%$ at the peak of dust emission. However, as grains grow to $\sim 10\mum$, the above decrease and increase of $p$ with $I/I_{\rm max}$ becomes less prominent. The obtained polarization degree then becomes a constant in the entire $\sim 400$ au region if the maximum size exceeds $ \sim 20\mum$, in which the alignment range is too large to feel the change of $a_{\rm align}$ with gas density and radiation field strength.

The upper right panel of Figure \ref{fig:P_I_amax} shows similar results as the upper left panel but for the model rIA of SPM grains with high $N_{\rm cl} = 10^{4}$. Beyond $\sim 400$ au, larger $a_{\rm max}$ induces higher $p$ as model PA because SPM grains with high embedded iron clusters can achieve perfect magnetic alignment by efficient MRAT alignment. As increasing intensity toward the center, the polarization degree obtained from all models of $a_{\rm max}$ decreases as a result of the reduced grain alignment efficiency (Figure \ref{fig:P_I_Ncl}). The increase of $a_{\rm max}$ from $1\mum$ to $2\mum$ helps to increase $p$ by about twice the order of magnitude due to the extension of the alignment range. However, the further growing of $a_{\rm max}$ from $2\mum$ to $100\mum$ does not clearly induce the increase of $p$ as model PA because the alignment range now covers large grains with inefficient IA by slow internal relaxation (Figure \ref{fig:aaj_highJ_amax} and \ref{fig:aaj_lowJ_amax}). Besides, as grains grow to above $50\mum$, VLGs within $100$ au are not able to be aligned with $\B$ (Figure \ref{fig:alar_amax}), inducing the slight reduction of $p$ with $a_{\rm max}$ (blue line).

The lower panels of Figure \ref{fig:P_I_amax} show results for model rIA with SPM grains with low $N_{\rm cl} = 100$ (left panel) and with PM grains (right panel). For all models of $a_{\rm max}$, $p$ always becomes smaller toward the protostar as a result of reducing the grain alignment efficiency, and the depolarization becomes more prominent for larger $a_{\rm max}$. The suppression of grain growth on $p$ is stronger for SPM grains with lower $N_{\rm cl}$ because of the dominance of large grains with slow internal relaxation. While for PM grains, it is caused by the alignment loss of VLGs around the protostar. For example, for SPM grains with low $N_{\rm cl} = 100$, $p$ at $\sim 200$ au will decrease $\sim 2$ times, from $p \sim 2\%$ for $a_{\rm max} = 1\mum$ to $p \sim 1\%$ for $a_{\rm max} = 100\mum$. For PM grains, the reduction is much more significant with the continuous reduction from $p\sim 0.3\%$ at $\sim 200$ au for $a_{\rm max} = 1\mum$ to $p\sim 0.01\%$ if grains grow to $a_{\rm max} = 100\mum$.

\begin{figure*}
\centering
    \includegraphics[width=\textwidth,height=\textheight,keepaspectratio]{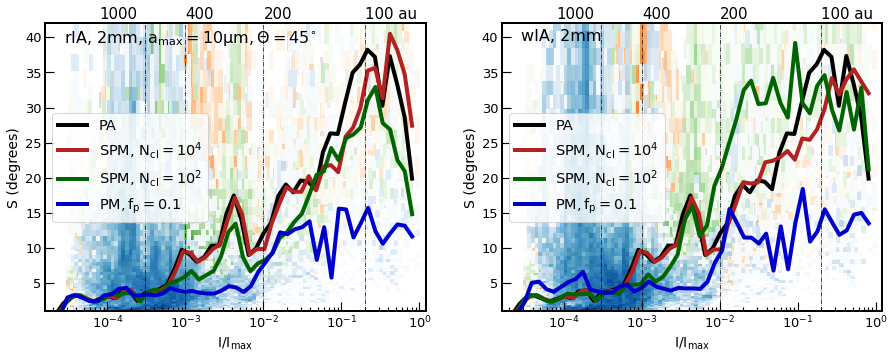}
 
    \caption{Effect of iron inclusions on the variation of the polarization angle dispersion function $S$ in the unit of degrees with normalized intensity $I/I_{\rm max}$ for model rIA (left panel) and model wIA (right panel), assuming $a_{\rm max} = 10\mum$ and $\Theta = 45^{\circ}$. The result from model PA is shown in the black line for comparison. In general, $S$ obtained from model PA, rIA, and wIA increases with increasing intensity due to the increased turbulence toward the center, then it decreases toward the protostar due to the formation of the well-ordered toroidal field inside the fast rotating protostellar disk (Figure \ref{fig:magnetic_field}). But in model rIA and wIA, $S$ obtained from aligned dust grains with lower levels of embedded inclusions is smaller than $S$ obtained from grains with higher embedded iron inclusions owing to the smaller area where they can trace $\B$-fields. Besides, the wrong interpretation of the polarization signal from wrong aligned dust grains (model wIA) also can induce extra dispersion in the polarization pattern compared with results in model PA.}
     \label{fig:S_I_Ncl}

\centering
 
    \includegraphics[width=\textwidth,height=\textheight,keepaspectratio]{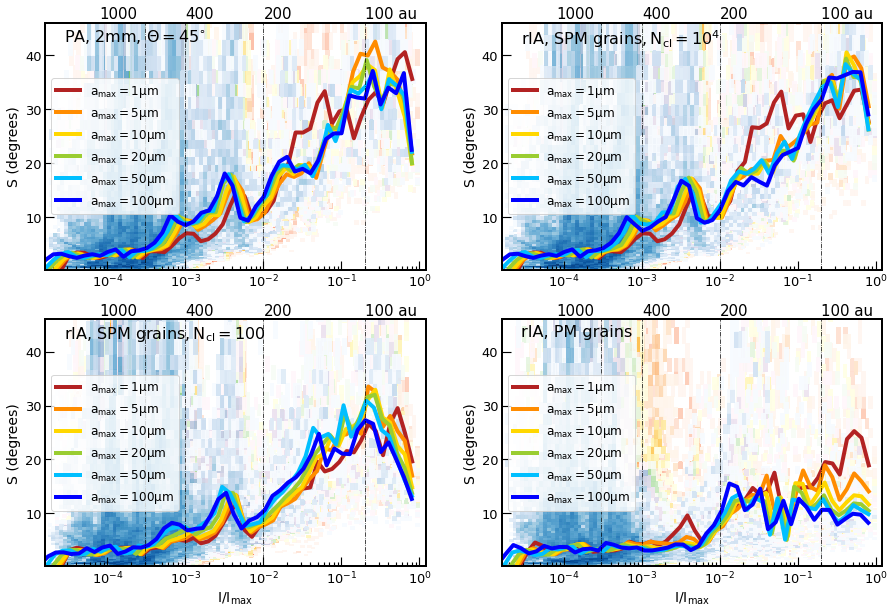}
    \caption{Effect of maximum grain size on the variation of $S$ with $I/I_{\rm max}$ obtained in model PA (upper left panel) and model rIA for SPM grains with high $N_{\rm cl} = 10^{4}$ (upper right panel), SPM grains with low $N_{\rm cl} = 100$ (lower left panel), and PM grains (lower right panel). In model rIA for SPM grains, $S$ obtained in the envelope increases with increasing $a_{\rm max}$ due to the increased amount of good tracer of magnetic fields. But in the inner region, $S$ decreases with increasing $a_{\rm max}$ because large grains now are not coupled well with $\B$fields, which reduces the information of $\B$-fields and turbulence carried inside dust polarization. The decrease of $S$ with grain growth is clearer for grains having smaller magnetic susceptibility.} 
     \label{fig:S_I_amax}
\end{figure*}

\subsection{On the slope of $p-I$}\label{sec:slope_p_I}
The slope of the $p-I$ is an important quantity for characterizing the effects of grain alignment, grain sizes, and B-fields on the observed polarization fraction. To connect above factors with the slope of $p-I$, we separate the $p-I$ relation at $\lambda = 2$ mm found in Sections \ref{sec:p_I_magnetoc} and \ref{sec:p_I_amax} into three segments which characterize the area of $>500$ au (envelope scale), $200-500$ au, and $<100$ au. Each segment is fitted with the power law $p \sim I^{\alpha}$ which $\alpha$ characterizes the slope of $p-I$ relation. The fitting is done by using the lmfit function in Python (\citealt{lmfit}). We show the details of our fitting in Appendix \ref{sec:fitting_slope}.
 
Figure \ref{fig:alpha_amax} shows the variation of $\alpha$ with $N_{\rm cl}$ for different maximum grain sizes from $a_{\rm max} = 1\mum$ to $a_{\rm max} = 100\mum$, considering model rIA. The left panel is for the slope of $p-I$ beyond $\sim 500$ au ($\alpha_{\rm > 500}$), the center panel is for the area within $\sim 200-500$ au ($\alpha_{\rm 200-500}$), and the right panel is for the disk scale within $200$ au ($\alpha_{\rm < 200}$). In general, the reduction of $p$ versus intensity in the entire protostellar core is shallower, i.e., lower negative $\alpha$, with increasing $N_{\rm cl}$ due to the enhanced grain alignment with magnetic fields by iron inclusions.

Beyond 500 au (left panel), for grains with low $N_{\rm cl} < 10$, $\alpha_{\rm > 500}$ is in the range of $[-0.4, -0.6]$ and it 
reduces continuously from $-0.4$ for $a_{\rm max} = 1\mum$ to $-0.6$ for $a_{\rm max} = 100\mu m$ causing by the misalignment of grains above $10\mum$ in the envelope (Figure \ref{fig:alar_amax}, upper panels). For grains with $N_{\rm cl} > 10$, $\alpha_{\rm > 500}$ becomes less negative, i.e., $\alpha_{\rm > 500}$ belongs to $[-0.25,-0.4]$, because of increasing the alignment degree of grains with magnetic fields as increasing the magnetic susceptibility (Figure \ref{fig:alar_amax}, upper panels). The value of $\alpha_{\geq 500}$ increases with increasing maximum grain sizes, from $\sim -0.4$ for $a_{\rm max} = 1\mum$ to $\sim -0.25$ for $a_{\rm max} = 100\mum$, owing to the extension of alignment range to larger sizes. Since SPM grains are able to have perfect alignment with $\B$ by MRAT alignment, turbulence and the projection effect of $\B$-fields are the major factors controlling the slope of $p-I$ in this envelope region (see the effect of iron inclusions and the inclination angle in Appendix \ref{sec:alpha_Theta}).

Moving toward the inner $200-500$ au region (center panel), the value of $\alpha_{\rm 200-500}$ for grains having low $N_{\rm cl} < 10$ decreases to $-0.8$ for $a_{\rm max} = 1\mum$ and even $-1$ for $a_{\rm max} = 100\mum$ due to the misalignment of micron-sized grains there. Grains with higher $N_{\rm cl} \geq 10$ show the similar range of $\alpha_{\rm 200-500}$ within $[0, -0.4]$ as found for $\alpha_{\rm > 500}$. However, $\alpha_{\rm 200-500}$ becomes more negative for larger maximum grain size owing to the presence of VLGs with inefficient IA by slow internal relaxation (Figures \ref{fig:aaj_highJ_amax} and \ref{fig:aaj_lowJ_amax}).
 
Inside the disk within $\sim 200$ au (right panel), $\alpha_{\rm < 200}$ for all values of $N_{\rm cl}$ and $a_{\rm max}$ decreases significantly to smaller values due to the reduced grain alignment efficiency in dense environments. For grains with low $N_{\rm cl}<10$, one gets very low $\alpha \sim -0.8$ up to $-1$ due to the alignment loss of dust grains with magnetic fields around the protostar. For grains with higher $N_{\rm cl} \geq 10$, $\alpha_{\leq 200}$ significantly reduces to $\sim [-0.5-0.7]$ and it decreases with increasing $a_{\rm max}$ owing to the change of the external alignment mechanism from MRAT to RATs and the increased amount of micron-sized grains with slow internal relaxation. For example, for SPM grains with $N_{\rm cl} \sim 10 - 10^{3}$, the slope of their $p-I$ relation changes from $\alpha_{\leq 200} \sim -0.6$ for $a_{\rm max} = 1\mum$ to $\alpha_{\leq 200} \sim -0.8$ if grains grow to $a_{\rm max} = 100\mum$.

\section{Effect of Iron Inclusions and Grain growth on $S$ and $p\times S$}\label{sec:iron_S_pS}
Next, we move to study the influence of iron inclusions and grain growth on the variation of the polarization angle dispersion function $S$ and the grain alignment efficiency $p\times S$ with intensity.

\subsection{$S-I/I_{\rm max}$ relation}\label{sec:S_I}
The left panel of Figure \ref{fig:S_I_Ncl} shows the variation of $S$ in unit of degrees with intensity obtained at 2 mm for model PA (black curve) and model rIA, assuming different grain magnetic properties and $a_{\rm max} = 10\mum$. For both model PA and model rIA, the angle dispersion function generally increases from $S \sim 3$ degrees at $\sim 1000$ au to peak in the outer edge of the disk of $\sim 100$ au then decreases toward the central region. The rise of $S$ in the envelope is caused by the increased turbulence in the contact areas between different infalling gas flows and between infalling and outflowing material. And the decrease of $S$ in the disk could be understood by the formation of the well-ordered toroidal $\B$-fields around the protostar wrapped by the fast rotating protostellar disk. However, taking into account the realistic model of grain alignment, $S$ seems not only to depend on the distortion of magnetic fields by turbulence but also depends on the grain magnetic properties. Particularly, $S$ for PM grains only slightly increases from $S \sim 3$ degrees in the envelope to a constant $S \sim 5$ degrees inside the disk, but $S$ for SPM grains with $N_{\rm cl} = 10^{4}$ increases significantly from $S \sim 3$ degrees in the envelope to the peak at $S \sim 25$ degrees in the outer edge of the disk and decreases to $S \sim 10$ degrees near the protostar as model PA. The variation of $S$ with $N_{\rm cl}$ is caused by the difference in the area where dust polarization can trace magnetic fields. In detail, PM grains only can trace $\B$-fields beyond 500 au (Figure \ref{fig:pol_map_2mm_disk_rIA}, second row), thus, their angle dispersion function only can reflect the turbulence level inside the envelope, which explains why $S$ only slightly increases by few degrees toward the center. In contrast, polarized dust emission from SPM grains with higher $N_{\rm cl}$ carries more information on magnetic fields and turbulence levels in the innermost region of the core (Figure \ref{fig:pol_map_2mm_disk_rIA}, third and fourth rows), inducing the clearer reflection of the variation of $S$ with $I/I_{\rm max}$ as model PA.

The right panel of Figure \ref{fig:S_I_Ncl} shows similar results as the left panel but for model wIA. Generally, the polarization angle dispersion function increases with increasing intensity and shows higher values for grains containing higher levels of iron inclusions. However, $S$ obtained from model wIA is slightly higher than results from model rIA. This additional distortion in the polarization pattern could be explained by the superposition of polarization signal from grains with the right and wrong IA along the LOS. In addition, $S$ obtained inside the disk for SPM grains with $N_{\rm cl} = 10^{4}$ is even higher than model PA, i.e., the maximum $S \sim 33$ degrees of SPM grains and $S \sim 20$ degrees for model PA, that the exceed $S$ arises from the area where polarization vectors $\P$ changes suddenly from $\P \perp \B$ (emission by grains with the right IA) to $\P \| \B$ (emission by grains with the wrong IA) (Figure \ref{fig:pol_map_2mm_disk_wIA}, third and fourth rows).

 \begin{figure*}
\centering
    \includegraphics[width=\textwidth,height=\textheight,keepaspectratio]{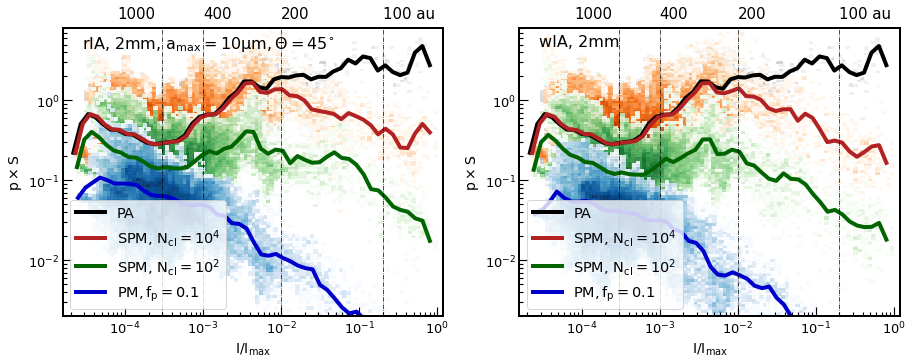}
    \caption{Variation of the grain alignment efficiency determined by the multiple between the polarization fraction $p$ with angle dispersion function $S$ $p\times S$ with normalized intensity $I/I_{\rm max}$ obtained from model PA (black line), model rIA (color lines, left panel) and wIA (right panel) for different grain magnetic properties, assuming $a_{\rm max} = 10\mum$, and $\Theta = 45^{\circ}$. In contrast to the rise of $p\times S$ with increasing intensity toward the central region as model PA, model rIA and wIA clearly show the reduction of $p\times S$ with intensity toward the protostar as the evidence of the reduced grain alignment efficiency by increasing gas randomization. The value of $p\times S$ is smaller and the reduction of $p\times S$ with $I/I_{\rm max}$ is stronger for grains containing lower levels of iron inclusions.} 
     \label{fig:PS_I_Ncl}

\centering
    \includegraphics[width=\textwidth,height=\textheight,keepaspectratio]{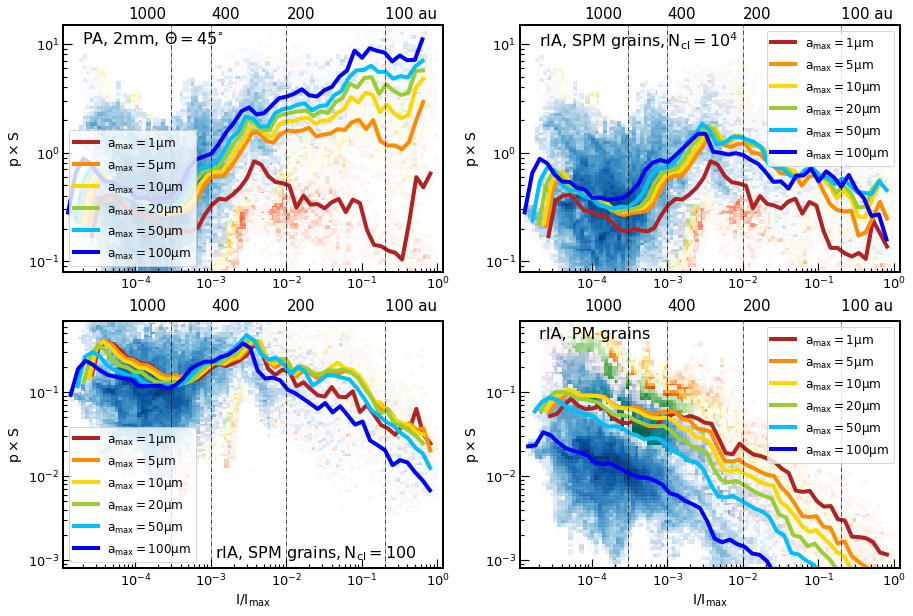}
    \caption{Effect of maximum grain size on the variation of $p\times S$ with normalized intensity obtained from model PA (upper left panel) and model rIA for SPM grains with high $N_{\rm cl} = 10^{4}$ (left panel), SPM grains with $N_{\rm cl} = 100$ (lower left panel), and for PM grains (lower right panel). For SPM grains, $p\times S$ obtained in the envelope increases with increasing $a_{\rm max}$ as found in model PA due to the extension of amount of grains with efficient magnetic alignment. However, $p\times S$ obtained in the inner region decreases with increasing $a_{\rm max}$ because of the enhanced amount of VLGs which are inefficiently aligned with magnetic fields. The reduction of the grain alignment degree with grain growth becomes more significant for grains having lower levels of iron inclusions.} 
     \label{fig:PS_I_amax}
\end{figure*}

Figure \ref{fig:S_I_amax} shows the comparison between the effect of grain growth on $S-I$ relation for model PA (upper left panel) and model rIA. The upper right panel is for SPM grains with high $N_{\rm cl} = 10^{4}$, the lower left panel is for SPM grains with low $N_{\rm cl} = 100$, and the lower right panel is for PM grains. In model PA, $S$ obtained from models with higher $a_{\rm max}$ is higher because the $\B$-fields tangling by turbulence is reflected better with the increasing amount of good tracer from perfectly aligned dust grains. For model rIA, increasing $a_{\rm max}$ also induces higher $S$ as model PA for SPM grains. However, this feature only happens in the envelope scale beyond $\sim 200$ au where large grains can have efficient magnetic alignment by MRAT mechanism. Moving toward the inner 200 au region, $S$ decreases with increasing $a_{\rm max}$ due to the reduced area where dust grains can trace $\B$-fields (similar to the case of PM grains, Figure \ref{fig:S_I_Ncl}). The reduced information of turbulence contained inside dust polarization becomes stronger for grains containing lower levels of iron inclusions (lower panels).

\subsection{$p(\%)\times S-I/I_{\rm max}$ relation}\label{sec:pxS_I}
Figure \ref{fig:PS_I_Ncl} shows the effect of iron inclusions on the variation of $p\times S$ with $I/I_{\rm max}$ at $\lambda = 2$ mm for model PA (black line) and model rIA (left panel) and model wIA (right panel), assuming $a_{\rm max} = 10\mum$. $p$ here is the polarization fraction, and $S$ is in unit of degrees. In model PA, $p\times S$ slightly decreases with $I/I_{\rm max}$ beyond $\sim 1000$ au then increases continuously toward the center owing to the significant rise of $S$ with intensity (Figure \ref{fig:S_I_Ncl}). The reduction of $p\times S$ in the envelope implies the subdominant effect of turbulence over the projection effect of magnetic fields on decreasing the observed polarization degree (Figure \ref{fig:P_I_Ncl}). And the rise of $p\times S$ in the inner region could be interpreted as the sign of increasing alignment degree around the protostar. However, taking the realistic model of grain alignment into account, one can see that within $\sim 200$ au, $p\times S$ clearly reduces toward the center for both PM and SPM grains, implying the significant reduction of the grain alignment efficiency around the protostar. The reduction of $p\times S $ versus $I/I_{\rm max}$ is more prominent for grains with lower magnetic susceptibility. The variation of $p\times S$ shares similar tendencies between model rIA (left panel) and wIA (right panel), but model wIA reveals a slightly steeper slope of $p\times S$ with $I/I_{\rm max}$ because of lower $p$ induced by the co-existence of grains with the right and wrong IA (Figures \ref{fig:P_I_Ncl}, right panel).

Figure \ref{fig:PS_I_amax} shows the difference in the effect of grain growth on the variation of $p\times S$ with $I/I_{\rm max}$ for model PA (upper left panel) and model rIA (remaining panels). In model PA, the increased maximum grain size increases values of $p\times S$ in the entire $\sim 1000$ au due to the rise of $p$ and $S$ with $a_{\rm max}$ (Figures \ref{fig:P_I_amax} and \ref{fig:S_I_amax}, upper left panel). At $ < 400$ au, one can see that for $a_{\rm max} = 1\mum$, $p\times S$ decreases from $\sim 400$ au to $\sim 100$ au as a result of the narrow alignment range by high gas density, i.e., larger $a_{\rm align}$, then it increases again toward the protostar position because small grains now can be aligned with $\B$ by efficient RATs, i.e., smaller $a_{\rm align}$ (Figure \ref{fig:align}). As $a_{\rm max}$ increases, this feature disappears due to the broadening of the alignment range to larger sizes. And starting with $a_{\rm max} \geq 10\mum$, $p \times S$ even rises with intensity within the 400 au region, which could be interpreted as the sign of the effective magnetic alignment of dust grains around the protostar (Figure \ref{fig:PS_I_Ncl}, black line). 
 
However, in model rIA, $p\times S$ obtained from all models of grain magnetic properties and maximum grain size, generally, always declines toward the center due to the reduction of the grain alignment efficiency toward the protostar. Beyond $\sim 400$ au, $p\times S$ obtained from the model of SPM grains (upper right and lower left panels) increases with increasing $a_{\rm max}$ as model PA because they have efficient alignment with $\B$ by MRAT mechanism. But for PM grains, $p\times S$ declines with $a_{\rm max}$ because of the inefficient alignment of VLGs by RATs in the envelope scale. Moving toward the central 400 au, grain growth induces the steeper reduction of $p\times S$ with intensity, reflecting well the weak alignment degree of almost aligned dust grains there. The suppression of grain growth on the quantity $p\times S$ is stronger for grains containing the lower amount of iron inclusions.

\begin{figure*}
\centering
    \includegraphics[width=\textwidth,height=\textheight,keepaspectratio]{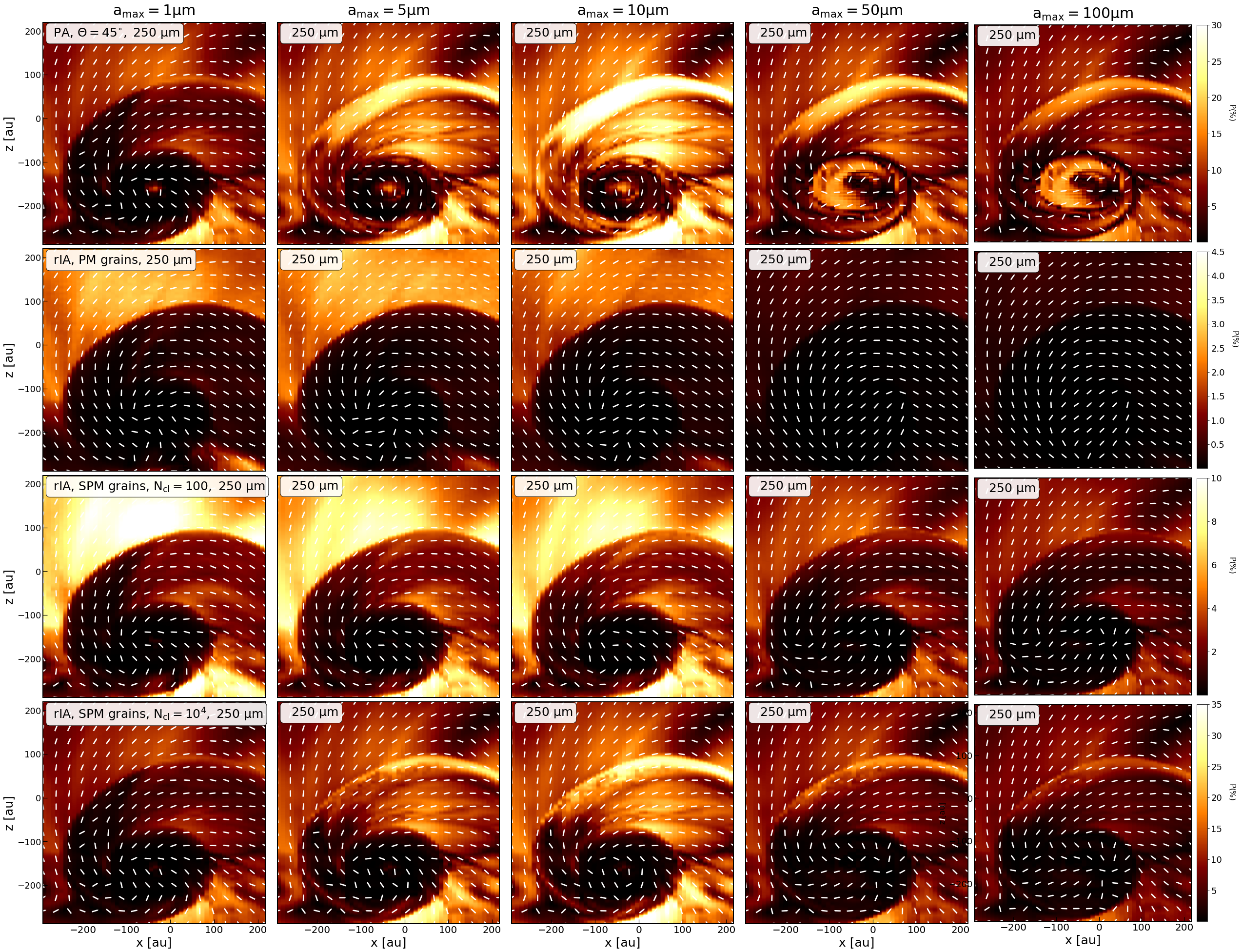}
    \caption{Effect of grain size and grain magnetic properties on the inferred magnetic field map obtained in the inner 500 au region at $250\mum$. The color code shows the polarization degree $p(\%)$, and white segments show the $\B$ vectors obtained by rotating the polarization vector by $90^{\circ}$. The upper row shows results for model PA, the second to fourth rows are for model rIA with different grain magnetic properties, the left column is for $a_{\rm max} = 1\mum$, and the right column is for $a_{\rm max} = 100\mum$. In model PA, the inferred magnetic field direction in the inner 200 au changes from parallel to perpendicular to the disk minor axis when the maximum grain size exceeds $a_{\rm max} \geq 50\mum$. The change in $\B$ vectors results from the change in the polarization mechanism from dichroic emission (at optically thin wavelengths at 2mm) to dichroic extinction  (at optically thick disk at $250\mum$). In model rIA, only SPM grains with high $N_{\rm cl} = 10^{4}$ can reveal the change in the polarization pattern caused by dichroic extinction at submillimeter wavelengths. PM and SPM grains with $N_{\rm cl}  =100$ do not reveal this feature because VLGs above $10\mum$ are not able to be aligned with $\B$inside the disk.} 
     \label{fig:pol_map_250um_rIA}
\end{figure*}

 \begin{figure*}
\centering
    \includegraphics[width=\textwidth,height=\textheight,keepaspectratio]{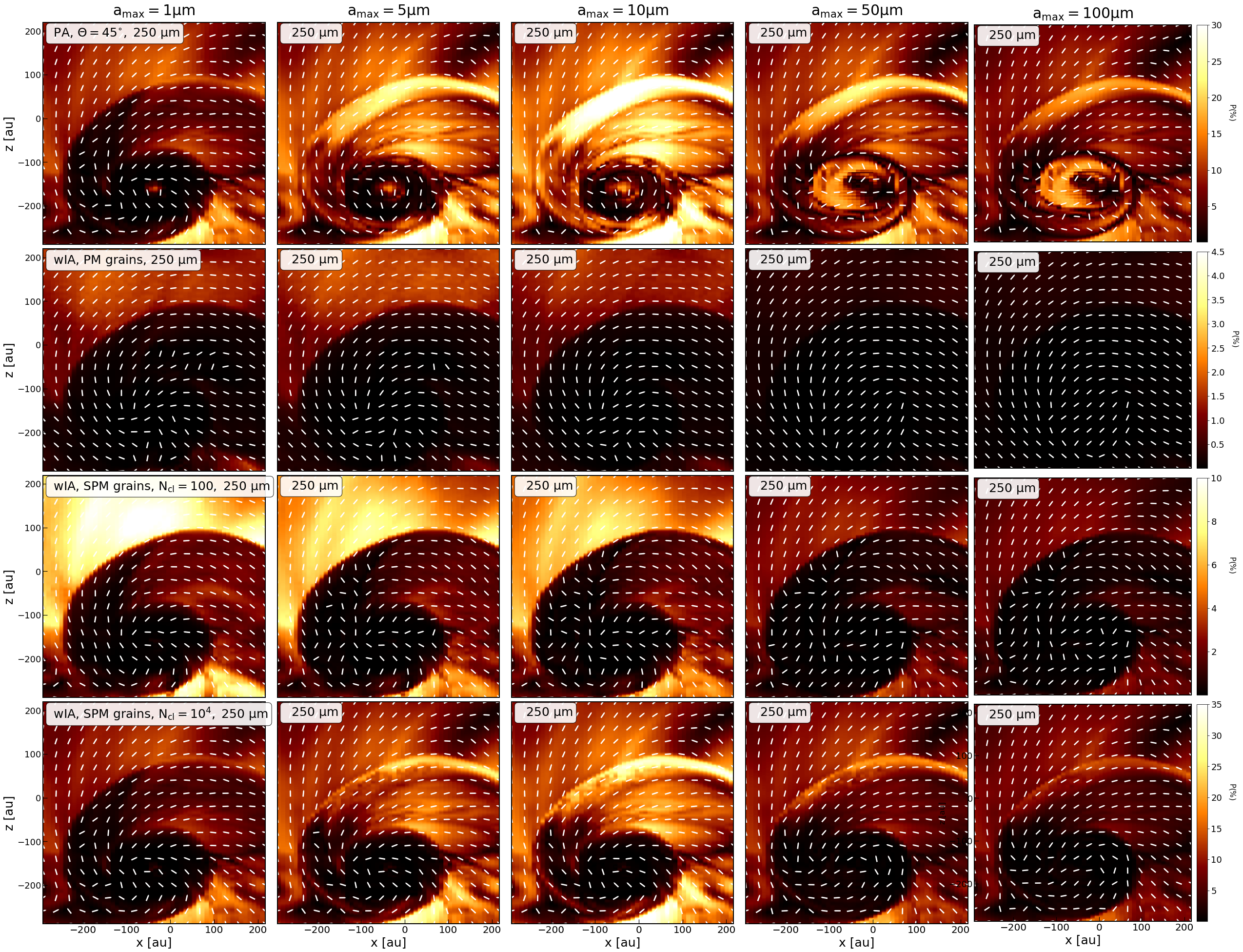}
    \caption{Similar results as Figure \ref{fig:pol_map_250um_rIA} but for model wIA. Obviously, only SPM grains with high $N_{\rm cl}$ can reveal in detail the change of magnetic fields from the inner part of the envelope to the disk scale because they can have the magnetic alignment around the protostar. However, the inferred magnetic fields at $250\mum$ inside the disk are similar to results from 2mm (Figure \ref{fig:pol_map_2mm_disk_wIA}), with $\B$vectors still perpendicular to the disk minor axis for all values of $a_{\rm max}$ from $1\mum$ to $100\mum$. This polarization signal originates from the emission of grains with the wrong IA, which is not strongly suppressed by the extinction of VLGs with the wrong IA as in the case of model rIA. } 
     \label{fig:pol_map_250um_wIA}
\end{figure*}

\begin{figure*}
\centering
 \includegraphics[width=\textwidth,height=\textheight,keepaspectratio]{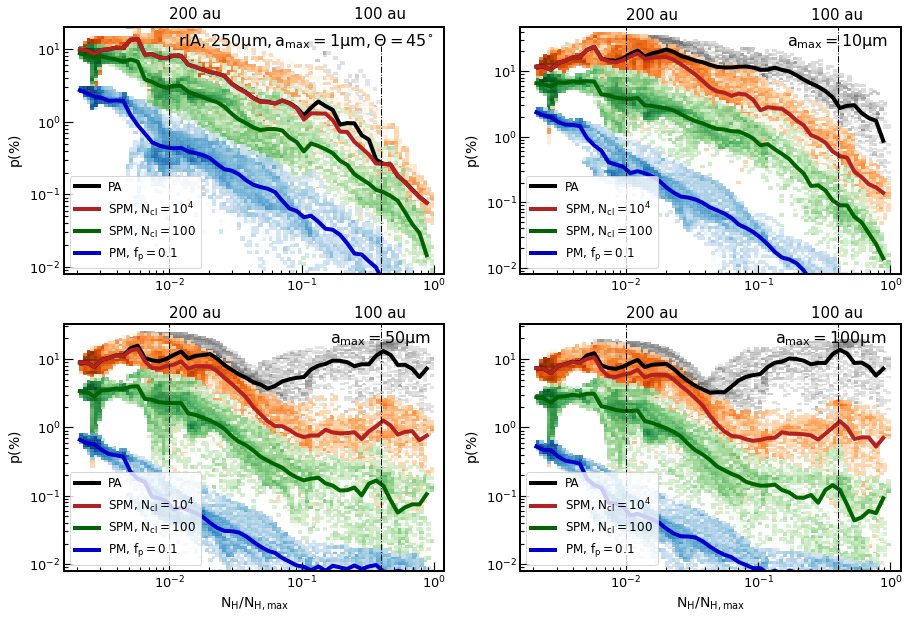}
     
    \caption{Comparison about the variation of $p(250\mum)$ with normalized column density $N_{\rm H}/N_{\rm H,0}$ between model PA and model rIA for $a_{\rm max} = 1\mum$, $a_{\rm max} = 10\mum$, $a_{\rm max} = 50\mum$, and $a_{\rm max} = 100\mum$, from left to right, from top to bottom, respectively. In model PA, for $a_{\rm max} \sim 5\mum$, $p(250\mum)$ decreases with increasing $N_{\rm H}$ due to the extinction of large grains. As $a_{\rm max}$ increases to $50 - 100\mum$, $p(250\mum)$ decreases then increases again to the center due to the change of polarization mechanism from dichroic emission to dichroic extinction (Figure \ref{fig:pol_map_250um_rIA}, upper row). However, taking into account the realistic model of grain alignment, $p(250\mum)$ just simply decreases toward the center, and the reduction of $p(250\mum)$ with $N_{\rm H}$ is stronger than model PA due to the significant reduction of the grain alignment in the innermost region of the protostar. The reduction of $p$ with $N_{\rm H}$ in model rIA is stronger for grains containing lower levels of iron inclusions.} 
     \label{fig:p_nh_250um}
\end{figure*}

\section{Effects of Dichroic Extinction Polarization}\label{sec:extinction}
Lastly, we move to study the properties of dust polarization at optically thick wavelengths where dichroic extinction becomes important. Indeed, the disk becomes optically thick at $450\mum$  (see the optical depth in the inner 500 au in Appendix \ref{sec:optical_depth}), but here we show results obtained at $250\mum$ because this wavelength shows clearer imprint of dichroic extinction on dust polarization. The results for $\lambda = 450\mum$ are in Appendix \ref{sec:450um}. We first study the effect of iron inclusions and maximum grain size on the inferred magnetic fields from dust polarization within the inner 500 au region in Section \ref{sec:pol_map_disk_250um}, and then on the variation of $p$ obtained from Section \ref{sec:pol_map_disk_250um} with gas column density $N_{\rm H}$ in section \ref{sec:p_nh_250um}. We choose $N_{\rm H}$ to be the parameter here instead of the intensity of dust emission as in Section \ref{sec:p_I} because the polarization signal originating from dichroic extinction is more sensitive with gas density than intensity. Thus, it is easier to understand how the density structure of the disk affects the observed polarization fraction.

\subsection{Polarization map}\label{sec:pol_map_disk_250um}
\subsubsection{Model rIA}
Figure \ref{fig:pol_map_250um_rIA} shows the effect of the dust model and maximum grain size on the magnetic field map inferred from dust polarization at $250\mum$. The color code shows the polarization degree, and white segments show the inferred magnetic field direction by rotating the polarization vector $\P$ by $90^{\circ}$, other setups are similar as in Figure \ref{fig:pol_map_2mm_disk_rIA}.

In model PA (first row), from $a_{\rm max} = 1\mum$ to $a_{\rm max} = 10\mum$, dust polarization observed at $250\mum$ reveals the similar inferred magnetic field map as detected at optically thin 2 mm (Figure \ref{fig:pol_map_2mm_disk_rIA}, first row), with $\B$ vectors follow the spiral pattern for $a_{\rm max} = 1\mum$, and the change from the spiral field to the pinched field along the disk minor axis for $a_{\rm max} \sim 5-10\mum$. When grains grow to above $50\mum$, one gets the similar inferred spiral $\B$-fields within 200-500 au as cases of $a_{\rm max} = 1-10\mum$, but $\B$ vectors change $90^{\circ}$ from parallel to be perpendicular to the disk minor axis in the inner 200 au region. The $90^{\circ}$ flipping of $\B$ vectors in the disk is caused by the change in the polarization mechanism from dichroic emission to dichroic extinction, which is activated by the presence of aligned VLGs above $50\mum$ inside the optically thick disk. In the term of the polarization degree, $p$ slightly declines from the outer edge of the disk at 200 au to the trough at about 100 au from the protostar and increases again toward the center, which corresponds to the transition of the polarization mechanism from dichroic emission (beyond 100 au) to dichroic extinction (within 100 au).
 
For model rIA, PM and SPM grains with low $N_{\rm cl} = 100$ (second and third rows) generally cannot reveal in detail the morphology of magnetic fields inside the disk as obtained at 2mm due to the alignment loss inside the disk. But for SPM grains with high $N_{\rm cl} = 10^{4}$ (fourth row), one can clearly obtain the change of $\B$-fields within 500 au driven by gas dynamic and the $90^{\circ}$ flipping of $\B$-fields caused by dichroic extinction as grains grow from $a_{\rm max} = 1-10\mum$ to $a_{\rm max} = 50-100\mum$ as found in model PA. However, one does not clearly see the decrease and increase of $p$ with the disk structure as model PA because of the weak coupling of VLGs with $\B$ within 100 au around the protostar. 

\subsubsection{Model wIA}
Figure \ref{fig:pol_map_250um_wIA} shows similar results as Figure \ref{fig:pol_map_250um_rIA} but for model wIA. Similar to model rIA, only SPM grains with high $N_{\rm cl} = 10^{4}$ (fourth row) can reveal the change of $\B$-fields driven by gas dynamic and grain growth within 500 au. However, when grains grow to $a_{\rm max} = 50-100\mum$, surprisingly, the inferred magnetic field inside the disk does not change from parallel to perpendicular to the disk minor axis as model PA even though VLGs here are large enough to activate the effect of dichroic extinction at submillimeter wavelengths. The polarization signal in this case still comes from the emission of VLGs with the wrong IA (Figure \ref{fig:pol_map_2mm_disk_wIA}, lower right panel), which could be explained by the weak extinction of  VLGs induced by the co-existence of absorber with right and wrong IA along the LOS.

\subsection{Polarization degree and gas column density}\label{sec:p_nh_250um}
The upper left panel of Figure \ref{fig:p_nh_250um} shows the comparison of the relation $p(250\mum)-N_{\rm H}$ obtained in the inner 500 au region between model PA (black line) and model rIA (colors lines) with different grain magnetic properties, assuming $a_{\rm max} = 1\mum$. In model PA, the polarization degree declines continuously from $p \sim 20\%$ at 500 au to $p \sim 1\%$ in the densest region inside the disk due to the narrower of the alignment range (similar to results from 2mm, Figure \ref{fig:P_I_amax}, case $a_{\rm max} = 1\mum$). In model rIA, the polarization degree produced from SPM grains with high $N_{\rm cl} = 10^{4}$ share similar behavior with $N_{\rm H}$ and values as model PA. SPM grains with lower $N_{\rm cl}= 100$ and PM grains also show the reduction of $p$ with increasing $N_{\rm H}$, but with smaller polarization degree due to the inefficient alignment of sub-micron grains with $\B$ around the protostar.

The upper right panel of Figure \ref{fig:p_nh_250um} shows similar results as the upper left panel but for $a_{\rm max} = 10\mum$. In the PA model, $p$ obtained in the inner 500 au region generally be higher than the case of $a_{\rm max} = 1\mum$ due to the extension of the alignment range. However, they still show the decline of $p$ with $N_{\rm H}$ owing to the dichroic extinction of aligned micron-sized inside the disk. In model rIA, one also obtains the reduction of $p$ with gas column density, but with a much steeper slope than model PA due to the additional contribution from the low alignment degree of micron-sized grains with $\B$ here. Grains with lower embedded iron inclusions produce smaller $p(\%)$ and stronger declination of $p$ with $N_{\rm H}$ due to their weaker alignment degree with magnetic fields.  

The lower left panel of Figure \ref{fig:p_nh_250um} shows similar results as the upper right panel but for 
$a_{\rm max} = 50\mum$. In model PA, the polarization fraction at $250\mum$ decreases from $p \sim 10\%$ at 500 au to $p \sim 5\%$ when $N_{\rm H} \sim 0.05/N_{\rm H,max}$ ($\sim 150$ au) then increases again to $p\sim 10\%$ when $N_{\rm H} \sim 0.5N_{\rm H,max}$ ($\sim 100$ au). The decrease and increase of $p$ with gas column density from 500 to $\sim 100$ au are driven by the increased extinction efficiency of aligned VLGs, which changes the polarization mechanism from dichroic emission in the outer edge of the disk to dichroic extinction inside the disk (Figure \ref{fig:pol_map_250um_rIA}, first row). Within $ \leq 100$ au, $p(250\mum)$ slightly decreases again from $p \sim 10\%$ to $p \sim 7\%$ with increasing $N_{\rm H}/N_{\rm H,max}$ due to the trap of polarization signal by high gas-radiation interactions. In model rIA for SPM grains with $N_{\rm cl} = 10^{4}$, one also sees the decrease and increase of $p(250\mum)$ with increasing $N_{\rm H}$ as model PA. However, the rise of $p$ with $N_{\rm H}$ within $\sim 150$ au is not prominent as model PA because, in reality, VLGs are not well coupled with $\B$ enough to clearly activate the effect of dichroic extinction inside the disk. For grains with lower values of $N_{\rm cl}$, $p(250\mum)$ just simply decreases continuously  with increasing gas column density due to the misalignment of VLGs within $\sim 200$ au around the protostar. The tendency of $p-N_{\rm H}$ and the values of $p$ for all realistic models of grain magnetic properties are the same as grains grow from $a_{\rm max} = 50\mum$ to $a_{\rm max} = 100\mum$ (lower right panel).

\section{Discussion}\label{sec:discuss}
In this section, we will discuss more details about our findings and how to connect dust physics with observations via synthetic modeling of polarized dust emission.

\subsection{Grain alignment in protostellar envelopes and disks}\label{sec:discuss_align}
 
The alignment of grains with magnetic fields is the key to using polarized dust emission to trace magnetic fields and study dust physics. However, in contrast to the well-determined behavior of aligned dust grains in ISM and MCs (\citealt{Stefan_2017}), the strong gas randomization in dense protostellar environments would reduce the magnetic alignment of grains, which hinders the usage of dust polarization within thousands au around the protostar   \citep{Hoang_2022}.

The detailed picture of grain alignment efficiency in protostellar cores and disks is revealed numerically in studies by \cite{Hoang_2022}, \cite{Hoang+2022} and in following synthetic modeling of polarized dust emission using the analytical model of the Bok globule by \cite{Giang_2022}. The magnetic alignment in dense protostellar environments is found to be size- and spatial-dependence, with VLGs of size $\geq 10\mum$ requiring embedded iron inclusions for having efficient internal and external alignment around the protostar. Results obtained in the realistic MHD simulation of the self-collapsing low-mass core in Section \ref{sec:align_range} are consistent with previous findings. Particularly, the maximum alignment size of PM grains is limited to $\sim 10\mum$ in protostellar envelopes and $\sim 1\mum$ inside the disk owing to the limitation of slow Larmor precession (Figure \ref{fig:amaxJB}, left panel). For grains below $a_{\rm max,JB}^{\rm Lar}$, RAT is the primary external alignment mechanism of PM grains with typical $f_{\rm high-J} \sim 0.25$ (Figures \ref{fig:adg_1}). PM grains at high-\textit{J} can have efficient IA due to their suprathermal rotation (Figure \ref{fig:aaj_highJ}), but grains at low-\textit{J} always have slow internal relaxation regardless of their position within 1000 au around the protostar (Figures \ref{fig:aaj_lowJ}).In contrast, considering $30\%$ of iron locked inside dust grains under the cluster form, all SPM grains with $N_{\rm cl} = 10^{4}$ from $a_{\rm align}$ (Figure \ref{fig:align}) up to $100\mum$ will have perfect magnetic alignment beyond $\sim 200$ au owing to efficient MRAT mechanism (Figures \ref{fig:amaxJB}, \ref{fig:aaj_highJ}, and \ref{fig:adg_1}, right panel).  However, within $\sim 100$ au around the protostar, SPM grains of size $\geq 40\mum$ cannot be aligned with $\B$, and grains above $\geq 20\mum$ will be aligned with $\B$ by only RATs due to strong gas randomization effect. The internal alignment of SPM grains inside the disk is also inefficient, given the maximum size for fast internal relaxation at high- and low-\textit{J} attractors of $\sim 20\mum$ and $\sim 1\mum$, respectively (\citealt{Hoang+2022}, \citealt{Giang_2022}). Indeed, \cite{Giang_2022} found that within 100 au around the protostar, VLGs at high-\textit{J} can have fast internal relaxation owing to their rapid rotation rate driven by efficient RATs. However, with the very high gas density up to $n_{\rm H} \sim 10^{9} - 10^{11}\cm^{-3}$ (Figure \ref{fig:gas_density}), gas randomization inside the disk is strong enough to eliminate the enhanced Barnett relaxation by high rotational rate, forcing grains to have slow internal relaxation at high-\textit{J} in our simulated protostellar disk. 

The internal alignment efficiency of large grains inside the protostellar envelope and disk will be improved (\citealt{Hoang+2022}) if considering both nuclear relaxation (\citealt{Lazarian_1999}, \citealt{Lazarian_Hoang_2008}) and inelastic relaxation (\citealt{Lazarian_Efroisky_1999}, \citealt{Efroimky_2000}). As shown in \cite{Hoang+2022}, the timescale of inelastic relaxation is $\tau_{\rm iER} \sim a^{11/2} St^{-3}$ (\citealt{Efroimky_2000}) and Barnett relaxation is $\tau_{\rm BR} \sim a^{7}St^{-2}$, given $St = J/J_{\rm th}$ the suprathermal rotation number (\citealt{Hoang+2022}) which characterizes the ratio of the grain angular momentum over the grain thermal angular momentum. As $\tau_{\rm iER}$ increases with grain sizes slower, decreases with increasing the grain rotational rate faster, and does not depend on the grain magnetic properties as Barnett relaxation, inelastic relaxation can help both PM and SPM grains above $1\mum$ in size to achieve fast internal relaxation at both high and low-\textit{J} attractors. Especially, inelastic relaxation can lead SPM grains with $N_{\rm cl} \geq 10^{2}$ of size $10-100\mum$ inside the disk to have efficient IA at high-\textit{J} owing to their high rotation rate around the protostar. Similarly, nuclear relaxation can dissipate rotational energy faster than Barnett relaxation by a factor of $\sim 10^{6}$ due to its stronger Barnett-equivalent magnetic fields (\citealt{Lazarian_Hoang_2008}). However, due to the saturation of the magnetic response inside suprathermal grains, nuclear relaxation may be only important in driving the internal alignment of large micron-sized grains at low-\textit{J} attractors. The rotation rate suitable for operating nuclear relaxation will be extended with the presence of iron inclusions (\citealt{Lazarian_Hoang_2008}, \citealt{Lazarian_Hoang_2019}), which consequently strengthens the IA for more large SPM grains at low-\textit{J} in the inner envelope and inside the disk.  Besides the two mechanisms mentioned above, irregular grains (not oblate grains as in our model) can also undergo internal alignment by the Amplitude Barnett Relaxation mechanism (ANR, \citealt{Lazarian_Hoang_2019}). ANR dissipates rotational energy via the change in the amplitude of $\J$ during the IA period of wobbling grains. Therefore, it can play some role in improving the IA of large grains facing slow internal relaxation by Barnett relaxation. We expect that it can work for both grains at high and low-\textit{J} because the grain wobbling always appears (\citealt{Lazarian_1994}) as a consequence of the rotational dissipation-induced fluctuation inside them. Given the complex dependence of inelastic relaxation, nuclear relaxation, and ANR on grain rotational rate, grain magnetic susceptibility, and also inelasticity (for inelastic relaxation scenario), these new features must be incorporated into POLARIS to precisely quantify their realistic strength in driving the internal alignment of VLGs in protostellar environments.

Besides MRAT mechanism, the external alignment of dust grains also can be improved with higher $f_{\rm high-J}$ if grains at low-\textit{J} can be lifted to stably aligned with $\B$ at high-\textit{J} by the gas bombardment (\citealt{Hoang_Lazarian_2008}). This process takes about $5 - 10$ time of the gas damping timescale $\tau_{\rm gas}$, but given small $\tau_{\rm gas}$ inside protostellar disks, gas bombardment may play an important role in enhancing the net magnetic alignment of dust grains within few hundred au around the protostar. The increased amount of grains at $f_{\rm high-J}$ not only increases the observed degree of polarization but also strengthens the origin of dust polarization from magnetically aligned dust grains (see Section \ref{sec:other_polarization}). Furthermore, they will validate the perpendicular orientation of the dust polarization $\P$ vectors with $\B$, given grains at high-\textit{J} always tend to have the right IA regardless of fast or slow internal relaxation. This issue is very important (see below Section \ref{sec:discuss_iron_pol_pattern}) in inferring $\B$ field morphology inside the disk. 

So in conclusion, considering Barnett relaxation as the major internal alignment mechanism of dust grains, it is impossible to assume the fast internal relaxation for all aligned dust grains inside the protostellar envelope and disk. The alignment size range is not only determined by the minimum alignment size by RATs, $a_{\rm align}$, but also by the rate of Larmor precession compared to the gas randomization, $a_{\rm max,JB}^{\rm Lar}$. Besides, the external alignment mechanism, RATs or MRAT, will be size-dependent and is controlled by the grain magnetic susceptibility, local gas density, and magnetic field strength inside the core. Theoretically, the alignment mechanism will determine the observed degree of dust polarization (Section \ref{sec:p_I}). The internal alignment determines the orientation of polarization vectors with local magnetic fields, and the alignment range determines the area where dust polarization can trace $\B$-fields (Section \ref{sec:iron_pol_map}). Therefore, detailed modeling of grain alignment with their magnetic properties is required to maximize the usage of dust polarization in probing magnetic fields and dust physics around the protostar.

 \subsection{The reliability of dust polarization in tracing magnetic fields in protostellar envelopes and disks}\label{sec:discuss_polarization_pattern}
 
\subsubsection{Inferred magnetic fields from dust polarization}\label{sec:discuss_iron_pol_pattern}
In Section \ref{sec:pol_map_envelope}, we found that in general, PM and SPM grains can trace the large-scale magnetic field in the envelope of $\sim 500-1000$ au scale because they can be aligned with $\B$ there. For SPM grains, most of them have fast internal relaxation and tend to be aligned with $\B$ at high-\textit{J} attractors. As a result, their polarization vectors are always perpendicular to the orientation of magnetic fields. For PM grains, although $\sim 75\%$ of grains will be aligned with $\B$ at low-\textit{J} attractors by RATs, the net polarization signal obtained in the envelope still be dominant by the emission of $25\%$ of grains with perfect right internal alignment and perfect external alignment at high-\textit{J} (Figure \ref{fig:adg_1_amax}, upper panels). Thus, one can confidently rotate polarization vectors $\P$ by $90^{\circ}$ to infer the magnetic field morphology in the envelope regardless of the magnetic properties of dust grains.

In the protostellar disk, the situation is different. In Section \ref{sec:pol_map_disk}, we show that PM grains cannot trace $\B$-fields inside the disk because grains above $1\mum$ cannot be aligned with $\B$ in this area. Their observed polarization signal thus only provides magnetic field information of foreground envelope, whose polarization vectors are always perpendicular to $\B$. On the contrary, polarized dust emission from SPM grains can be used to trace $\B$ fields inside the disk because micron-sized and VLGs can have magnetic alignment in this region with the help of iron inclusions. The detailed morphology of magnetic fields around the protostar increases with increasing amount of iron inclusions locked inside SPM grains. For example, when $\Theta = 45^{\circ}$ and $a_{\rm max} = 10-100\mum$, SPM grains with $N_{\rm cl} = 100$ only can capture the spiral pattern of magnetic fields beyond $\sim 200$ au from the protostar (Figure \ref{fig:pol_map_2mm_disk_rIA}, third row). While SPM grains with higher $N_{\rm cl} = 10^{4}$ have the capability to trace the transition of $\B$-fields, from the spiral pattern beyond 200 au to the uniform field along the disk minor axis in the inner 200 au region as expected in model PA (Figure \ref{fig:pol_map_2mm_disk_rIA}, fourth row). 

However, the alignment of SPM grains with $\B$ in the disk accidentally activates the problem related to the alignment direction of grains having slow internal relaxation with magnetic fields. As shown in Section \ref{sec:aaJ} (see also Figures \ref{fig:aaj_highJ_amax} and \ref{fig:aaj_lowJ_amax}), almost micron-sized grains above $1\mum$ inside the disk will have slow internal relaxation at both high and low-\textit{J} attractors. Since RATs play the major role in aligning large SPM grains within the disk with $f_{\rm high-J} = 0.25$ (Figures \ref{fig:adg_0.5_amax} and \ref{fig:adg_1_amax}), emission from grains at low-\textit{J} may dominant over grains at high-\textit{J} to decide the net polarization pattern. In contrast to grains with fast internal relaxation whose alignment direction is well-determined to be perpendicular with $\B$, the alignment direction of grains with slow internal relaxation is still not well understood. As shown in \cite{Hoang_Lazarian_2009}, grains at high-\textit{J} may still have right IA due to the spinning torque of RATs, but grains at low-\textit{J} may have right or wrong IA, with unclear conditions supporting their orientation in space. We show Section \ref{sec:pol_map_disk} the polarization pattern obtained in two separate cases. If all grains with slow internal relaxation have the right IA, they will produce $\P \perp \B$, and rotating $\P$ by $90^{\circ}$ is reasonable, yielding the correct magnetic field map inside the disk as obtained in model PA (Figure \ref{fig:pol_map_2mm_disk_rIA}, fourth row). However, if grains with slow internal relaxation at low-\textit{J} have the wrong IA, emission from aligned dust grains at low-\textit{J} will dominant and produce the net polarization vectors $\P \| \B$. Consequently, rotating $\P$ by $90^{\circ}$ induces the wrong inferred pattern of magnetic fields in the disk (Figure \ref{fig:pol_map_2mm_disk_wIA}, fourth row).

   \begin{table}
  \centering
         \caption{Effects of maximum grain size and dust model on the recovery rate of the entire protostellar core}
  \begin{tabular} {c|ccc}
  \hline 
\textbf{Model} & \textbf{PM} & \textbf{SPM, $N_{\rm cl} = 100$} & \textbf{SPM, $N_{\rm cl} = 10^{4}$} \\
  \hline 
PA, $a_{\rm max} = 10\mum$   &  \multicolumn{3}{c}{$83.39\%$}\\ 
rIA                          &   $72.31\%$   & $83.51\%$  & $84.92\%$ \\
wIA                          &   $64.30\%$   & $78.02\%$  & $83.25\%$  \\

PA, $a_{\rm max} = 100\mum$   &    \multicolumn{3}{c}{$83.76\%$} \\
rIA                           &    $71.83\%$    & $82.46\%$   & $83.53\%$  \\
wIA                           &    $69.26\%$    & $81.25\%$   & $82.92\%$  \\
  \hline 

    \label{tab:recover_rate}
    \end{tabular}
\end{table}

   \begin{table}
  \centering
         \caption{Recovery rate in concentric rings focusing in the envelope and disk scale for different grain magnetic properties, assuming $a_{\rm max} = 10\mum$}
  \begin{tabular} {ccccc}
  \hline 
\textbf{Model} & \textbf{Grain type} & \textbf{$<100$ au} & \textbf{100-500 au} & \textbf{500-2000 au}  \\
  \hline 
PA   &                            & 93.37$\%$   & 90.90$\%$     & 82.72$\%$       \\
 
rIA  & PM                         & 44.58$\%$   & 59.39$\%$     & 73.70$\%$       \\ 
     & SPM, $N_{\rm cl} = 100$    & 54.82$\%$   & 82.70$\%$     & 82.59$\%$        \\ 
     & SPM, $N_{\rm cl} = 10^{4}$ & 78.61$\%$   & 89.60$\%$     & 82.72$\%$        \\ 

wIA  & PM                         & 43.37$\%$   & 56.32$\%$     & 71.19$\%$        \\ 
     & SPM, $N_{\rm cl} = 100$    & 45.78$\%$   & 75.39$\%$     & 82.26$\%$      \\ 
     & SPM, $N_{\rm cl} = 10^{4}$ & 32.83$\%$   & 86.44$\%$     & 82.72$\%$    \\ 
  
  \hline 

    \label{tab:recover_rate_ring_a100}
    \end{tabular}
\end{table}

So in conclusion, we can freely rotate polarization vectors of dust polarization obtained beyond $\sim 200$ au by $90^{\circ}$ to infer the magnetic field morphology in the protostellar envelope, regardless of examining the grain magnetic properties. But for the polarization signal obtained within the inner $200$ au, careful examination of its origin must be done before further investigating the magnetic field information embedded in dust polarization. {As discussed in the above paragraph, we can estimate the region where dust polarization trace $\B$ fields (envelope or disk) based on the grain magnetic properties (we will discuss this issue in the following Section \ref{sec:magnetic_prop_amax}). But in terms of the orientation between $\P$ and $\B$, we indeed do not have a clear answer due to the undetermined alignment direction of large micron-sized grains with slow internal relaxation with $\B$ inside the disk. We refer readers to Section \ref{sec:alignment_direction} where we discuss the method to constrain the grain alignment direction inside the disk via multi-wavelength observations based on our synthetic results obtained in Sections \ref{sec:pol_map_disk} and \ref{sec:pol_map_disk_250um}. However, as discussed in Section \ref{sec:discuss_align}, more SPM grains above $\sim 1\mum$ may have fast internal relaxation at both high and low-\textit{J} if we consider both nuclear relaxation and inelastic relaxation effects. The higher amount of grains with fast internal relaxation will reduce the uncertainty in rotating $\P$ to infer $\B$. However, the detailed impact of the mentioned internal relaxation mechanisms on aligned dust grains is not quantified and should be addressed in future studies to provide a definitive answer to this question. 

\subsubsection{Recovery rate}\label{sec:recovery_rate}
To quantify how well dust polarization can capture the orientation of protostellar magnetic fields, \cite{Valdivia_2022} calculate the magnetic field angle difference between MHD simulations and synthetic observations of dust polarization in different cores with different initial setups of magnetic fields and turbulence. They conclude that dust polarization from aligned dust grains is the robust tracer of magnetic fields in $\sim 500 - 1000$ au scale, with the ability to recover $\sim 90\%$ of magnetic field orientation in this region. However, this paper considers standard RAT theory with the constant $f_{\rm high-J} = 0.25$ and assumes that all grains from $a_{\rm align} - a_{\rm max} = 20\mum$ have fast internal relaxation and can be aligned with $\B$. As discussed in Section \ref{sec:discuss_align}, this assumption is not always valid, even in the protostellar envelope. Therefore, even polarized dust emission generally can trace $\B$-fields in the envelope, how accurately is this method must be re-examined with the new grain alignment model shown in Section \ref{sec:align_range}.
 
Following \cite{Valdivia_2022}, we calculate the angle difference $\Delta \phi$ between the integrated magnetic field from the MHD simulation shown in the right panel of Figure \ref{fig:magnetic_field} and the inferred magnetic fields from dust polarization $\phi_{\rm B,syn}$ derived by rotating $\P$ by $90^{\circ}$. Here, we consider the observed direction is along the face-on direction ($\Theta = 0^{\circ}$) and polarized dust emission can trace magnetic fields if $\Delta \phi \leq 20^{\circ}$. The recovery rate is then defined as the ratio of the area where dust polarization can capture $\B$-field orientation over the studied area. 

Table \ref{tab:recover_rate} shows the recovery rate of dust polarization from PM grains (first column), SPM grains with low $N_{\rm cl} = 100$ (second column), and SPM grains with high $N_{\rm cl} = 10^{4}$ (third column). Three first rows show results obtained from model PA, rIA, and wIA with $a_{\rm max} = 10\mum$, and the next three rows are for the model with $a_{\rm max} = 100\mum$. In general, the recovery rate is smaller when we consider the realistic model of grain alignment in protostellar environments. In particular, PM grains only can cover $\sim 70-75\%$ of magnetic fields in both the envelope and disk due to their weak alignment degree with magnetic fields. In contrast, SPM grains can recover $80 - 90\%$ of $\B$ within $\sim 2500$ au around the protostar depending on the amount of iron inside dust grains. The recovery rate slightly declines with increasing $a_{\rm max}$ due to the increased amount of VLGs with inefficient magnetic alignment. This parameter is also smaller for model wIA as a result of the superposition of polarization signal from grains with the right and wrong IA along the LOS. 
 
Table \ref{tab:recover_rate_ring_a100} shows similar results as Table \ref{tab:recover_rate} but for different concentric rings of thickness from 500-2000au (envelope scale), 100-500 au (inner envelope), and within 100 au (the disk scale), assuming $a_{\rm max} = 100\mum$. In general, the accuracy of $\B$-field morphology inferred from dust polarization decreases toward the center and decreases with decreasing amount of iron inclusions embedded inside aligned dust grains. In particular, PM grains can recover $\sim 70\%$ beyond $> 500$ au and $\sim 45\%$ in the disk, while SPM grains can recover higher percentages of $ \sim 80-90\%$ in the envelope but only $\sim 50- 78\%$ inside the disk. The recovery rate declines with increasing maximum grain sizes, and it is smaller if grains with slow internal relaxation at low-\textit{J} have wrong alignment direction. 

However, one can see that for SPM grains with $N_{\rm cl} = 10^{4}$ in model wIA, the recovery rate within $<100$ au is very small of $\sim 30\%$ due to the wrong interpretation of dust polarization radiating from wrong aligned grains (Figure \ref{fig:pol_map_2mm_disk_wIA}, lower right panel). This problem thus emphasizes the importance of accurately determining the alignment direction of grains with $\B$-fields before deciding how to get $\B$ vectors from $\P$.

 \subsection{Origin of the polarization hole}\label{sec:polarization_hole}
One of the long-standing puzzles toward protostellar environments is the reduction of the polarization degree toward the central protostar, named the polarization hole (\citealt{Henning_2001}, \citealt{Gigart_2006}, \citealt{Hull_2014}, \citealt{Cox_2018}, \citealt{Galametz_2018}, \citealt{Kwon_2019}, \citealt{Ko_2020}). By fitting the relation of $p-I$ with the power law $p\sim I^{\alpha}$, the depolarization appears to vary case by case, from the low $\alpha \sim -0.4$ in the 1000 au scale of L1448 IRS2 (\citealt{Kwon_2019}), to $\sim -0.6$ in Bok globule (\citealt{Henning_2001}), to extremely high $\alpha \sim -0.97$ in NGC 2024 FIR 5 (\citealt{Lai_2002}) and $\sim -1$ for the inner region of L1448 IRS2 (\citealt{Kwon_2019}). The mechanism driving the depolarization effect is still unclear, but it is usually assigned to the geometric origin, such as the projection effect of magnetic fields (\citealt{Kataoka_2012}) and the tangling of $\B$ by turbulence; or the physical origin, such as the decrease of grain alignment efficiency due to gas randomization (\citealt{Hoang_Lazarian_2016a}, \citealt{Brauer_2016}); the change in dust population and composition (\citealt{Brauer_2016}); or the extinction of VLGs (\citealt{Brauer_2016}, \citealt{Liu_et_al_2016}, \citealt{Ko_2020}, \citealt{Liu_2021}).  However, the above studies and discussions assume the alignment model in the diffuse medium, which is not valid in protostellar environments (Discussion \ref{sec:discuss_align}). Recently, \cite{Giang_2022} indicated that the alignment loss of PM grains, and the inefficient internal and external alignment of SPM grains inside the protostellar disk, are the major origins causing the polarization hole in both optically thin and optically thick wavelengths of low-mass Class 0/I YSOs. However, our previous study used the uniform magnetic field model and a lower gas density profile than the results obtained in MHD simulations and observations of protostellar disks. Thus, this conclusion may overestimate the effect of grain alignment efficiency and underestimate the effect of turbulence, magnetic field geometry, gas density, and dichroic extinction on producing the polarization hole. 

By post-processing the realistic MHD simulation of the protostellar core and disk with POLARIS, in Section \ref{sec:p_I}, for the fixed value of $\Theta = 45^{\circ}$, one can see that in general, the reduced grain alignment efficiency suppresses the polarization degree much stronger than the effects of turbulence and magnetic field morphology, which is especially clear in the inner 200 au region (Figures \ref{fig:P_I_Ncl} and \ref{fig:P_I_amax}). The low grain alignment degree is also directly implied via the continuous decrease of quantity $p\times S$ with intensity at $<200$ au (Figures \ref{fig:PS_I_Ncl} and \ref{fig:PS_I_amax}).  

To understand in detail the mechanism behind the behavior of $p$ with intensity, we find the slope $\alpha$ in three separate regions, beyond 500au (envelope scale), between 200 and 500au, and within 200au (disk scale) (Section \ref{sec:slope_p_I}). We found that for $\Theta = 45^{\circ}$, the slope beyond $\sim 200$ au for SPM grains with $N_{\rm cl} = 10^{3} - 10^{4}$ is shallow with $p \sim I^{-0.2} - I^{-0.3}$ (Figure \ref{fig:alpha_inclination_angle}, left and central panels) and the corresponding polarization degree is rather high with $p\sim 10-40\%$ as found in model PA (Figure \ref{fig:P_I_Ncl}). The high $p \geq 10\%$ obtained in the envelope scale implies that SPM grains there could achieve perfect alignment with $\B$ by MRAT alignment. Under this condition, the depolarization effect beyond 200 au is totally induced by the projection effect of magnetic fields on the POS and the $\B$ field tangling by turbulence. In contrast, for SPM grains with lower $N_{\rm cl} \leq 10^{2}$ and for PM grains, their polarization-intensity curve is much steeper with $p \sim I^{-0.4} - I^{-0.6}$ and the corresponding polarization fraction is low of $p \sim 1 - 10\%$. The latter term implies that most of the grains inside the envelope still have magnetic alignment, but with inefficient IA, and they are only aligned with $\B$ by RATs. In this case, the reduction of the IA efficiency of aligned dust grains is the primary reason inducing the steep reduction of $p$ with intensity beyond $\sim 200$ au. In Appendix \ref{sec:alpha_Theta}, we change the inclination angle $\Theta$ from the face-on to the edge-on direction, and come with the same conclusion. 
 
Moving into the inner 200 au, grain alignment efficiency dominants turbulence and projection effect of magnetic fields on producing the depolarization, regardless of  grain magnetic properties, maximum grain size, and inclination angles (Figures \ref{fig:alpha_amax} and \ref{fig:alpha_inclination_angle}, right panel). In particular, for PM grains and SPM grains with low $N_{\rm cl} < 10$, the alignment loss around the protostar is the primary reason causing the reduction of $p$ with intensity in this region, characterized by the sleep slope $\alpha \sim -0.8$ to $\alpha \sim -1$. For SPM grains with higher $N_{\rm cl} > 10^{2}$ which can have the magnetic alignment in the disk scale, the reduced IA efficiency and the transition from MRAT alignment to RATs inside the disk are the main reasons producing the polarization hole within 200 au. As the maximum grain size increases, the amount of grains with inefficient IA and the possibility of grains losing their alignment with $\B$ increases. Consequently, grain growth further emphasizes the impact of inefficient grain alignment on producing the polarization hole within $\sim 200$ au around the protostar. This effect does not only work effectively at optically thin wavelengths, but also be more effective than dichroic extinction in producing the polarization hole at optically thick wavelengths (see below Section \ref{sec:dichroic_extinction}). Given that micron-sized grains above $\sim 1\mum$ are very easier to have inefficient IA by slow internal relaxation and even misalignment with $\B$ in protostellar disks, the deficient magnetic alignment of dust grains should be the key to producing the depolarization effect obtained in protostellar disks instead of the effect of magnetic field morphology, turbulence, and inclination angle, which vary with protostellar cores and observation conditions.

\subsection{Effect of dichroic extinction in producing $90^{\circ}$ flipping of the polarization pattern }\label{sec:dichroic_extinction}
As discussed in \cite{Brauer_2016}, dichroic extinction from aligned VLGs above $10\mum$ can reduce polarized thermal emission at submillimeter wavelengths and replace dichroic emission to become the dominant polarization mechanism in the optically thick disk. The change in the polarization mechanism induces the $90^{\circ}$ flipping of the polarization pattern from optically thin to optically thick wavelengths, as detected in the inner 100 au region of IRAS4A  (\citealt{Ko_2020}) and OMC-2/MMS 6 (\citealt{Liu_2021}). However, this explanation is based on the assumption that VLGs have efficient magnetic alignment even in very dense disks with $n_{\rm H} \sim 10^{9} - 10^{11}\cm^{-3}$, which is proved to be not valid in recent studies of \cite{Hoang_2022}, \citealt{Hoang+2022}, \citealt{Giang_2022}, and Section \ref{sec:discuss_align}.

In Section \ref{sec:p_nh_250um}, we show that at optically thick wavelengths of $250\mum$, dichroic extinction can cause the depolarization in the inner 200 au region if grains grow to $\sim 10\mum$. But taking into account the realistic alignment state of VLGs of  $\sim 10\mum$, the weak internal and external alignment of VLGs inside the disk compared with envelope grains induces much steeper reduction of $p(\%)$ toward the center. Therefore, even at optically thick wavelengths, the inefficient alignment of VLGs still be the priority in producing the polarization hole inside the disk (Section \ref{sec:polarization_hole}).

With the presence of VLGs above $50\mum$, we show Section \ref{sec:pol_map_disk_250um} that in model PA, dichroic extinction can cause the $90^{\circ}$ flipping of the polarization pattern between optically thin wavelengths of 2mm and optically thick wavelengths of $250\mum$ as expected in \cite{Brauer_2016}) (Figures \ref{fig:pol_map_2mm_disk_rIA} and \ref{fig:pol_map_250um_rIA}, upper panels). However, taking the realistic alignment degree of VLGs inside the disk into account, the flipping of polarization pattern only can be activated if VLGs are SPM with $N_{\rm cl} \sim 10^{4}$ and all aligned dust grains have right IA (Figures \ref{fig:pol_map_2mm_disk_rIA} and \ref{fig:pol_map_250um_rIA}, bottom panels). If SPM grains with high $N_{\rm cl} \sim 10^{4}$ have the wrong IA at low-\textit{J}, the polarization flipping does not appear as changing observed wavelengths from millimeter to submillimeter (Figure \ref{fig:pol_map_250um_wIA}, bottom panels). The independence of the polarization pattern with wavelengths is also found for SPM grains with lower $N_{\rm cl} \leq 10^{2}$ and for PM grains (Figure \ref{fig:pol_map_250um_wIA}) owing to the misalignment of large micron-sized grains with $\B$ inside the disk (Figure \ref{fig:alar_amax}, upper panels). Therefore, it is impossible to assign dichroic extinction to the origin of the polarization hole detected at optically thick wavelengths and the $90^{\circ}$ flipping of $\P$ with wavelengths without carefully examining the magnetic properties and alignment direction of grains with $\B$. 
  
\subsection{Probing grain properties with multiwavelength dust polarization}

\begin{figure*}
\centering
\includegraphics[width=\textwidth,height=\textheight,keepaspectratio]{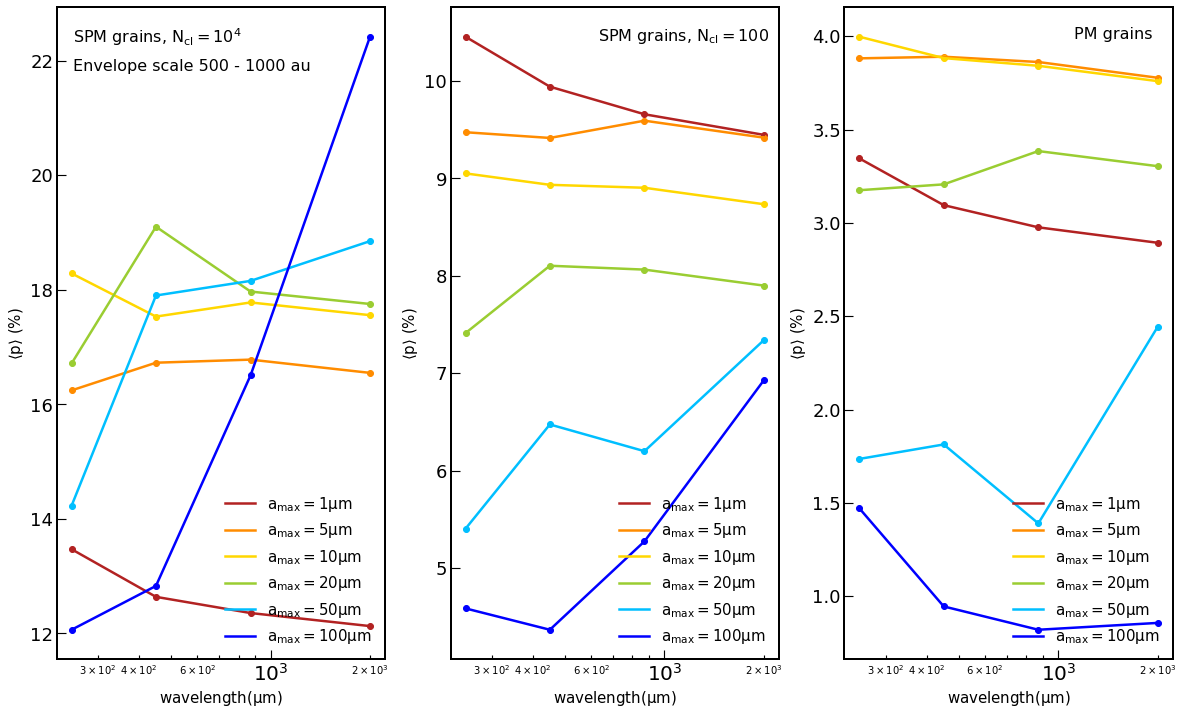}
     
    \caption{Left panel: variation of the mean polarization degree beyond 500 au from $250\mum$ to 2 mm for different maximum grain sizes, assuming SPM grains with high $N_{\rm cl} = 10^{4}$. Central and right panels: similar results as the left panel but for SPM grains with low $N_{\rm cl} = 100$ and for PM grains. One can see that $p$ clearly decreases with increasing wavelengths if the maximum size is small of $a_{\rm max} \sim 1\mum$ due to the weak emission of small grains at millimeter wavelengths. As the maximum grain size increases, the decrease of $p$ with $\lambda$ is weaker and $p$ then increases toward millimeter wavelengths if $a_{\rm max}\geq 50\mum$. The tendency of the polarization curve is similar for both SPM and PM grains, but the rise of $p$ with wavelengths for PM grains is not prominent as in the case of SPM grains due to the weak alignment of VLGs in protostellar environments.} 
     \label{fig:p_lambda_500au}
\end{figure*}

\subsubsection{Constrain of the grain alignment direction and $\P-\B$ orientation}\label{sec:alignment_direction}
We discuss in Section \ref{sec:discuss_iron_pol_pattern} that dust polarization from SPM grains with high $N_{\rm cl} \sim 10^{4}$ can trace magnetic fields inside the disk. However, large micron-sized and VLGs inside the disk tend to have slow internal relaxation at both high and low-\textit{J} attractors (Section \ref{sec:discuss_align}), inducing the unclear alignment direction between dust grains and $\B$ (\citealt{Hoang_Lazarian_2009}). As a result, understanding the grain alignment direction is the key to accurately inferring magnetic field morphology from dust polarization obtained in the inner $\sim 100$ au region. This issue is also important to quantify the effect of dichroic extinction and analyze the origin behind the $90^{\circ}$ flipping of the polarization pattern with wavelengths as obtained in several disks of Class 0/I YSOs (Section \ref{sec:dichroic_extinction}). 

As discussed in Section \ref{sec:dichroic_extinction}, the presence of grains with the right and wrong IA affects differently the variation of the polarization pattern inside the disk with wavelengths, given grains can grow to $\geq 50\mum$ and have high amount of iron inclusions of $N_{\rm cl} \sim 10^{4}$. We will discuss the constraints of $a_{\rm max}$ and $N_{\rm cl}$ in the following section. Assuming specific combinations of $a_{\rm max}\geq 50\mum$ and $N_{\rm cl} \sim 10^{4}$, we found that $\P$ inside the disk can flip $90^{\circ}$ as changing the observed wavelength from optically thin to optically thick if almost aligned VLGs have the right IA. Conversely, the polarization pattern inside the disk is consistent between optically thick and thin wavelengths if VLGs have the wrong IA. Thus, the search for polarization flipping between optically thin and thick wavelengths may help us to constrain the alignment direction of dust grains. If the polarization pattern is consistent with wavelengths, $\P$ vectors already imply $\B$ vectors because grains with the wrong IA are aligned with their longest axis along the magnetic field direction. If the flipping happens, almost aligned dust grains are aligning their longest axis perpendicular to $\B$. Consequently, rotating $\P$ vectors obtained at optically thin wavelengths by $90^{\circ}$ or keeping $\P$ vectors at optically thick wavelengths will refer us to the correct map of $\B$ fields around the protostar. In this case, the change of the polarization mechanism from dichroic emission to dichroic extinction is the origin behind the polarization flipping.

Another sign for inferring the grain alignment direction with $\B$ is the $90^{\circ}$ flipping of $\P$ at optically thin wavelengths (\citealt{Giang_2022}). This flipping is expected to occur when the dominant emission source changes from VLGs with the wrong IA  to micron-sized grains with the right IA.  In our simulation, the $90^{\circ}$ flipping of $\P$ at optically thin 2mm appears along the North-East direction beyond $\sim 200$ au for SPM grains with moderate $N_{\rm cl} = 100$ (Figure \ref{fig:pol_map_2mm_disk_rIA}). However, this feature is not clear and rather hard to recognize given the complex infall and accretion gas flow toward the disk. Inside the protostellar disk, no clear flipping of the polarization pattern between optically thin wavelengths of 2mm, $870\mum$, and $450\mum$ (Sections \ref{sec:pol_map_disk},  \ref{sec:870um}, and \ref{sec:450um}) is found in model wIA.  We suggest the high disk density in our simulation is responsible for the disappearance of the polarization flipping at optically thin wavelengths, given the large amount of grains with the wrong IA within $\sim 200$ au. In this case, inferring the grain alignment direction and the orientation of $\P$ and $\B$ via optically thin wavelengths is impossible. The polarization flipping at optically thin wavelengths may occur if nuclear relaxation and inelastic relaxation are taken into account (Section \ref{sec:discuss_align}). In this case, one can recognize the presence of VLGs with the wrong IA, which produces  $\P \parallel \B$ at longer wavelengths, and micron-sized grains with the right IA, which produces $\P \perp \B$ at shorter wavelengths (\citealt{Giang_2022}). However, if their efficiency is too high, they may completely eliminate the presence of grains with slow internal relaxation inside the disk, leading to the consistency of $\P$ at optically thin wavelengths and deactivating the usability of this method. This motivates future studies that take into account all well-known internal alignment mechanisms to accurately understand the internal alignment of large grains inside protostellar environments. Nevertheless, the search for polarization flipping at optically thin wavelengths is another way that we suggest to constrain the alignment direction of dust grains around the protostar.

We note that the exact degree of grain alignment at low-J attractors for the case of slow internal relaxation is not quantified yet (\citealt{Hoang_Lazarian_2009}). In our study, we consider that grains with slow internal relaxation always have low alignment degree due to the thermal wobbling inside grains (\citealt{Lazarian_1994}, \citealt{Lazarian_Roberge_1997}, \citealt{Weingartner_Draine_2003}, \citealt{Hoang_Lazarian_2008}), characterized by low values of $Q_{\rm X}^{\rm low-J}$ and $Q_{\rm X}^{\rm high-J}$ (Table \ref{tab:parameter}). However, \cite{Hoang_Lazarian_2009} indicated that there still be some configurations that support efficient 'right' alignment between $\J$ and $\hat{a}_{1}$ for grains having slow internal relaxation at high-\textit{J}, i.e., if the symmetric axis initially makes a small angle with $\J$. The recent multi-wavelength study of dust polarization in HL Tau by \cite{Thang_2023} also indicates that the combination of self-scattering and polarized dust emission from VLGs with efficient 'wrong' alignment at low-\textit{J} can reproduce the wavelength-dependence polarization pattern of HL Tau. However, which conditions can support and maintain the efficient 'right' or 'wrong' alignment of grains with slow internal relaxation inside the disk, and whether such grains have an efficient or inefficient internal alignment degree are unclear. Therefore, searching for the existence of grains with the wrong IA around the protostar is very important to study the dynamics and properties of grains with slow internal relaxation. Indeed, the amount of grains with slow internal relaxation inside the protostellar disk may be smaller than our expectation (Section \ref{sec:discuss_align}), and we may not see the imprint of grains with the wrong IA in observed polarized dust emission. But even in this case, multi-wavelength observations are still necessary for confirming the alignment degree and alignment direction of grains inside the disk environments.

\subsubsection{Constraining the grain magnetic properties and grain growth}\label{sec:magnetic_prop_amax}
Given the positive correlation between the amount of iron inclusions inside dust grains and the produced polarization degree, \citealt{Giang_2022} suggests a new advantage of dust polarization in constraining the grain magnetic properties. However, as illustrated in Figures \ref{fig:P_I_Ncl} and \ref{fig:P_I_amax}, the polarization degree obtained in protostellar envelopes and disks is controlled by both the grain magnetic properties and the maximum grain size. $p(\%)$ still maintains a simple positive correlation with $N_{\rm cl}$, but it behaves differently with the maximum grain size depending on the grain magnetic properties. In particular, for SPM grains with high $N_{\rm cl} \sim 10^{4}$, $p(\%)$ increases with increasing $a_{\rm max}$ in the envelope scale, but decreases with $a_{\rm max}$ inside the disk owing to the reduction of the IA and the change of the external alignment of VLGs. For SPM grains with lower $N_{\rm cl} < 10^{2}$ and for PM grains, $p(\%)$ within 1000 au to the protostar is smaller as $a_{\rm max}$ increases due to the reduction of the IA and the increase in amount of randomized grains in the central region. The coupling between the increase/decrease of $p(\%)$ with $N_{\rm cl}$ and $a_{\rm max}$ makes the possibility for probing iron inclusions inside dust grains via dust polarization more complicated, unless $a_{\rm max}$ is already constrained previously.
 
Photometry toward the inner envelope and the protostellar disk of Class 0/I YSOs frequently reports the low dust opacity index of $\beta \sim 1$ (\citealt{Kwon_2009}, \citealt{Miotello_2014}, \citealt{Galametz_2019}, \citealt{Cacciapuoti_2023a}), implying the existence of VLGs above $10\mum$ within few hundreds au around the protostar. The presence of VLGs in the inner envelope is also indirectly revealed via the high dust polarization degree above $p \geq 5\%$ (\citealt{Valdivia_2019}, \citealt{Valentin_2023b}), and the rise of $p(\%)$ toward millimeter wavelengths (\citealt{Valdivia_2019}, \citealt{Yen_2020}). However, the previous studies on the effect of grain growth on dust polarization by \cite{Valdivia_2019} and \cite{Valentin_2023b} assumed fast internal relaxation for all aligned dust grains and did not consider the alignment loss of VLGs inside the protostellar envelope and disk. Given the complex alignment state of VLGs around the protostar, it is unclear whether the grain growth still holds the positive correlation of $p(\%)$ and wavelengths. We revisit the variation of the mean polarization degree $\langle p \rangle(\%)$ found in the envelope beyond $500$ au to the protostar with wavelengths from $250\mum$ to 2 mm for different maximum grain size from $a_{\rm max} = 1\mum$ to $a_{\rm max} = 100\mum$ in Figure \ref{fig:p_lambda_500au}. The left panel shows results for SPM grains with high $N_{\rm cl} = 10^{4}$, the central panel is for SPM grains with low $N_{\rm cl} = 100$, and the right panel is for PM grains. The results for $p-\lambda$ relation in the inner envelope and protostellar disk are shown in Appendix \ref{sec:p_lambda}. Excluding the complex variation of $\langle p \rangle$ with $a_{\rm max}$ and $N_{\rm cl}$, one can see that for both SPM and PM grains, $\langle p \rangle$ decreases toward longer wavelengths if the maximum size is small of $a_{\rm max} = 1\mum$. As $a_{\rm max}$ increases, the decrease of $\langle p \rangle$ with $\lambda$ becomes weaker because of increasing polarized dust emission from large grains at long wavelengths, then $\langle p \rangle$ changes to increase with increasing wavelengths as grains grow to $a_{\rm max} \geq 50\mum$. In contrast, for PM grains, $p$ tends to decrease toward longer wavelengths even when grains grow to above $50\mum$ owning to the alignment loss of PM grains with sizes above $10\mum$ there (Figure \ref{fig:alar_amax}, upper panels). The results for the inner 500 au region are generally consistent with the tendency of $\langle p \rangle -\lambda$ for different $a_{\rm max}$ as found in the envelope scale (see Appendix \ref{sec:p_lambda}). Thus, we confirm that the maximum grain size (or the grain growth activities) could be recognized by observing multiwavelength dust polarization from submillimeter to millimeter wavelengths (\citealt{Valdivia_2019}).

Given the first constraint on the maximum grain size, the grain magnetic properties (including amount of iron inclusions locked inside dust grains) can be estimated by fitting the observed polarization degree with simulated results from different dust models. Otherwise, two quantities will be the free parameters. In this case, we suggest looking into the polarization spectrum (\citealt{Valdivia_2019}, Figure \ref{fig:p_lambda_500au}), the overall degree of dust polarization (Figure \ref{fig:P_I_Ncl}), and the slope of $p-I$ (Section \ref{sec:slope_p_I}) to have the first constraints on $a_{\rm max}$  and grain magnetic properties. Then, the accurate values of the above quantities can be found by confronting the synthetic dust polarization degree from POLARIS with observational data. Based on the consistency between grain properties - grain alignment degree - and synthetic dust polarization implemented inside the updated POLARIS version, now we can accurately extract both information on magnetic properties and grain growth inside the star-forming region. One can have a look at \cite{Ngoc_2023} and \cite{Akshaya_2023} who carry the first numerical interpretation of grain magnetic properties and maximum grain size via dust polarization in the massive filament G11.11$-$0.12 and the Galactic center.
 
\subsection{Effects of grain alignment on the B-field strength measured by the DCF technique}

Dust polarization is a popular tool to measure the magnetic field strength via the David-Chandrasekhar-Fermi (DCF) method (\citealt{David_1951}, \citealt{chandraseshar_1953}). The key assumptions of the DCF technique include (1) the balance between turbulent kinetic energy and turbulent magnetic energy, and (2) the polarization angle dispersion describes the turbulent component of the magnetic field in the POS. Obviously, this technique assumes uniform alignment throughout the cloud and neglects the effect of grain properties. Therefore, the DCF technique can provide a reasonable measurement of the B-field strength for the cloud, as given by
\bea
B_{\rm POS}=\sqrt{4\pi \rho}\frac{\sigma_{V}}{\sigma_{\psi}}
\ena,
where $\rho$ is the gas mass density, $\sigma_{V}$ is the turbulent velocity dispersion, and $\sigma_{\psi}$ is the standard deviation of the polarization angle, which is equivalent to $S$. Note that various improvements have been proposed to increase the accuracy of the DCF method, but the role of the key parameter, $\sigma_{\psi}$, remains unchanged (see e.g., \citealt{Pattle_2022}, \citealt{Lazarian.2022} and references therein). 
 
However, our results in Section \ref{sec:S_I} indicate that grain alignment efficiency also significantly affects the polarization angle dispersion function, $S$, which depends on the grain magnetic properties and maximum grain size. In particular, grains with low iron inclusions are not aligned with $\B$ in the inner $\sim 200$ au region. Consequently, their dust polarization only contains magnetic field and turbulence information in the envelope of thousands au scale from the protostar, producing low $S$ in the entire observed region (Figure \ref{fig:S_I_Ncl}). In contrast, grains with high iron can probe magnetic fields and turbulence in the deeper regions, reflecting more accurately the strength of turbulence inside the core (Figure \ref{fig:S_I_Ncl}). However, even in this case, interpreting wrong the origin of polarized dust emission can induce extra dispersion on polarization angle than reality (Figure \ref{fig:S_I_Ncl}, right panel). Besides, grain growth in the envelope can induce higher $S$ in the envelope for grains with high iron inclusions (Figure \ref{fig:S_I_amax}, upper panels). But for a few hundred au around the protostar, the increasing maximum grain size clearly induces lower $S$ than reality (model PA) due to the narrower alignment area (Figure \ref{fig:S_I_amax}, lower panels). The decrease/increase of S with iron inclusions can affect the estimated strength of the magnetic field in protostellar cores, especially in the central region. Since the magnetic field strength is an important parameter for understanding the role of $\B$-fields in regulating star formation together with other parameters such as gravity, rotational energy, thermal and non-thermal energy, further studies about the effect of iron inclusions on the prefactor of the DCF method should be done to accurately estimate the value of $\B$ in such dense environments.

\subsection{Other external alignment mechanisms and self-scattering}\label{sec:other_polarization}
 \subsubsection{Self-dust scattering of thermal dust emission}
So far, we have shown that dust polarization can be used to trace magnetic fields and probe dust physics inside the protostellar disk if VLGs inside the disk are SPM with high amount of iron inclusions (Sections \ref{sec:discuss_iron_pol_pattern} and \ref{sec:magnetic_prop_amax}). However, the scattering between VLGs in this region and thermal dust emission also can produce dust polarization at submillimeter wavelengths, known as self-scattering (\citealt{Kataoka_2015}, \citealt{Yang_2016}). Self-scattering is widely suggested as another dust polarization source (\citealt{Lam_2021}) 
given that its uniform polarization pattern along the inclined disk minor axis has been detected in many protostellar disks of Class 0/I YSOs (\citealt{Cox_2018}). As shown in Section \ref{sec:p_I}, both PM grains and SPM grains with low $N_{\rm cl} < 10^{2}$ produce negligible polarization degree $p<<1\%$ inside the disk because of their misalignment with magnetic fields within few hundreds au around the protostar. In this case, self-scattering may play a major role in producing dust polarization with $p \sim 1\%$ inside the disk of Class 0/I YSOs (\citealt{Lam_2021}). In contrast, for SPM grains with high $N_{\rm cl} \sim 10^{4}$, polarized dust emission can also produce $p \geq 1\%$ inside the disk at submillimeter wavelengths. However, how the outcome from the superposition of dust polarization from two distinct mechanisms looks like is unclear. Given many objects whose polarization pattern and polarization degree inside the disk cannot be simply explained by only thermal emission of magnetically aligned dust grains or self-scattering (\citealt{Kwon_2019}, \citealt{Takahashi_2019}), further synthetic modeling of dust polarization combining the above mechanisms simultaneously must be performed to diagnose their operation conditions and how to recognize them via observational data. 
 
\subsubsection{Grain alignment along the radiation direction}\label{sec:krat}
Within the RAT paradigm, grains may also be aligned along the radiation direction (i.e., k$-$RAT alignment, \citealt{Lazarian_2007}, \citealt{Lazarian_Hoang_2007}) when the radiative precession is faster than Larmor precession and gas randomization. Unlike Larmor precession which is not affected by the grain rotational rate, the radiative precession is strongly suppressed as grains rotate suprathermally (\citealt{Lazarian_Hoang_2007}). Consequently, k$-$RAT tends to be effective for micron-sized grains at low-\textit{J} inside the disk (\citealt{Hoang+2022}) which always have slow internal relaxation. Considering the situation that SPM grains at high-\textit{J} can be aligned with $\B$ inside the disk, if the size range and the alignment degree of grains with k$-$RAT is much smaller than those having magnetic alignment, k$-$RAT signal may either be hidden or just cause additional distortion in the observed polarization pattern. In contrast, k$-$RAT signal from grains at low-\textit{J} can dominate B$-$RAT signal from grain at high-\textit{J} as seen in results obtained in model wIA (Figure \ref{fig:pol_map_2mm_disk_wIA}). Consequently, the observed dust polarization pattern will tell us about the radiation field direction instead of the magnetic field direction around the protostar. In another situation where VLGs cannot have the magnetic alignment inside the disk (PM or SPM grains with low $N_{\rm cl} \leq 100$), k$-$RAT may become the major source of dust polarization in this region if the radiative precession is faster than the gas randomization. However, their emission may be hidden behind the strong dust polarization from foreground grains which mostly have efficient IA at high-\textit{J} attractors. Furthermore, since the alignment direction and alignment degree of grains with slow internal relaxation at low-\textit{J} are not well studied, whether we can recognize hints of radiative alignment from observations remains unclear.

\subsubsection{MEchanical torques, B$-$MET and v$-$MET alignment}\label{sec:mets}
Besides RAT, dust grains also can be spun up and then aligned with $\B$ by MEchanical Torques (METs) resulting from the drift of dust grains inside the accretion gas (\citealt{Hoang+2018}, \citealt{Hoang_2020}), or B$-$MET alignment. As the RAT efficiency is weakened in protostellar disk owing to the strong attenuation of the stellar radiation field strength and its high isotropic degree (Figure \ref{fig:U_gamma_Td}, left panel), METs may play an important role in improving the magnetic alignment in this region as expected in \cite{Hoang+2022}. As grains can rotate faster with the additional spinning torques from METs, more submicron-grains can be aligned with $\B$, and more large grains can have fast internal relaxation at high-\textit{J} attractors. Consequently, one will have more confidence in assigning the observed dust polarization to the emission of magnetically aligned dust grains and can confidently rotate $\P$ by $90^{\circ}$ to infer $\B$. METs are also expected to work effectively in strong turbulent environments where the grain drift relative to molecular gas becomes significant. However, given the strong tangling of $\B$ fields by turbulence, whether we can recover full advantage of dust polarization inside the disk is unclear. Indeed, the basic properties of METs are not universal and have not yet been quantified in detail as RATs (\citealt{Hoang+2018}). \cite{Reissl_2023} shows that the MET efficiency can change up to several magnitudes as changing grain shapes and sizes. However, given their potentially important role in spinning up and driving the external alignment of VLGs, the detailed study of the magnetic alignment driven by METs is very important to accurately understand the grain alignment in such dynamic, turbulent disk environments.

Besides B$-$MET, METs also can lead grains to be aligned with the gas flow, or v$-$MET alignment (\citealt{Lazarian_Hoang_2007}). This alignment will happen if the mechanical precession is faster than Larmor precession, radiative precession, and gas randomization. Similar to k$-$RAT, v$-$MET prefers grains at low-\textit{J} attractors and may be visible if their polarization signal dominates emission from magnetically aligned dust grains at high-\textit{J}. The prediction of polarization patterns from different alignment mechanisms and alignment directions of dust grains is quantitatively described in \cite{Hoang+2022}. However, given the dual role of RATs/METs in enhancing magnetic alignment for grains at high-\textit{J} and producing k$-$RAT/v$-$MET for grains at low-\textit{J} inside the disk, how actual they control the picture of grain alignment in this region remain unclear. Moreover,  considering the superposition of dust polarization produced from more than two distinct alignment sources, future incorporation of k$-$RAT and METs into POLARIS must be done to exactly quantify the contribution of each mechanism on the net polarization signal. This work will set the catalog for us to apply to interpret observational data

\subsubsection{Factors suppressing external alignment of dust grains}
In our study, gas randomization is the major mechanism distorting the magnetic alignment of dust grains. But as discussed in Sections \ref{sec:krat} and \ref{sec:mets}, as grains at low-\textit{J} can subject to both the Larmor precession, radiative precession, and mechanical precession, the dual precession of the grain magnetic moment around more than one alignment direction also causes additional decrease of external alignment of dust grains (\citealt{Lazarian_Hoang_2019}). This issue is more important as grains contain iron inclusions (\citealt{Lazarian_Hoang_2019}), thus emphasizing the importance of understanding whether k$-$RAT and v$-$MET can coexist with B$-$RAT and B$-$RAT inside the disk. Besides, the saturation of the magnetic response due to the fast rotation also can limit the effect of magnetic relaxation in driving the external magnetic alignment of grains (\citealt{Lazarian_Hoang_2019}). In our study, grains around low-mass protostar do not rotate fast enough to reach the saturation point of the magnetic susceptibility. But around high-mass protostar where grains receive efficient RATs by stronger radiation field and are even expected to be disrupted by RAdiative Torque Disruption (\citealt{Hoang_2019}), this issue may happen that limits the MRAT alignment on VLGs around the protostar.

\section{Summary}\label{sec:summary}
In this study, we post-process an MHD simulation of a protostellar core with our updated POLARIS code to study the effects of grain magnetic properties and maximum grain sizes on synthetic dust polarization from magnetically aligned dust grains. Our main findings are summarized as follows:

 \begin{enumerate}

 \item We found that only PM grains below $10\mum$ can have the magnetic alignment beyond $\sim 200$ au by RATs, but most of them have inefficient internal alignment (IA) by slow internal relaxation. In contrast, SPM grains with a high amount of iron inclusions can have perfect magnetic alignment in the envelope by an efficient MRAT mechanism. However, in the disk scale of $\sim 200$ au around the protostar, very large SPM grains (VLGs) above $40\mum$ are not aligned with $\B$ and large micron-sized grains mostly have inefficient IA at both high and low-\textit{J} attractors due to slow internal relaxation. In contrast to SPM grains in the envelope which can be aligned with $\B$ by MRAT mechanism, RAT is the major external alignment of SPM grains inside the disk with low $f_{\rm high-J} \sim 0.25$. The internal alignment degree of VLGs in the envelope and disk may improve if we take into account the full effect of Barnett relaxation, nuclear relaxation, and inelastic relaxation.
  
\item We found a positive correlation between the degree of polarization $p$ and the level of iron inclusions locked inside dust grains. Besides, for grains with high embedded iron inclusions, $p$ observed in the envelope tends to increase with grain growth, $a_{\rm max}$. However, toward the inner region, $p$ only slightly increases with the maximum size up to $a_{\rm max} = 20\mum$, and a further increase in $a_{\rm max}$ decreases the polarization degree, $p$. For grains with low embedded iron inclusions, grain growth tends to reduce $p$, and this anti-correlation trend becomes more prominent for grains with lower magnetic susceptibility.

\item We found that for SPM grains with high iron inclusions, turbulence and geometrical effect of magnetic fields is the major origin driving the depolarization effect in the envelope scale with the weak relation $p\sim I^{-0.3}$. In the inner 500 au region, the depolarization is mainly caused by the reduced IA and the change of external alignment from MRAT to RATs, resulting in the relation $p \sim I^{-0.7}$. For grains with a lower level of iron inclusions, the reduction of the IA efficiency is the origin behind the depolarization in thousands au scale, with $p\sim I^{-0.6}$, and the alignment loss is the origin of the polarization hole found in the disk scale with very steep relation $p \sim I^{-1}$. 
 
\item  We found the increase in the polarization angle dispersion function, $S$, with increasing the grain magnetic susceptibility due to the broadening in the region where grains can be aligned with magnetic fields. Grain growth also affects $S$. Further numerical studies on the effects of grain alignment and growth on $S$ are needed to achieve accurate measurements of the $\B$-field strength using the DCF method. 

\item Dust polarization is a robust tool for tracing the magnetic field orientation in the envelope scale, with their polarization vectors always being $\P \perp \B$. In the inner $200$ au around the protostar, only dust polarization from SPM grains with high embedded iron inclusions can be used to infer again $\B$ field orientation around the protostar. However, the alignment direction of grains with $\B$ must be examined carefully before deciding how to get $\B$ from $\P$. 

\item We found that the polarization spectrum produced by both PM and SPM grains tends to rise toward millimeter wavelengths if grains grow to above $a_{\rm max} \geq 10\mum$, which could be treated as a sign of grain growth in protostellar environments. The strong positive correlation between $p$ and iron inclusions opens the powerful method for us to constrain the level of iron locked inside dust grains via dust polarization. In addition, we suppose that the alignment direction of grains with $\B$ inside the disk can be constrained by searching for the $90^{\circ}$ flipping of the polarization pattern within $\sim 200$ au from optically thin to optically thick wavelengths.

\item We found that at optically thick submillimeter wavelengths, the inefficient alignment of VLGs with $\B$ dominates the effect of dichroic extinction in producing the polarization hole obtained within the disk scale of $\sim 200$ au. In addition, dichroic extinction only can become the major source of polarization and produce the $90^{\circ}$ flipping of the polarization pattern between optically thin and thick wavelengths inside the disk only if 1) grains grow to above $a_{\rm max} \geq 50\mum$, 2) they are SPM with a high amount of iron clusters, and 3) almost aligned dust grains at both high and low-\textit{J} attractors have the right IA.
 
 \end{enumerate}
   
\section*{Acknowledgements}
We thank the anonymous referee for insightful comments that helped improve our paper. We thank Jeong-Gyu Kim for stimulating discussions and Ka Ho Lam for sharing with us the MHD simulation datacube. We thank the members of the Vietnam Astrophysics Research Network (VARNET) for various useful discussions and comments. T.H. is supported by the National Research Foundation of Korea (NRF) grant funded by the Korean government (MSIT (No. 2019R1A2C1087045).  

 
\section*{Data Availability}
The data underlying this article will be shared on reasonable request to the corresponding author.

\bibliographystyle{mnras}
\bibliography{main.bib} 

\begin{thebibliography}{}
\makeatletter
\relax
\def\mn@urlcharsother{\let\do\@makeother \do\$\do\&\do\#\do\^\do\_\do\%\do\~}
\def\mn@doi{\begingroup\mn@urlcharsother \@ifnextchar [ {\mn@doi@} {\mn@doi@[]}}
\def\mn@doi@[#1]#2{\def\@tempa{#1}\ifx\@tempa\@empty \href {http://dx.doi.org/#2} {doi:#2}\else \href {http://dx.doi.org/#2} {#1}\fi \endgroup}
\def\mn@eprint#1#2{\mn@eprint@#1:#2::\@nil}
\def\mn@eprint@arXiv#1{\href {http://arxiv.org/abs/#1} {{\tt arXiv:#1}}}
\def\mn@eprint@dblp#1{\href {http://dblp.uni-trier.de/rec/bibtex/#1.xml} {dblp:#1}}
\def\mn@eprint@#1:#2:#3:#4\@nil{\def\@tempa {#1}\def\@tempb {#2}\def\@tempc {#3}\ifx \@tempc \@empty \let \@tempc \@tempb \let \@tempb \@tempa \fi \ifx \@tempb \@empty \def\@tempb {arXiv}\fi \@ifundefined {mn@eprint@\@tempb}{\@tempb:\@tempc}{\expandafter \expandafter \csname mn@eprint@\@tempb\endcsname \expandafter{\@tempc}}}

\bibitem[\protect\citeauthoryear{{Akshaya} \& {Hoang}}{{Akshaya} \& {Hoang}}{2023}]{Akshaya_2023}
{Akshaya} M.~S.,  {Hoang} T.,  2023, \mn@doi [\mnras] {10.1093/mnras/stad1246}, \href {https://ui.adsabs.harvard.edu/abs/2023MNRAS.522.4196A} {522, 4196}

\bibitem[\protect\citeauthoryear{{Allen}, {Li}  \& {Shu}}{{Allen} et~al.}{2003}]{Allen_2003}
{Allen} A.,  {Li} Z.-Y.,   {Shu} F.~H.,  2003, \mn@doi [\apj] {10.1086/379243}, \href {https://ui.adsabs.harvard.edu/abs/2003ApJ...599..363A} {599, 363}

\bibitem[\protect\citeauthoryear{{Andersson}, {Lazarian}  \& {Vaillancourt}}{{Andersson} et~al.}{2015}]{Anderson_2015}
{Andersson} B.~G.,  {Lazarian} A.,   {Vaillancourt} J.~E.,  2015, \mn@doi [\araa] {10.1146/annurev-astro-082214-122414}, \href {https://ui.adsabs.harvard.edu/abs/2015ARA&A..53..501A} {53, 501}

\bibitem[\protect\citeauthoryear{{Bally}}{{Bally}}{2016}]{Bally_2016}
{Bally} J.,  2016, \mn@doi [\araa] {10.1146/annurev-astro-081915-023341}, \href {https://ui.adsabs.harvard.edu/abs/2016ARA&A..54..491B} {54, 491}

\bibitem[\protect\citeauthoryear{Barnett}{Barnett}{1915}]{Barnett_1915}
Barnett S.~J.,  1915, \mn@doi [Phys. Rev.] {10.1103/PhysRev.6.239}, 6, 239

\bibitem[\protect\citeauthoryear{{Basu} \& {Mouschovias}}{{Basu} \& {Mouschovias}}{1994}]{Basu_1994}
{Basu} S.,  {Mouschovias} T.~C.,  1994, \mn@doi [\apj] {10.1086/174611}, \href {https://ui.adsabs.harvard.edu/abs/1994ApJ...432..720B} {432, 720}

\bibitem[\protect\citeauthoryear{{Bich Ngoc} et~al.,}{{Bich Ngoc} et~al.}{2023}]{Ngoc_2023}
{Bich Ngoc} N.,  et~al., 2023, \mn@doi [arXiv e-prints] {10.48550/arXiv.2302.10543}, \href {https://ui.adsabs.harvard.edu/abs/2023arXiv230210543B} {p. arXiv:2302.10543}

\bibitem[\protect\citeauthoryear{{Brauer}, {Wolf}  \& {Reissl}}{{Brauer} et~al.}{2016}]{Brauer_2016}
{Brauer} R.,  {Wolf} S.,   {Reissl} S.,  2016, \mn@doi [\aap] {10.1051/0004-6361/201527546}, \href {https://ui.adsabs.harvard.edu/abs/2016A&A...588A.129B} {588, A129}

\bibitem[\protect\citeauthoryear{{Cacciapuoti} et~al.,}{{Cacciapuoti} et~al.}{2023}]{Cacciapuoti_2023a}
{Cacciapuoti} L.,  et~al., 2023, \mn@doi [\aap] {10.1051/0004-6361/202346204}, \href {https://ui.adsabs.harvard.edu/abs/2023A&A...676A...4C} {676, A4}

\bibitem[\protect\citeauthoryear{Chandrasekhar \& Fermi}{Chandrasekhar \& Fermi}{1953}]{chandraseshar_1953}
Chandrasekhar S.,  Fermi E.,  1953, \apj, 118, 116

\bibitem[\protect\citeauthoryear{{Ching}, {Lai}, {Zhang}, {Girart}, {Qiu}  \& {Liu}}{{Ching} et~al.}{2017}]{Ching_2017}
{Ching} T.-C.,  {Lai} S.-P.,  {Zhang} Q.,  {Girart} J.~M.,  {Qiu} K.,   {Liu} H.~B.,  2017, \mn@doi [\apj] {10.3847/1538-4357/aa65cc}, \href {https://ui.adsabs.harvard.edu/abs/2017ApJ...838..121C} {838, 121}

\bibitem[\protect\citeauthoryear{{Cox} et~al.,}{{Cox} et~al.}{2015}]{Cox_2015}
{Cox} E.~G.,  et~al., 2015, \mn@doi [\apjl] {10.1088/2041-8205/814/2/L28}, \href {https://ui.adsabs.harvard.edu/abs/2015ApJ...814L..28C} {814, L28}

\bibitem[\protect\citeauthoryear{Cox, Harris, Looney, Li, Yang, Tobin  \& Stephens}{Cox et~al.}{2018}]{Cox_2018}
Cox E.~G.,  Harris R.~J.,  Looney L.~W.,  Li Z.-Y.,  Yang H.,  Tobin J.~J.,   Stephens I.,  2018, \mn@doi [The Astrophysical Journal] {10.3847/1538-4357/aaacd2}, 855, 92

\bibitem[\protect\citeauthoryear{{Dapp}, {Basu}  \& {Kunz}}{{Dapp} et~al.}{2012}]{Dapp_2012}
{Dapp} W.~B.,  {Basu} S.,   {Kunz} M.~W.,  2012, \mn@doi [\aap] {10.1051/0004-6361/201117876}, \href {https://ui.adsabs.harvard.edu/abs/2012A&A...541A..35D} {541, A35}

\bibitem[\protect\citeauthoryear{{Davidson} et~al.,}{{Davidson} et~al.}{2014}]{Davidson_2014}
{Davidson} J.~A.,  et~al., 2014, \mn@doi [\apj] {10.1088/0004-637X/797/2/74}, \href {https://ui.adsabs.harvard.edu/abs/2014ApJ...797...74D} {797, 74}

\bibitem[\protect\citeauthoryear{Davis~Jr \& Greenstein}{Davis~Jr \& Greenstein}{1951}]{David_1951}
Davis~Jr L.,  Greenstein J.~L.,  1951, \apj, 114, 206

\bibitem[\protect\citeauthoryear{{Dolginov} \& {Mitrofanov}}{{Dolginov} \& {Mitrofanov}}{1976}]{Dolginov_1976}
{Dolginov} A.~Z.,  {Mitrofanov} I.~G.,  1976, \mn@doi [\apss] {10.1007/BF00640010}, \href {https://ui.adsabs.harvard.edu/abs/1976Ap&SS..43..291D} {43, 291}

\bibitem[\protect\citeauthoryear{{Draine}}{{Draine}}{1996}]{Draine_1996}
{Draine} B.~T.,  1996, in {Roberge} W.~G.,  {Whittet} D. C.~B.,  eds,  Astronomical Society of the Pacific Conference Series Vol. 97, Polarimetry of the Interstellar Medium. p.~16 (\mn@eprint {arXiv} {astro-ph/9603053})

\bibitem[\protect\citeauthoryear{{Draine} \& {Weingartner}}{{Draine} \& {Weingartner}}{1996}]{Draine_Weingartner_1996}
{Draine} B.~T.,  {Weingartner} J.~C.,  1996, \mn@doi [\apj] {10.1086/177887}, \href {https://ui.adsabs.harvard.edu/abs/1996ApJ...470..551D} {470, 551}

\bibitem[\protect\citeauthoryear{{Duffin} \& {Pudritz}}{{Duffin} \& {Pudritz}}{2009}]{Duffin_2009}
{Duffin} D.~F.,  {Pudritz} R.~E.,  2009, \mn@doi [\apjl] {10.1088/0004-637X/706/1/L46}, \href {https://ui.adsabs.harvard.edu/abs/2009ApJ...706L..46D} {706, L46}

\bibitem[\protect\citeauthoryear{{Efroimsky}}{{Efroimsky}}{2000}]{Efroimky_2000}
{Efroimsky} M.,  2000, \mn@doi [Journal of Mathematical Physics] {10.1063/1.533216}, \href {https://ui.adsabs.harvard.edu/abs/2000JMP....41.1854E} {41, 1854}

\bibitem[\protect\citeauthoryear{{Fiedler} \& {Mouschovias}}{{Fiedler} \& {Mouschovias}}{1993}]{Fiedler_1993}
{Fiedler} R.~A.,  {Mouschovias} T.~C.,  1993, \mn@doi [\apj] {10.1086/173193}, \href {https://ui.adsabs.harvard.edu/abs/1993ApJ...415..680F} {415, 680}

\bibitem[\protect\citeauthoryear{{Galametz} et~al.,}{{Galametz} et~al.}{2018}]{Galametz_2018}
{Galametz} M.,  et~al., 2018, \mn@doi [\aap] {10.1051/0004-6361/201833004}, \href {https://ui.adsabs.harvard.edu/abs/2018A&A...616A.139G} {616, A139}

\bibitem[\protect\citeauthoryear{{Galametz}, {Maury}, {Valdivia}, {Testi}, {Belloche}  \& {Andr{\'e}}}{{Galametz} et~al.}{2019}]{Galametz_2019}
{Galametz} M.,  {Maury} A.~J.,  {Valdivia} V.,  {Testi} L.,  {Belloche} A.,   {Andr{\'e}} P.,  2019, \mn@doi [\aap] {10.1051/0004-6361/201936342}, \href {https://ui.adsabs.harvard.edu/abs/2019A&A...632A...5G} {632, A5}

\bibitem[\protect\citeauthoryear{{Galli} \& {Shu}}{{Galli} \& {Shu}}{993a}]{Galli_1993a}
{Galli} D.,  {Shu} F.~H.,  1993a, \mn@doi [\apj] {10.1086/173305}, \href {https://ui.adsabs.harvard.edu/abs/1993ApJ...417..220G} {417, 220}

\bibitem[\protect\citeauthoryear{{Galli} \& {Shu}}{{Galli} \& {Shu}}{993b}]{Galli_1993b}
{Galli} D.,  {Shu} F.~H.,  1993b, \mn@doi [\apj] {10.1086/173306}, \href {https://ui.adsabs.harvard.edu/abs/1993ApJ...417..243G} {417, 243}

\bibitem[\protect\citeauthoryear{{Galli}, {Lizano}, {Shu}  \& {Allen}}{{Galli} et~al.}{2006}]{Galli_2006}
{Galli} D.,  {Lizano} S.,  {Shu} F.~H.,   {Allen} A.,  2006, \mn@doi [\apj] {10.1086/505257}, \href {https://ui.adsabs.harvard.edu/abs/2006ApJ...647..374G} {647, 374}

\bibitem[\protect\citeauthoryear{Giang, Hoang, Kim  \& Tram}{Giang et~al.}{2023}]{Giang_2022}
Giang N.~C.,  Hoang T.,  Kim J.-G.,   Tram L.~N.,  2023, \mn@doi [Monthly Notices of the Royal Astronomical Society] {10.1093/mnras/stad020}, 520, 3788

\bibitem[\protect\citeauthoryear{{Girart}, {Crutcher}  \& {Rao}}{{Girart} et~al.}{1999}]{Gigart_1999}
{Girart} J.~M.,  {Crutcher} R.~M.,   {Rao} R.,  1999, \mn@doi [\apjl] {10.1086/312345}, \href {https://ui.adsabs.harvard.edu/abs/1999ApJ...525L.109G} {525, L109}

\bibitem[\protect\citeauthoryear{{Girart}, {Rao}  \& {Marrone}}{{Girart} et~al.}{2006}]{Gigart_2006}
{Girart} J.~M.,  {Rao} R.,   {Marrone} D.~P.,  2006, \mn@doi [Science] {10.1126/science.1129093}, \href {https://ui.adsabs.harvard.edu/abs/2006Sci...313..812G} {313, 812}

\bibitem[\protect\citeauthoryear{{Gon{\c{c}}alves}, {Galli}  \& {Girart}}{{Gon{\c{c}}alves} et~al.}{2008}]{Goncalves_2008}
{Gon{\c{c}}alves} J.,  {Galli} D.,   {Girart} J.~M.,  2008, \mn@doi [\aap] {10.1051/0004-6361:200810861}, \href {https://ui.adsabs.harvard.edu/abs/2008A&A...490L..39G} {490, L39}

\bibitem[\protect\citeauthoryear{Gouellec et~al.,}{Gouellec et~al.}{2019}]{Valentin_2019}
Gouellec V. J. M.~L.,  et~al., 2019, \mn@doi [The Astrophysical Journal] {10.3847/1538-4357/ab43c2}, 885, 106

\bibitem[\protect\citeauthoryear{{Greenberg}}{{Greenberg}}{1968}]{Greenberg_1968}
{Greenberg} J.~M.,  1968, in {Middlehurst} B.~M.,  {Aller} L.~H.,  eds, , Nebulae and Interstellar Matter.
p.~221

\bibitem[\protect\citeauthoryear{{Hennebelle} \& {Ciardi}}{{Hennebelle} \& {Ciardi}}{2009}]{Hennebelle_2009}
{Hennebelle} P.,  {Ciardi} A.,  2009, \mn@doi [\aap] {10.1051/0004-6361/200913008}, \href {https://ui.adsabs.harvard.edu/abs/2009A&A...506L..29H} {506, L29}

\bibitem[\protect\citeauthoryear{{Hennebelle} \& {Fromang}}{{Hennebelle} \& {Fromang}}{2008}]{Hennebelle_2008}
{Hennebelle} P.,  {Fromang} S.,  2008, \mn@doi [\aap] {10.1051/0004-6361:20078309}, \href {https://ui.adsabs.harvard.edu/abs/2008A&A...477....9H} {477, 9}

\bibitem[\protect\citeauthoryear{{Henning}, {Wolf}, {Launhardt}  \& {Waters}}{{Henning} et~al.}{2001}]{Henning_2001}
{Henning} T.,  {Wolf} S.,  {Launhardt} R.,   {Waters} R.,  2001, \mn@doi [\apj] {10.1086/323362}, \href {https://ui.adsabs.harvard.edu/abs/2001ApJ...561..871H} {561, 871}

\bibitem[\protect\citeauthoryear{{Herranen}, {Lazarian}  \& {Hoang}}{{Herranen} et~al.}{2021}]{Herranen_2021}
{Herranen} J.,  {Lazarian} A.,   {Hoang} T.,  2021, \mn@doi [\apj] {10.3847/1538-4357/abf096}, \href {https://ui.adsabs.harvard.edu/abs/2021ApJ...913...63H} {913, 63}

\bibitem[\protect\citeauthoryear{{Ho}, {Moran}  \& {Lo}}{{Ho} et~al.}{2004}]{Ho_2014}
{Ho} P. T.~P.,  {Moran} J.~M.,   {Lo} K.~Y.,  2004, \mn@doi [\apjl] {10.1086/423245}, \href {https://ui.adsabs.harvard.edu/abs/2004ApJ...616L...1H} {616, L1}

\bibitem[\protect\citeauthoryear{{Hoang}}{{Hoang}}{2020}]{Hoang_2020}
{Hoang} T.,  2020, \mn@doi [Galaxies] {10.3390/galaxies8030052}, \href {https://ui.adsabs.harvard.edu/abs/2020Galax...8...52H} {8, 52}

\bibitem[\protect\citeauthoryear{Hoang}{Hoang}{2022}]{Hoang_2022}
Hoang T.,  2022, \mn@doi [The Astrophysical Journal] {10.3847/1538-4357/ac5408}, 928, 102

\bibitem[\protect\citeauthoryear{Hoang \& Lazarian}{Hoang \& Lazarian}{2008}]{Hoang_Lazarian_2008}
Hoang T.,  Lazarian A.,  2008, \mn@doi [Monthly Notices of the Royal Astronomical Society] {10.1111/j.1365-2966.2008.13249.x}, 388, 117

\bibitem[\protect\citeauthoryear{Hoang \& Lazarian}{Hoang \& Lazarian}{2009}]{Hoang_Lazarian_2009}
Hoang T.,  Lazarian A.,  2009, \mn@doi [The Astrophysical Journal] {10.1088/0004-637x/697/2/1316}, 697, 1316

\bibitem[\protect\citeauthoryear{{Hoang} \& {Lazarian}}{{Hoang} \& {Lazarian}}{2014}]{Hoang_Lazarian_2014}
{Hoang} T.,  {Lazarian} A.,  2014, \mn@doi [\mnras] {10.1093/mnras/stt2240}, \href {https://ui.adsabs.harvard.edu/abs/2014MNRAS.438..680H} {438, 680}

\bibitem[\protect\citeauthoryear{Hoang \& Lazarian}{Hoang \& Lazarian}{2016a}]{Hoang_Lazarian_2016b}
Hoang T.,  Lazarian A.,  2016a, \mn@doi [\apj] {10.3847/0004-637x/821/2/91}, 821, 91

\bibitem[\protect\citeauthoryear{{Hoang} \& {Lazarian}}{{Hoang} \& {Lazarian}}{2016b}]{Hoang_Lazarian_2016a}
{Hoang} T.,  {Lazarian} A.,  2016b, \mn@doi [\apj] {10.3847/0004-637X/831/2/159}, \href {https://ui.adsabs.harvard.edu/abs/2016ApJ...831..159H} {831, 159}

\bibitem[\protect\citeauthoryear{Hoang, Cho  \& Lazarian}{Hoang et~al.}{2018}]{Hoang+2018}
Hoang T.,  Cho J.,   Lazarian A.,  2018, \mn@doi [The Astrophysical Journal] {10.3847/1538-4357/aa9edc}, 852, 129

\bibitem[\protect\citeauthoryear{{Hoang}, {Tram}, {Lee}  \& {Ahn}}{{Hoang} et~al.}{2019}]{Hoang_2019}
{Hoang} T.,  {Tram} L.~N.,  {Lee} H.,   {Ahn} S.-H.,  2019, \mn@doi [Nature Astronomy] {10.1038/s41550-019-0763-6}, \href {https://ui.adsabs.harvard.edu/abs/2019NatAs...3..766H} {3, 766}

\bibitem[\protect\citeauthoryear{{Hoang}, {Tram}, {Minh Phan}, {Giang}, {Phuong}  \& {Dieu}}{{Hoang} et~al.}{2022}]{Hoang+2022}
{Hoang} T.,  {Tram} L.~N.,  {Minh Phan} V.~H.,  {Giang} N.~C.,  {Phuong} N.~T.,   {Dieu} N.~D.,  2022, \mn@doi [\aj] {10.3847/1538-3881/ac9af5}, \href {https://ui.adsabs.harvard.edu/abs/2022AJ....164..248H} {164, 248}

\bibitem[\protect\citeauthoryear{{Hull} \& {Zhang}}{{Hull} \& {Zhang}}{2019}]{Hull_2019}
{Hull} C. L.~H.,  {Zhang} Q.,  2019, \mn@doi [Frontiers in Astronomy and Space Sciences] {10.3389/fspas.2019.00003}, \href {https://ui.adsabs.harvard.edu/abs/2019FrASS...6....3H} {6, 3}

\bibitem[\protect\citeauthoryear{Hull et~al.,}{Hull et~al.}{2014}]{Hull_2014}
Hull C. L.~H.,  et~al., 2014, \mn@doi [The Astrophysical Journal Supplement Series] {10.1088/0067-0049/213/1/13}, 213, 13

\bibitem[\protect\citeauthoryear{{Hull} et~al.,}{{Hull} et~al.}{2017a}]{Hull_2017a}
{Hull} C. L.~H.,  et~al., 2017a, \mn@doi [\apjl] {10.3847/2041-8213/aa71b7}, \href {https://ui.adsabs.harvard.edu/abs/2017ApJ...842L...9H} {842, L9}

\bibitem[\protect\citeauthoryear{{Hull} et~al.,}{{Hull} et~al.}{2017b}]{Hull_2017b}
{Hull} C. L.~H.,  et~al., 2017b, \mn@doi [\apj] {10.3847/1538-4357/aa7fe9}, \href {https://ui.adsabs.harvard.edu/abs/2017ApJ...847...92H} {847, 92}

\bibitem[\protect\citeauthoryear{{Jones} \& {Spitzer}}{{Jones} \& {Spitzer}}{1967}]{Jones_Spitzer_1967}
{Jones} R.~V.,  {Spitzer} Lyman J.,  1967, \mn@doi [\apj] {10.1086/149086}, \href {https://ui.adsabs.harvard.edu/abs/1967ApJ...147..943J} {147, 943}

\bibitem[\protect\citeauthoryear{{Joos}, {Hennebelle}  \& {Ciardi}}{{Joos} et~al.}{2012}]{Joos_2012}
{Joos} M.,  {Hennebelle} P.,   {Ciardi} A.,  2012, \mn@doi [\aap] {10.1051/0004-6361/201118730}, \href {https://ui.adsabs.harvard.edu/abs/2012A&A...543A.128J} {543, A128}

\bibitem[\protect\citeauthoryear{Kataoka, Machida  \& Tomisaka}{Kataoka et~al.}{2012}]{Kataoka_2012}
Kataoka A.,  Machida M.~N.,   Tomisaka K.,  2012, \mn@doi [The Astrophysical Journal] {10.1088/0004-637X/761/1/40}, 761, 40

\bibitem[\protect\citeauthoryear{{Kataoka}, {Tanaka}, {Okuzumi}  \& {Wada}}{{Kataoka} et~al.}{2013}]{Kataoka_2013}
{Kataoka} A.,  {Tanaka} H.,  {Okuzumi} S.,   {Wada} K.,  2013, \mn@doi [\aap] {10.1051/0004-6361/201322151}, \href {https://ui.adsabs.harvard.edu/abs/2013A&A...557L...4K} {557, L4}

\bibitem[\protect\citeauthoryear{{Kataoka} et~al.,}{{Kataoka} et~al.}{2015}]{Kataoka_2015}
{Kataoka} A.,  et~al., 2015, \mn@doi [\apj] {10.1088/0004-637X/809/1/78}, \href {https://ui.adsabs.harvard.edu/abs/2015ApJ...809...78K} {809, 78}

\bibitem[\protect\citeauthoryear{{Ko}, {Liu}, {Lai}, {Ching}, {Rao}  \& {Girart}}{{Ko} et~al.}{2020}]{Ko_2020}
{Ko} C.-L.,  {Liu} H.~B.,  {Lai} S.-P.,  {Ching} T.-C.,  {Rao} R.,   {Girart} J.~M.,  2020, \mn@doi [\apj] {10.3847/1538-4357/ab5e79}, \href {https://ui.adsabs.harvard.edu/abs/2020ApJ...889..172K} {889, 172}

\bibitem[\protect\citeauthoryear{{Krumholz} \& {Federrath}}{{Krumholz} \& {Federrath}}{2019}]{Krumholz_2019}
{Krumholz} M.~R.,  {Federrath} C.,  2019, \mn@doi [Frontiers in Astronomy and Space Sciences] {10.3389/fspas.2019.00007}, \href {https://ui.adsabs.harvard.edu/abs/2019FrASS...6....7K} {6, 7}

\bibitem[\protect\citeauthoryear{Krumholz, Crutcher  \& Hull}{Krumholz et~al.}{2013}]{Krumholz_2013}
Krumholz M.~R.,  Crutcher R.~M.,   Hull C. L.~H.,  2013, \mn@doi [The Astrophysical Journal Letters] {10.1088/2041-8205/767/1/L11}, 767, L11

\bibitem[\protect\citeauthoryear{{Krumholz} et~al.,}{{Krumholz} et~al.}{2014}]{Krumholz_2014}
{Krumholz} M.~R.,  et~al., 2014, in {Beuther} H.,  {Klessen} R.~S.,  {Dullemond} C.~P.,   {Henning} T.,  eds, Protostars and Planets VI. pp 243--266 (\mn@eprint {arXiv} {1401.2473}), \mn@doi{10.2458/azu_uapress_9780816531240-ch011}

\bibitem[\protect\citeauthoryear{{Kwon}, {Looney}, {Mundy}, {Chiang}  \& {Kemball}}{{Kwon} et~al.}{2009}]{Kwon_2009}
{Kwon} W.,  {Looney} L.~W.,  {Mundy} L.~G.,  {Chiang} H.-F.,   {Kemball} A.~J.,  2009, \mn@doi [\apj] {10.1088/0004-637X/696/1/841}, \href {https://ui.adsabs.harvard.edu/abs/2009ApJ...696..841K} {696, 841}

\bibitem[\protect\citeauthoryear{Kwon, Stephens, Tobin, Looney, Li, van~der Tak  \& Crutcher}{Kwon et~al.}{2019}]{Kwon_2019}
Kwon W.,  Stephens I.~W.,  Tobin J.~J.,  Looney L.~W.,  Li Z.-Y.,  van~der Tak F. F.~S.,   Crutcher R.~M.,  2019, \mn@doi [The Astrophysical Journal] {10.3847/1538-4357/ab24c8}, 879, 25

\bibitem[\protect\citeauthoryear{{Lai}, {Crutcher}, {Girart}  \& {Rao}}{{Lai} et~al.}{2002}]{Lai_2002}
{Lai} S.-P.,  {Crutcher} R.~M.,  {Girart} J.~M.,   {Rao} R.,  2002, \mn@doi [\apj] {10.1086/338336}, \href {https://ui.adsabs.harvard.edu/abs/2002ApJ...566..925L} {566, 925}

\bibitem[\protect\citeauthoryear{Lam, Li, Chen, Tomida  \& Zhao}{Lam et~al.}{2019}]{Lam_2019}
Lam K.~H.,  Li Z.-Y.,  Chen C.-Y.,  Tomida K.,   Zhao B.,  2019, \mn@doi [Monthly Notices of the Royal Astronomical Society] {10.1093/mnras/stz2436}, 489, 5326

\bibitem[\protect\citeauthoryear{{Lam}, {Chen}, {Li}, {Yang}, {Cox}, {Looney}  \& {Stephens}}{{Lam} et~al.}{2021}]{Lam_2021}
{Lam} K.~H.,  {Chen} C.-Y.,  {Li} Z.-Y.,  {Yang} H.,  {Cox} E.~G.,  {Looney} L.~W.,   {Stephens} I.,  2021, \mn@doi [\mnras] {10.1093/mnras/stab2105}, \href {https://ui.adsabs.harvard.edu/abs/2021MNRAS.507..608L} {507, 608}

\bibitem[\protect\citeauthoryear{{Lazarian}}{{Lazarian}}{1994}]{Lazarian_1994}
{Lazarian} A.,  1994, \mn@doi [\mnras] {10.1093/mnras/268.3.713}, \href {https://ui.adsabs.harvard.edu/abs/1994MNRAS.268..713L} {268, 713}

\bibitem[\protect\citeauthoryear{{Lazarian}}{{Lazarian}}{2007}]{Lazarian_2007}
{Lazarian} A.,  2007, \mn@doi [\jqsrt] {10.1016/j.jqsrt.2007.01.038}, \href {https://ui.adsabs.harvard.edu/abs/2007JQSRT.106..225L} {106, 225}

\bibitem[\protect\citeauthoryear{{Lazarian} \& {Draine}}{{Lazarian} \& {Draine}}{1999}]{Lazarian_1999}
{Lazarian} A.,  {Draine} B.~T.,  1999, \mn@doi [\apjl] {10.1086/312137}, \href {https://ui.adsabs.harvard.edu/abs/1999ApJ...520L..67L} {520, L67}

\bibitem[\protect\citeauthoryear{{Lazarian} \& {Efroimsky}}{{Lazarian} \& {Efroimsky}}{1999}]{Lazarian_Efroisky_1999}
{Lazarian} A.,  {Efroimsky} M.,  1999, \mn@doi [\mnras] {10.1046/j.1365-8711.1999.02235.x}, \href {https://ui.adsabs.harvard.edu/abs/1999MNRAS.303..673L} {303, 673}

\bibitem[\protect\citeauthoryear{{Lazarian} \& {Hoang}}{{Lazarian} \& {Hoang}}{2007}]{Lazarian_Hoang_2007}
{Lazarian} A.,  {Hoang} T.,  2007, \mn@doi [\mnras] {10.1111/j.1365-2966.2007.11817.x}, \href {https://ui.adsabs.harvard.edu/abs/2007MNRAS.378..910L} {378, 910}

\bibitem[\protect\citeauthoryear{{Lazarian} \& {Hoang}}{{Lazarian} \& {Hoang}}{2008}]{Lazarian_Hoang_2008}
{Lazarian} A.,  {Hoang} T.,  2008, \mn@doi [\apjl] {10.1086/586706}, \href {https://ui.adsabs.harvard.edu/abs/2008ApJ...676L..25L} {676, L25}

\bibitem[\protect\citeauthoryear{Lazarian \& Hoang}{Lazarian \& Hoang}{2019}]{Lazarian_Hoang_2019}
Lazarian A.,  Hoang T.,  2019, \mn@doi [The Astrophysical Journal] {10.3847/1538-4357/ab3d39}, 883, 122

\bibitem[\protect\citeauthoryear{Lazarian \& Hoang}{Lazarian \& Hoang}{2021}]{Lazarian_Hoang_2021}
Lazarian A.,  Hoang T.,  2021, \mn@doi [The Astrophysical Journal] {10.3847/1538-4357/abd02c}, 908, 12

\bibitem[\protect\citeauthoryear{Lazarian \& Roberge}{Lazarian \& Roberge}{1997}]{Lazarian_Roberge_1997}
Lazarian A.,  Roberge W.,  1997, The Astrophysical Journal, 484, 230

\bibitem[\protect\citeauthoryear{{Lazarian}, {Andersson}  \& {Hoang}}{{Lazarian} et~al.}{2015}]{Lazarian_2015}
{Lazarian} A.,  {Andersson} B.~G.,   {Hoang} T.,  2015, in , Polarimetry of Stars and Planetary Systems.
p.~81

\bibitem[\protect\citeauthoryear{{Lazarian}, {Yuen}  \& {Pogosyan}}{{Lazarian} et~al.}{2022}]{Lazarian.2022}
{Lazarian} A.,  {Yuen} K.~H.,   {Pogosyan} D.,  2022, \mn@doi [\apj] {10.3847/1538-4357/ac6877}, \href {https://ui.adsabs.harvard.edu/abs/2022ApJ...935...77L} {935, 77}

\bibitem[\protect\citeauthoryear{{Le Gouellec} et~al.,}{{Le Gouellec} et~al.}{2020}]{Valentin_2020}
{Le Gouellec} V.~J.~M.,  et~al., 2020, \mn@doi [\aap] {10.1051/0004-6361/202038404}, \href {https://ui.adsabs.harvard.edu/abs/2020A&A...644A..11L} {644, A11}

\bibitem[\protect\citeauthoryear{{Le Gouellec}, {Maury}  \& {Hull}}{{Le Gouellec} et~al.}{2023a}]{Valentin_2023a}
{Le Gouellec} V.~J.~M.,  {Maury} A.~J.,   {Hull} C.~L.~H.,  2023a, \mn@doi [\aap] {10.1051/0004-6361/202244865}, \href {https://ui.adsabs.harvard.edu/abs/2023A&A...671A.167L} {671, A167}

\bibitem[\protect\citeauthoryear{{Le Gouellec}, {Maury}  \& {Hull}}{{Le Gouellec} et~al.}{2023b}]{Valentin_2023b}
{Le Gouellec} V.~J.~M.,  {Maury} A.~J.,   {Hull} C.~L.~H.,  2023b, \mn@doi [\aap] {10.1051/0004-6361/202244865}, \href {https://ui.adsabs.harvard.edu/abs/2023A&A...671A.167L} {671, A167}

\bibitem[\protect\citeauthoryear{Li, Goodman, Sridharan, Houde, Li, Novak  \& Tang}{Li et~al.}{2014}]{Li_2014}
Li H.,  Goodman A.,  Sridharan T.,  Houde M.,  Li Z.-Y.,  Novak G.,   Tang K.~S.,  2014, Protostars and Planets VI, pp 101--123

\bibitem[\protect\citeauthoryear{Liu}{Liu}{2021}]{Liu_2021}
Liu H.~B.,  2021, \mn@doi [The Astrophysical Journal] {10.3847/1538-4357/abf8b6}, 914, 25

\bibitem[\protect\citeauthoryear{{Liu} et~al.,}{{Liu} et~al.}{2016}]{Liu_et_al_2016}
{Liu} H.~B.,  et~al., 2016, \mn@doi [\apj] {10.3847/0004-637X/821/1/41}, \href {https://ui.adsabs.harvard.edu/abs/2016ApJ...821...41L} {821, 41}

\bibitem[\protect\citeauthoryear{{Lucy}}{{Lucy}}{1999}]{Lucy_1999}
{Lucy} L.~B.,  1999, \aap, \href {https://ui.adsabs.harvard.edu/abs/1999A&A...344..282L} {344, 282}

\bibitem[\protect\citeauthoryear{{Machida}, {Inutsuka}  \& {Matsumoto}}{{Machida} et~al.}{2011}]{Machida_2010}
{Machida} M.~N.,  {Inutsuka} S.-I.,   {Matsumoto} T.,  2011, \mn@doi [\pasj] {10.1093/pasj/63.3.555}, \href {https://ui.adsabs.harvard.edu/abs/2011PASJ...63..555M} {63, 555}

\bibitem[\protect\citeauthoryear{{Machida}, {Inutsuka}  \& {Matsumoto}}{{Machida} et~al.}{2014}]{Machida_2014}
{Machida} M.~N.,  {Inutsuka} S.-i.,   {Matsumoto} T.,  2014, \mn@doi [\mnras] {10.1093/mnras/stt2343}, \href {https://ui.adsabs.harvard.edu/abs/2014MNRAS.438.2278M} {438, 2278}

\bibitem[\protect\citeauthoryear{{Mathis}, {Rumpl}  \& {Nordsieck}}{{Mathis} et~al.}{1977}]{Mathis_1977}
{Mathis} J.~S.,  {Rumpl} W.,   {Nordsieck} K.~H.,  1977, \mn@doi [\apj] {10.1086/155591}, \href {https://ui.adsabs.harvard.edu/abs/1977ApJ...217..425M} {217, 425}

\bibitem[\protect\citeauthoryear{{Matthews}, {McPhee}, {Fissel}  \& {Curran}}{{Matthews} et~al.}{2009}]{Matthews_2009}
{Matthews} B.~C.,  {McPhee} C.~A.,  {Fissel} L.~M.,   {Curran} R.~L.,  2009, \mn@doi [\apjs] {10.1088/0067-0049/182/1/143}, \href {https://ui.adsabs.harvard.edu/abs/2009ApJS..182..143M} {182, 143}

\bibitem[\protect\citeauthoryear{Maury et~al.,}{Maury et~al.}{2018}]{Maury_2018}
Maury A.~J.,  et~al., 2018, Monthly Notices of the Royal Astronomical Society, 477, 2760

\bibitem[\protect\citeauthoryear{McKee \& Ostriker}{McKee \& Ostriker}{2007}]{Mckee_2007}
McKee C.~F.,  Ostriker E.~C.,  2007, \mn@doi [Annual Review of Astronomy and Astrophysics] {10.1146/annurev.astro.45.051806.110602}, 45, 565

\bibitem[\protect\citeauthoryear{{Mellon} \& {Li}}{{Mellon} \& {Li}}{2008a}]{Melon_2008}
{Mellon} R.~R.,  {Li} Z.-Y.,  2008a, \mn@doi [\apj] {10.1086/587542}, \href {https://ui.adsabs.harvard.edu/abs/2008ApJ...681.1356M} {681, 1356}

\bibitem[\protect\citeauthoryear{{Mellon} \& {Li}}{{Mellon} \& {Li}}{2008b}]{Mellon_2008}
{Mellon} R.~R.,  {Li} Z.-Y.,  2008b, \mn@doi [\apj] {10.1086/587542}, \href {https://ui.adsabs.harvard.edu/abs/2008ApJ...681.1356M} {681, 1356}

\bibitem[\protect\citeauthoryear{{Mellon} \& {Li}}{{Mellon} \& {Li}}{2009}]{Mellon_Li_2009}
{Mellon} R.~R.,  {Li} Z.-Y.,  2009, \mn@doi [\apj] {10.1088/0004-637X/698/1/922}, \href {https://ui.adsabs.harvard.edu/abs/2009ApJ...698..922M} {698, 922}

\bibitem[\protect\citeauthoryear{Mestel \& Spitzer~Jr}{Mestel \& Spitzer~Jr}{1956}]{Mestel_Spitzer_1956}
Mestel L.,  Spitzer~Jr L.,  1956, Monthly Notices of the Royal Astronomical Society, 116, 503

\bibitem[\protect\citeauthoryear{{Miotello}, {Testi}, {Lodato}, {Ricci}, {Rosotti}, {Brooks}, {Maury}  \& {Natta}}{{Miotello} et~al.}{2014}]{Miotello_2014}
{Miotello} A.,  {Testi} L.,  {Lodato} G.,  {Ricci} L.,  {Rosotti} G.,  {Brooks} K.,  {Maury} A.,   {Natta} A.,  2014, \mn@doi [\aap] {10.1051/0004-6361/201322945}, \href {https://ui.adsabs.harvard.edu/abs/2014A&A...567A..32M} {567, A32}

\bibitem[\protect\citeauthoryear{{Mouschovias} \& {Paleologou}}{{Mouschovias} \& {Paleologou}}{1979a}]{Mouschovias_1979}
{Mouschovias} T.~C.,  {Paleologou} E.~V.,  1979a, \mn@doi [\apj] {10.1086/157077}, \href {https://ui.adsabs.harvard.edu/abs/1979ApJ...230..204M} {230, 204}

\bibitem[\protect\citeauthoryear{{Mouschovias} \& {Paleologou}}{{Mouschovias} \& {Paleologou}}{1979b}]{Mouschovias_1980}
{Mouschovias} T.~C.,  {Paleologou} E.~V.,  1979b, \mn@doi [\apj] {10.1086/157077}, \href {https://ui.adsabs.harvard.edu/abs/1979ApJ...230..204M} {230, 204}

\bibitem[\protect\citeauthoryear{{Nakano} \& {Nakamura}}{{Nakano} \& {Nakamura}}{1978}]{Nakano_1978}
{Nakano} T.,  {Nakamura} T.,  1978, \pasj, \href {https://ui.adsabs.harvard.edu/abs/1978PASJ...30..671N} {30, 671}

\bibitem[\protect\citeauthoryear{Newville et~al.,}{Newville et~al.}{2021}]{lmfit}
Newville M.,  et~al., 2021, lmfit/lmfit-py: 1.0.3, \mn@doi{10.5281/zenodo.5570790}, \url {https://doi.org/10.5281/zenodo.5570790}

\bibitem[\protect\citeauthoryear{{Nguyen Tat}, {Diep}, {Hoang}, {Tram}, {Bich Ngoc}, {Phuong}  \& {Truong}}{{Nguyen Tat} et~al.}{2023}]{Thang_2023}
{Nguyen Tat} T.,  {Diep} P.~N.,  {Hoang} T.,  {Tram} L.~N.,  {Bich Ngoc} N.,  {Phuong} N.~T.,   {Truong} B.,  2023, \mn@doi [arXiv e-prints] {10.48550/arXiv.2401.00220}, \href {https://ui.adsabs.harvard.edu/abs/2024arXiv240100220N} {p. arXiv:2401.00220}

\bibitem[\protect\citeauthoryear{{Okuzumi}, {Tanaka}, {Kobayashi}  \& {Wada}}{{Okuzumi} et~al.}{2012}]{Okuzumi}
{Okuzumi} S.,  {Tanaka} H.,  {Kobayashi} H.,   {Wada} K.,  2012, \mn@doi [\apj] {10.1088/0004-637X/752/2/106}, \href {https://ui.adsabs.harvard.edu/abs/2012ApJ...752..106O} {752, 106}

\bibitem[\protect\citeauthoryear{{Pabst} et~al.,}{{Pabst} et~al.}{2019}]{Pabst_2019}
{Pabst} C.,  et~al., 2019, \mn@doi [\nat] {10.1038/s41586-018-0844-1}, \href {https://ui.adsabs.harvard.edu/abs/2019Natur.565..618P} {565, 618}

\bibitem[\protect\citeauthoryear{Pattle et~al.,}{Pattle et~al.}{2018}]{Pattle_2018}
Pattle K.,  et~al., 2018, \mn@doi [The Astrophysical Journal Letters] {10.3847/2041-8213/aac771}, 860, L6

\bibitem[\protect\citeauthoryear{{Pattle}, {Fissel}, {Tahani}, {Liu}  \& {Ntormousi}}{{Pattle} et~al.}{2022}]{Pattle_2022}
{Pattle} K.,  {Fissel} L.,  {Tahani} M.,  {Liu} T.,   {Ntormousi} E.,  2022, \mn@doi [arXiv e-prints] {10.48550/arXiv.2203.11179}, \href {https://ui.adsabs.harvard.edu/abs/2022arXiv220311179P} {p. arXiv:2203.11179}

\bibitem[\protect\citeauthoryear{{Planck Collaboration} et~al.,}{{Planck Collaboration} et~al.}{2015}]{PLanck_2015}
{Planck Collaboration} et~al., 2015, \mn@doi [\aap] {10.1051/0004-6361/201424082}, \href {https://ui.adsabs.harvard.edu/abs/2015A&A...576A.104P} {576, A104}

\bibitem[\protect\citeauthoryear{{Pudritz} \& {Norman}}{{Pudritz} \& {Norman}}{1983}]{Pudritz_1983}
{Pudritz} R.~E.,  {Norman} C.~A.,  1983, \mn@doi [\apj] {10.1086/161481}, \href {https://ui.adsabs.harvard.edu/abs/1983ApJ...274..677P} {274, 677}

\bibitem[\protect\citeauthoryear{{Purcell}}{{Purcell}}{1979}]{Purcell_1979}
{Purcell} E.~M.,  1979, \mn@doi [\apj] {10.1086/157204}, \href {https://ui.adsabs.harvard.edu/abs/1979ApJ...231..404P} {231, 404}

\bibitem[\protect\citeauthoryear{{Rao}, {Girart}, {Marrone}, {Lai}  \& {Schnee}}{{Rao} et~al.}{2009}]{Rao_2009}
{Rao} R.,  {Girart} J.~M.,  {Marrone} D.~P.,  {Lai} S.-P.,   {Schnee} S.,  2009, \mn@doi [\apj] {10.1088/0004-637X/707/2/921}, \href {https://ui.adsabs.harvard.edu/abs/2009ApJ...707..921R} {707, 921}

\bibitem[\protect\citeauthoryear{{Reissl}, {Wolf}  \& {Brauer}}{{Reissl} et~al.}{2016}]{Reissl_2016}
{Reissl} S.,  {Wolf} S.,   {Brauer} R.,  2016, \mn@doi [\aap] {10.1051/0004-6361/201424930}, \href {https://ui.adsabs.harvard.edu/abs/2016A&A...593A..87R} {593, A87}

\bibitem[\protect\citeauthoryear{Reissl, Seifried, Wolf, Banerjee  \& Klessen}{Reissl et~al.}{2017}]{Reissl_2017}
Reissl S.,  Seifried D.,  Wolf S.,  Banerjee R.,   Klessen R.~S.,  2017, \mn@doi [Astronomy and Astrophysics] {10.1051/0004-6361/201730408}, 603, A71

\bibitem[\protect\citeauthoryear{{Reissl}, {Guillet}, {Brauer}, {Levrier}, {Boulanger}  \& {Klessen}}{{Reissl} et~al.}{2020a}]{Reissl_2020}
{Reissl} S.,  {Guillet} V.,  {Brauer} R.,  {Levrier} F.,  {Boulanger} F.,   {Klessen} R.~S.,  2020a, \mn@doi [\aap] {10.1051/0004-6361/201937177}, \href {https://ui.adsabs.harvard.edu/abs/2020A&A...640A.118R} {640, A118}

\bibitem[\protect\citeauthoryear{{Reissl}, {Guillet}, {Brauer}, {Levrier}, {Boulanger}  \& {Klessen}}{{Reissl} et~al.}{2020b}]{Stefan_2017}
{Reissl} S.,  {Guillet} V.,  {Brauer} R.,  {Levrier} F.,  {Boulanger} F.,   {Klessen} R.~S.,  2020b, \mn@doi [\aap] {10.1051/0004-6361/201937177}, \href {https://ui.adsabs.harvard.edu/abs/2020A&A...640A.118R} {640, A118}

\bibitem[\protect\citeauthoryear{{Reissl}, {Meehan}  \& {Klessen}}{{Reissl} et~al.}{2023}]{Reissl_2023}
{Reissl} S.,  {Meehan} P.,   {Klessen} R.~S.,  2023, \mn@doi [\aap] {10.1051/0004-6361/202142528}, \href {https://ui.adsabs.harvard.edu/abs/2023A&A...674A..47R} {674, A47}

\bibitem[\protect\citeauthoryear{{Sadavoy} et~al.,}{{Sadavoy} et~al.}{2018}]{Sadavoy_2018}
{Sadavoy} S.~I.,  et~al., 2018, \mn@doi [\apj] {10.3847/1538-4357/aaef81}, \href {https://ui.adsabs.harvard.edu/abs/2018ApJ...869..115S} {869, 115}

\bibitem[\protect\citeauthoryear{Sadavoy et~al.,}{Sadavoy et~al.}{2019}]{Sadavoy_2019}
Sadavoy S.~I.,  et~al., 2019, \mn@doi [The Astrophysical Journal Supplement Series] {10.3847/1538-4365/ab4257}, 245, 2

\bibitem[\protect\citeauthoryear{{Santos-Lima}, {de Gouveia Dal Pino}  \& {Lazarian}}{{Santos-Lima} et~al.}{2012}]{Santos_2012}
{Santos-Lima} R.,  {de Gouveia Dal Pino} E.~M.,   {Lazarian} A.,  2012, \mn@doi [\apj] {10.1088/0004-637X/747/1/21}, \href {https://ui.adsabs.harvard.edu/abs/2012ApJ...747...21S} {747, 21}

\bibitem[\protect\citeauthoryear{{Santos-Lima}, {de Gouveia Dal Pino}  \& {Lazarian}}{{Santos-Lima} et~al.}{2013}]{Santos_2013}
{Santos-Lima} R.,  {de Gouveia Dal Pino} E.~M.,   {Lazarian} A.,  2013, \mn@doi [\mnras] {10.1093/mnras/sts597}, \href {https://ui.adsabs.harvard.edu/abs/2013MNRAS.429.3371S} {429, 3371}

\bibitem[\protect\citeauthoryear{{Seifried}, {Banerjee}, {Pudritz}  \& {Klessen}}{{Seifried} et~al.}{2015}]{Seifried_2015}
{Seifried} D.,  {Banerjee} R.,  {Pudritz} R.~E.,   {Klessen} R.~S.,  2015, \mn@doi [\mnras] {10.1093/mnras/stu2282}, \href {https://ui.adsabs.harvard.edu/abs/2015MNRAS.446.2776S} {446, 2776}

\bibitem[\protect\citeauthoryear{Seifried, Walch, Reissl  \& Ibáñez-Mejía}{Seifried et~al.}{2018}]{Seifried_2018}
Seifried D.,  Walch S.,  Reissl S.,   Ibáñez-Mejía J.~C.,  2018, \mn@doi [arXiv.org] {10.1093/mnras/sty2831}, astro-ph.GA

\bibitem[\protect\citeauthoryear{{Shu}}{{Shu}}{1992}]{Shu_1992}
{Shu} F.~H.,  1992, {The physics of astrophysics. Volume II: Gas dynamics.}

\bibitem[\protect\citeauthoryear{{Shu}, {Adams}  \& {Lizano}}{{Shu} et~al.}{1987}]{Shu_1987}
{Shu} F.~H.,  {Adams} F.~C.,   {Lizano} S.,  1987, \mn@doi [\araa] {10.1146/annurev.aa.25.090187.000323}, \href {https://ui.adsabs.harvard.edu/abs/1987ARA&A..25...23S} {25, 23}

\bibitem[\protect\citeauthoryear{{Stephens} et~al.,}{{Stephens} et~al.}{2013}]{Stephens_2013}
{Stephens} I.~W.,  et~al., 2013, \mn@doi [\apjl] {10.1088/2041-8205/769/1/L15}, \href {https://ui.adsabs.harvard.edu/abs/2013ApJ...769L..15S} {769, L15}

\bibitem[\protect\citeauthoryear{Takahashi, Machida, Tomisaka, Ho, Fomalont, Nakanishi  \& Girart}{Takahashi et~al.}{2019}]{Takahashi_2019}
Takahashi S.,  Machida M.~N.,  Tomisaka K.,  Ho P. T.~P.,  Fomalont E.~B.,  Nakanishi K.,   Girart J.~M.,  2019, \mn@doi [The Astrophysical Journal] {10.3847/1538-4357/aaf6ed}, 872, 70

\bibitem[\protect\citeauthoryear{{Tomida}, {Okuzumi}  \& {Machida}}{{Tomida} et~al.}{2015}]{Tomida_2015}
{Tomida} K.,  {Okuzumi} S.,   {Machida} M.~N.,  2015, \mn@doi [\apj] {10.1088/0004-637X/801/2/117}, \href {https://ui.adsabs.harvard.edu/abs/2015ApJ...801..117T} {801, 117}

\bibitem[\protect\citeauthoryear{{Tsukamoto}, {Iwasaki}, {Okuzumi}, {Machida}  \& {Inutsuka}}{{Tsukamoto} et~al.}{2015}]{Tsukamoto_2015}
{Tsukamoto} Y.,  {Iwasaki} K.,  {Okuzumi} S.,  {Machida} M.~N.,   {Inutsuka} S.,  2015, \mn@doi [\apjl] {10.1088/2041-8205/810/2/L26}, \href {https://ui.adsabs.harvard.edu/abs/2015ApJ...810L..26T} {810, L26}

\bibitem[\protect\citeauthoryear{{Tsukamoto} et~al.,}{{Tsukamoto} et~al.}{2022}]{Yusuke_2022}
{Tsukamoto} Y.,  et~al., 2022, \mn@doi [arXiv e-prints] {10.48550/arXiv.2209.13765}, \href {https://ui.adsabs.harvard.edu/abs/2022arXiv220913765T} {p. arXiv:2209.13765}

\bibitem[\protect\citeauthoryear{{Valdivia}, {Maury}, {Brauer}, {Hennebelle}, {Galametz}, {Guillet}  \& {Reissl}}{{Valdivia} et~al.}{2019}]{Valdivia_2019}
{Valdivia} V.,  {Maury} A.,  {Brauer} R.,  {Hennebelle} P.,  {Galametz} M.,  {Guillet} V.,   {Reissl} S.,  2019, \mn@doi [\mnras] {10.1093/mnras/stz2056}, \href {https://ui.adsabs.harvard.edu/abs/2019MNRAS.488.4897V} {488, 4897}

\bibitem[\protect\citeauthoryear{{Valdivia}, {Maury}  \& {Hennebelle}}{{Valdivia} et~al.}{2022}]{Valdivia_2022}
{Valdivia} V.,  {Maury} A.,   {Hennebelle} P.,  2022, \mn@doi [\aap] {10.1051/0004-6361/202243633}, \href {https://ui.adsabs.harvard.edu/abs/2022A&A...668A..83V} {668, A83}

\bibitem[\protect\citeauthoryear{{Weingartner} \& {Draine}}{{Weingartner} \& {Draine}}{2003}]{Weingartner_Draine_2003}
{Weingartner} J.~C.,  {Draine} B.~T.,  2003, \mn@doi [\apj] {10.1086/374597}, \href {https://ui.adsabs.harvard.edu/abs/2003ApJ...589..289W} {589, 289}

\bibitem[\protect\citeauthoryear{{Wurster}, {Bate}  \& {Price}}{{Wurster} et~al.}{2018}]{Wurster_2018}
{Wurster} J.,  {Bate} M.~R.,   {Price} D.~J.,  2018, \mn@doi [\mnras] {10.1093/mnras/sty2212}, \href {https://ui.adsabs.harvard.edu/abs/2018MNRAS.480.4434W} {480, 4434}

\bibitem[\protect\citeauthoryear{{Yang}}{{Yang}}{2021}]{Yang_2021}
{Yang} H.,  2021, \mn@doi [\apj] {10.3847/1538-4357/abebde}, \href {https://ui.adsabs.harvard.edu/abs/2021ApJ...911..125Y} {911, 125}

\bibitem[\protect\citeauthoryear{{Yang}, {Li}, {Looney}, {Cox}, {Tobin}, {Stephens}, {Segura-Cox}  \& {Harris}}{{Yang} et~al.}{2016}]{Yang_2016}
{Yang} H.,  {Li} Z.-Y.,  {Looney} L.~W.,  {Cox} E.~G.,  {Tobin} J.,  {Stephens} I.~W.,  {Segura-Cox} D.~M.,   {Harris} R.~J.,  2016, \mn@doi [\mnras] {10.1093/mnras/stw1253}, \href {https://ui.adsabs.harvard.edu/abs/2016MNRAS.460.4109Y} {460, 4109}

\bibitem[\protect\citeauthoryear{{Yen} et~al.,}{{Yen} et~al.}{2020}]{Yen_2020}
{Yen} H.-W.,  et~al., 2020, \mn@doi [\apj] {10.3847/1538-4357/ab7eb3}, \href {https://ui.adsabs.harvard.edu/abs/2020ApJ...893...54Y} {893, 54}

\bibitem[\protect\citeauthoryear{{Zielinski}, {Wolf}  \& {Brunngr{\"a}ber}}{{Zielinski} et~al.}{2021}]{Zielinski_2021}
{Zielinski} N.,  {Wolf} S.,   {Brunngr{\"a}ber} R.,  2021, \mn@doi [\aap] {10.1051/0004-6361/202039126}, \href {https://ui.adsabs.harvard.edu/abs/2021A&A...645A.125Z} {645, A125}

\makeatother
\end{thebibliography}

 \appendix
\section{Effect of maximum grain size on grain alignment}\label{sec:grain_alignment_amax}

\begin{figure*}
\centering
    \includegraphics[width=\textwidth,height=\textheight,keepaspectratio]{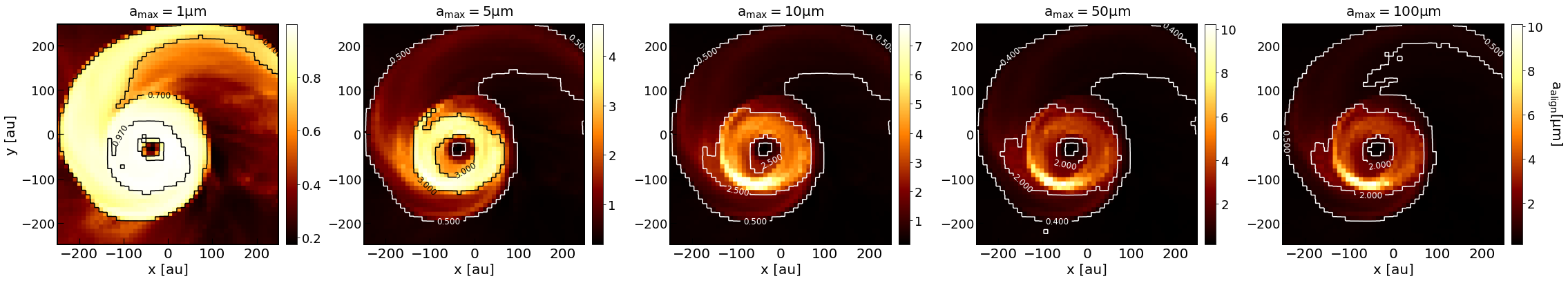}
   \caption{Spatial distribution of the density-weighted minimum alignment size $a_{\rm align}$ within the inner 500 au region for different maximum grain sizes from $a_{\rm max} = 1\mum$ (left panel) to $a_{\rm max} = 100\mum$ (right panel). For all values of $a_{\rm max}$,  $a_{\rm align}$ increases from the center toward the disk due to the reduced RAT efficiency (Figure \ref{fig:U_gamma_Td}), then it decreases outward due to the reduction of the gas density in the envelope scale. Besides, $a_{\rm align}$ increases when the maximum size increases from $1\mum$ to $5\mum$ as a result of the stronger attenuation of stellar radiation field by the presence of large grains. But for $a_{\rm max} \geq 10\mum$, $a_{\rm align}$ slightly decreases with increasing $a_{\rm max}$ due to the reduction in the amount of absorbers as the maximum sizes extend to VLGs.} 
     \label{fig:align_amax}
\end{figure*}

\begin{figure*}
\centering
   \includegraphics[width=\textwidth,height=\textheight,keepaspectratio]{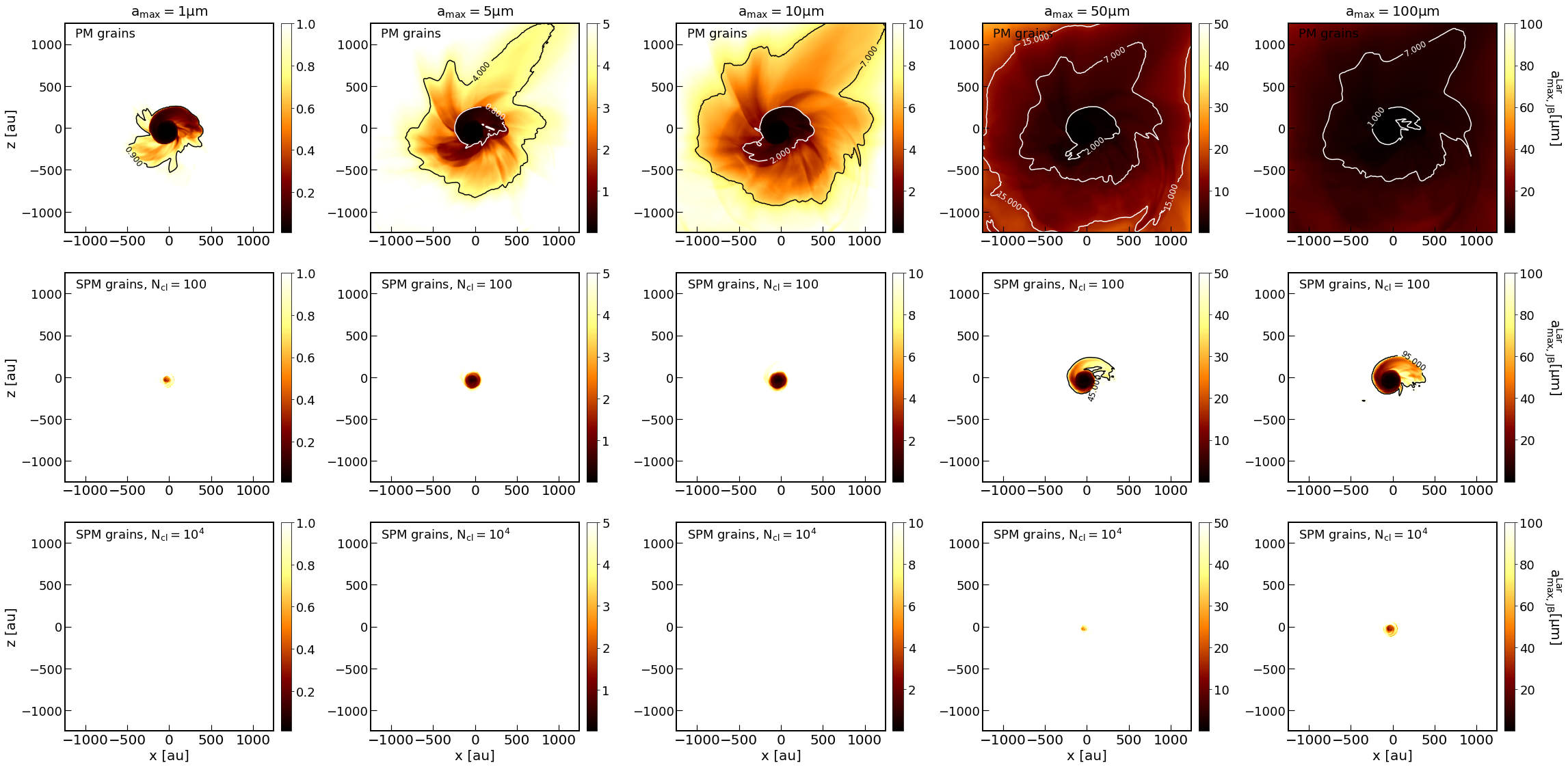}
    \caption{Spatial distribution of the weighted-density maximum alignment size $a_{\rm max,JB}^{\rm Lar}$ for different maximum grain sizes from $a_{\rm max} = 1\mum$ to $a_{\rm max} = 100\mum$, from left to right. The upper panels are for PM grains, the central panels are for SPM grains with low $N_{\rm cl} = 100$, and the lower panels are for SPM grains with high $N_{\rm cl } = 10^{4}$. The maximum alignment size decreases toward the center as increasing the gas randomization, but this problem could be improved if grains are SPM with higher amount of iron inclusions. However, even for SPM grains with high $N_{\rm cl} = 10^{4}$, VLGs above $a\geq 50\mum$ are not able to be aligned with $\B$in the protostellar disk due to their weak Larmor precession.} 
    \label{fig:alar_amax}
\end{figure*}

Figure \ref{fig:align_amax} shows the variation of the minimum alignment size $a_{\rm align}$ within 500 au region with different maximum grain sizes from $a_{\rm max} = 1\mum$ to $a_{\rm max}  =100\mum$, from left to right, respectively. For all values of $a_{\rm max}$, one can see that in general, $a_{\rm align}$ increases from the envelope toward the disk of $\sim 100$ au owing to the decrease of RAT efficiency with increasing gas density, then decreases again toward the protostar position due to increasing RAT efficiency with increasing stellar radiation field strength. The minimum alignment size in the inner 500 au region slightly changes with the maximum grain size following the change in terms of the optical depth. In particular, $a_{\rm align}$ inside the disk increases from $a_{\rm align} \sim 0.7-0.97\mum$ for $a_{\rm max} = 1\mum$ to $a_{\rm align} \sim 0.5 - 3\mum$ for $a_{\rm max} = 5\mum$, in which the removal of sub-micron aligned dust grains inside the disk is caused by the stronger extinction of stellar radiation field by the existence of micron-sized grains there. However, as grains grow to $a_{\rm max} \geq 10\mum$, $a_{\rm align}$ inside the disk slightly decreases owing to the reduced amount of absorber and scatterer around the protostar. 
 
Figure \ref{fig:alar_amax} shows the variation of the maximum alignment size for PM (upper panels), SPM grains with $N_{\rm cl} = 100$ (center panels), and SPM grains with $N_{\rm cl} = 10^{4}$ (lower panels) from $a_{\rm max} = 1\mum$ to $a_{\rm max} = 100\mum$. Obviously, large grains can be aligned with $\B$in the envelope due to their lower gas density. For example, almost all of grains up to $a_{\rm max} = 1 \sim 5\mum$ can be aligned with $\B$beyond $\sim 200$ au even when they are PM grains. However, for grains above $10\mum$, they must be SPM in order to have the magnetic alignment in this area. Moving toward the central region, the maximum alignment size reduces significantly as increasing gas randomization, for example, all PM grains above $0.4\mum$ are not able to be aligned with $\B$inside $\sim 200$ au. But if grains are SPM, the enhanced Larmor precession by iron inclusions can allow large grains up to $\sim 60\mum$ to have the magnetic alignment there, for instance, SPM grains have $N_{\rm cl} = 10^{4}$ and $\phi_{\rm sp} = 0.1$.

Figures \ref{fig:aaj_highJ_amax} and \ref{fig:aaj_lowJ_amax} shows the spatial distribution of $a_{\rm max,aJ}^{\rm high-J}$ and $a_{\rm max,aJ}^{\rm low-J}$ as the function of $a_{\rm max}$ and the grain magnetic properties. Large grains will have a higher potential to achieve fast internal relaxation in protostellar environments if they have higher magnetic susceptibility and are able to align with $\B$ at suprathermal rotation. However, the increased iron inclusions cannot help large grains inside protostellar disks to have fast internal relaxation at both high and low-\textit{J} attractors. For example, for SPM grains with $N_{\rm cl} = 10^{4}$, if $a_{\rm max} = 1\mum$ (lower left panel), large grains above $0.1\mum$ will have slow internal relaxation within the inner 200 au regions. Similarly, if grains grow to $a_{\rm max} = 10\mum$ or $a_{\rm max} = 100\mum$, grains above $2\mum$ and $20\mum$, respectively, will have slow internal relaxation there due to their weak Barnett relaxation caused by their higher inertia moment. 
  
\begin{figure*}
\centering
    \includegraphics[width=\textwidth,height=\textheight,keepaspectratio]{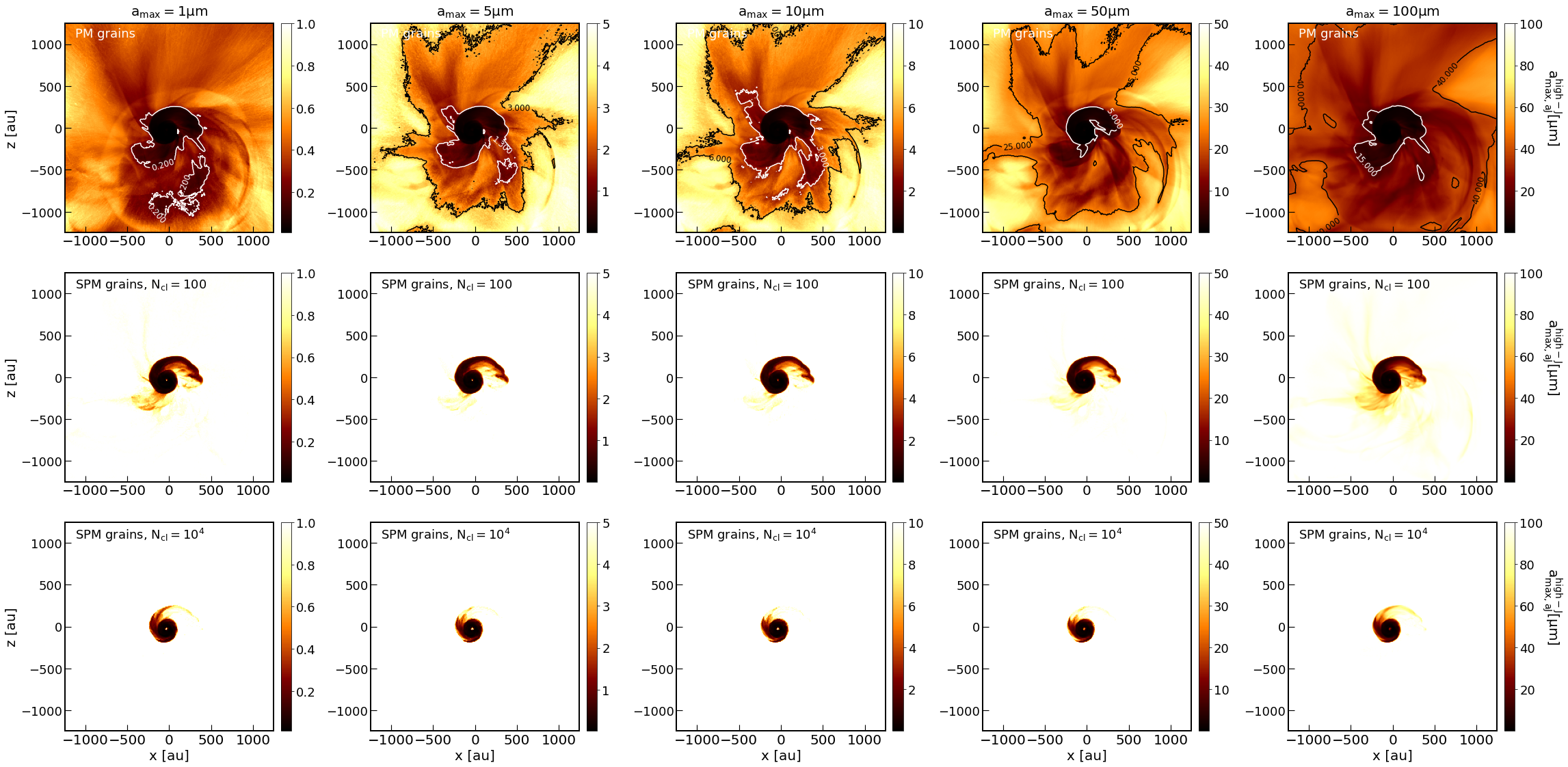}
    \caption{Effect of grain magnetic properties and maximum grain size on the distribution of the maximum size that grains having fast internal relaxation at high-\textit{J} attractors in the protostellar core, $a_{\rm max,aJ}^{\rm high-J}$. The value of $a_{\rm max,aJ}^{\rm high-J}$ decreases toward the center as increasing gas randomization but it can increase with increasing the level of iron inclusions inside dust grains. However, even SPM grains contain high $N_{\rm cl} = 10^{4}$, large grains still tend to have slow internal relaxation inside the disk with all values of $a_{\rm max}$, as a result of the strong gas randomization during their internal alignment stage.} 
     \label{fig:aaj_highJ_amax}
\centering
    \includegraphics[width=\textwidth,height=\textheight,keepaspectratio]{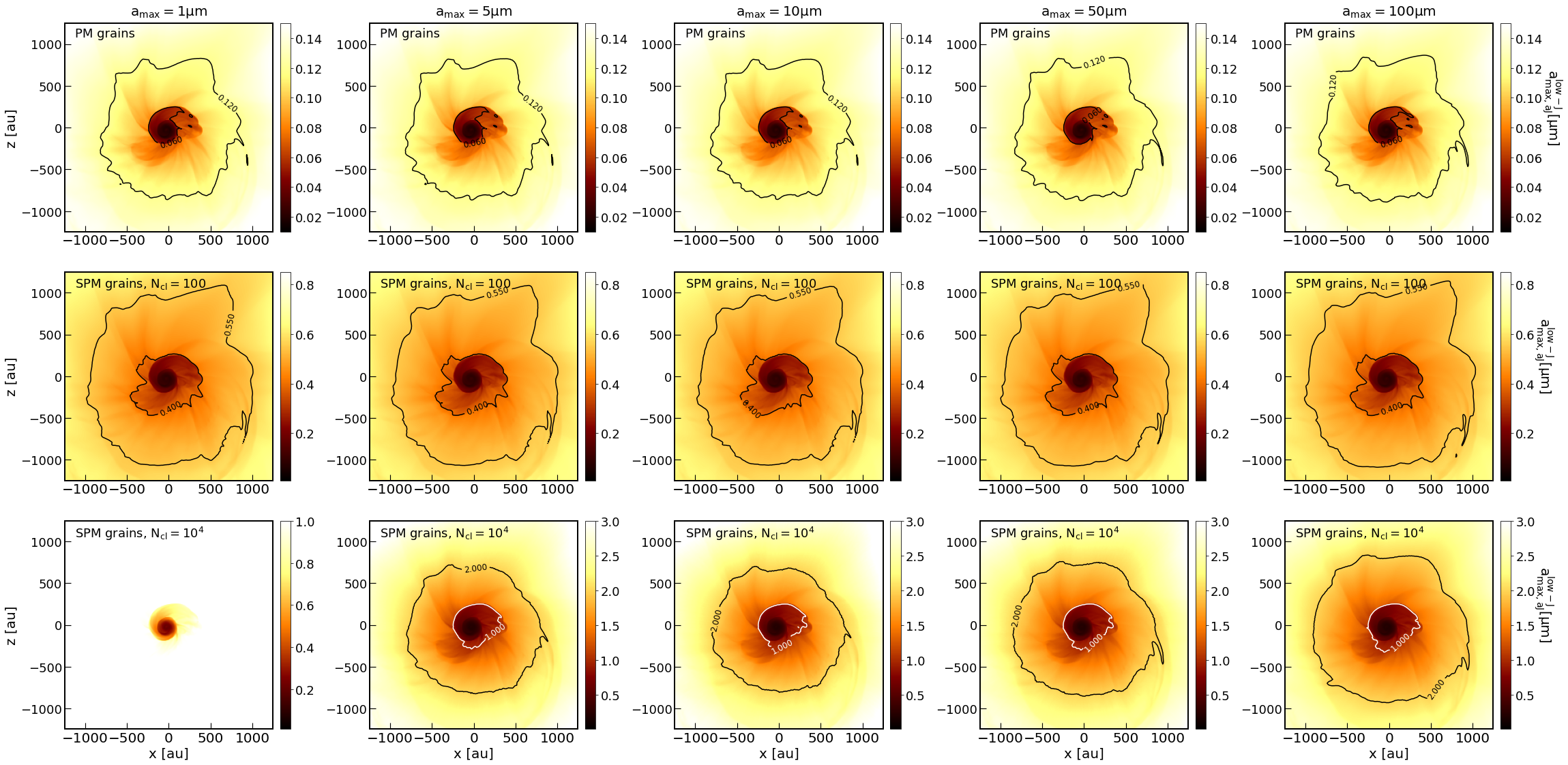}
    \caption{Similar results as Figure \ref{fig:aaj_highJ_amax} but for grains aligning with $\B$ at low-\textit{J} attractors, $a_{\rm max,aJ}^{\rm low-J}$. The range of grains having fast internal relaxation slightly increases with increasing amount of iron locked inside grains, but large grains above $3\mum$ always have slow internal relaxation at low-\textit{J} attractors in the entire protostellar core regardless of the difference in $a_{\rm max}$ and grain properties.} 
     \label{fig:aaj_lowJ_amax}
\end{figure*}

Figures \ref{fig:adg_0.5_amax} and \ref{fig:adg_1_amax} show the variation of the maximum size for which $50\%$ and $100\%$ of grains are aligned with $\B$ by MRAT alignment with different $a_{\rm max}$ and grain type. Similar to the results in Figure \ref{fig:adg_1}, PM grains will be aligned with $\B$ by only RATs in the entire protostellar core regardless of the difference in maximum grain size. For SPM grains with $N_{\rm cl} = 10^{4}$, all aligned dust grains above $a_{\rm align}$ in the envelope will have efficient magnetic alignment by MRAT alignment even if grains grow to $a_{\rm max} = 10-100\mum$. However, in the disk scale, large grains above $5\mum$ tend to be aligned with $\B$by RATs instead of the MRAT mechanism as a result of the weak magnetic relaxation efficiency in the high-density area. For example, for SPM grains with high $N_{\rm cl} = 10^{4}$, if grains grow to $a_{\rm max} = 10\mum$ or $a_{\rm max} = 100\mum$, the maximum size for grains aligning with $\B$ by MRAT mechanism is only $a_{\rm max,JB}^{\rm DG,0.5} \sim 4\mum$ and $a_{\rm max,JB}^{\rm DG,0.5} \sim 20\mum$ in the inner 100 au region to the protostar, respectively.

\begin{figure*}
\centering
    \includegraphics[width=\textwidth,height=\textheight,keepaspectratio]{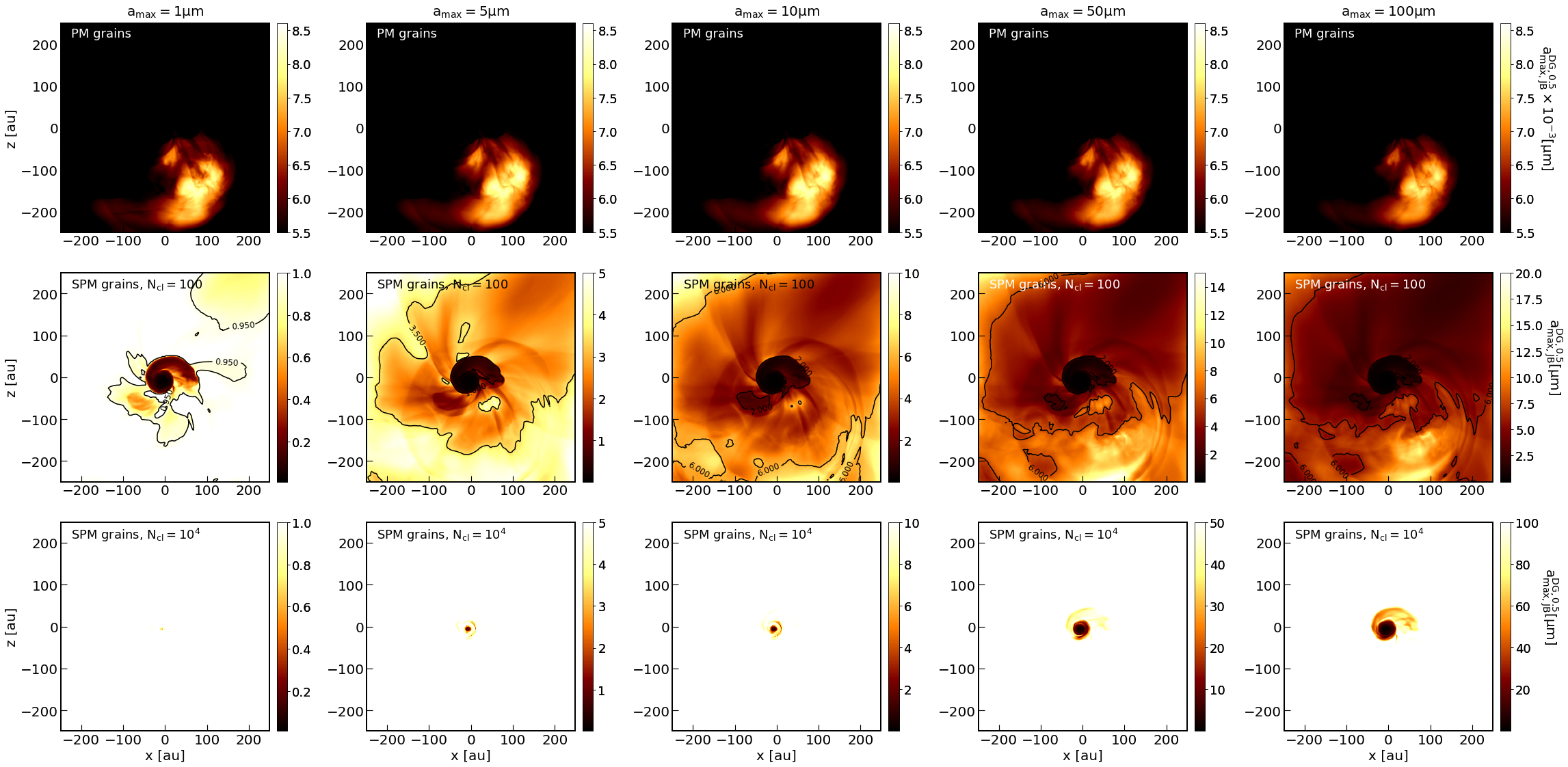}
    \caption{Effect of maximum grain size and the grain magnetic properties on the spatial distribution of the maximum size that $50\%$ of grains will be aligned with $\B$ at high-\textit{J} attractors by MRAT alignment, $a_{\rm max,JB}^{\rm DG,0.5}$. Similar to Figure \ref{fig:adg_1}, RAT is the major alignment mechanism for PM grains regardless of the maximum grain size, while MRAT alignment can drive the efficient external alignment for half of aligned SPM grains in the envelope with $\B$due to the enhanced magnetic relaxation by iron inclusions. However, the efficiency of MRAT alignment is not strong enough to push all micron-sized grains inside the disk to have $f_{\rm high-J} = 0.5$ due to the strong gas randomization there.} 
     \label{fig:adg_0.5_amax}
\end{figure*} 
\begin{figure*}
\centering
    \includegraphics[width=\textwidth,height=\textheight,keepaspectratio]{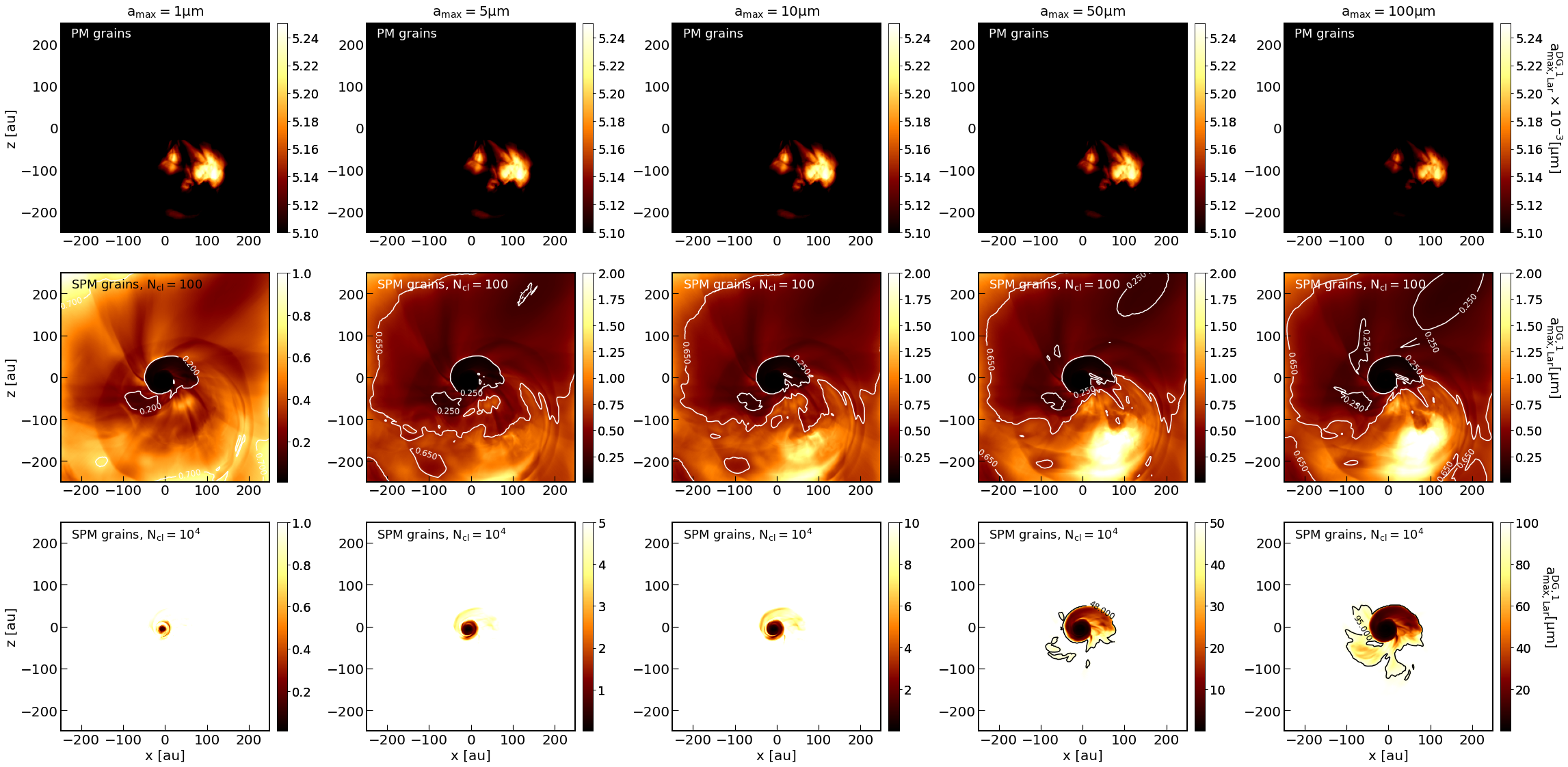}
    \caption{Similar results as Figure \ref{fig:adg_0.5_amax} but for the maximum size that $100\%$ of grains will be aligned with $\B$ at high-\textit{J} attractors by MRAT alignment, $a_{\rm max,JB}^{\rm DG,1}$. The range that grains can have $f_{\rm high-J} = 1$ by MRAT alignment increases with increasing levels of iron locked inside dust grains due to their enhanced magnetic relaxation strength by iron inclusions.}  
     \label{fig:adg_1_amax}
\end{figure*}

\section{Full map of inferred magnetic fields from dust polarization from model wIA}\label{sec:pol_envelope_wIA}
 
Figure \ref{fig:pol_map_2mm_wIA} shows the inferred magnetic fields obtained from model PA and model wIA with different grain magnetic properties and different maximum grain sizes from $a_{\rm max} = 1\mum$ to $a_{\rm max} = 100\mum$. Similar to Figure \ref{fig:pol_map_2mm_rIA}, both PM and SPM grains can infer again the large-scale spiral pattern of the magnetic field in the envelope as model PA. However, the polarization degree obtained from model wIA is smaller for grains containing lower levels of iron inclusions and for larger maximum grain size. In addition, $p$ obtained in model wIA is slightly smaller than the results from model rIA due to the self-suppression of polarized dust emission from grains with right and wrong IA.
 
\begin{figure*}
\centering
    \includegraphics[width=\textwidth,height=\textheight,keepaspectratio]{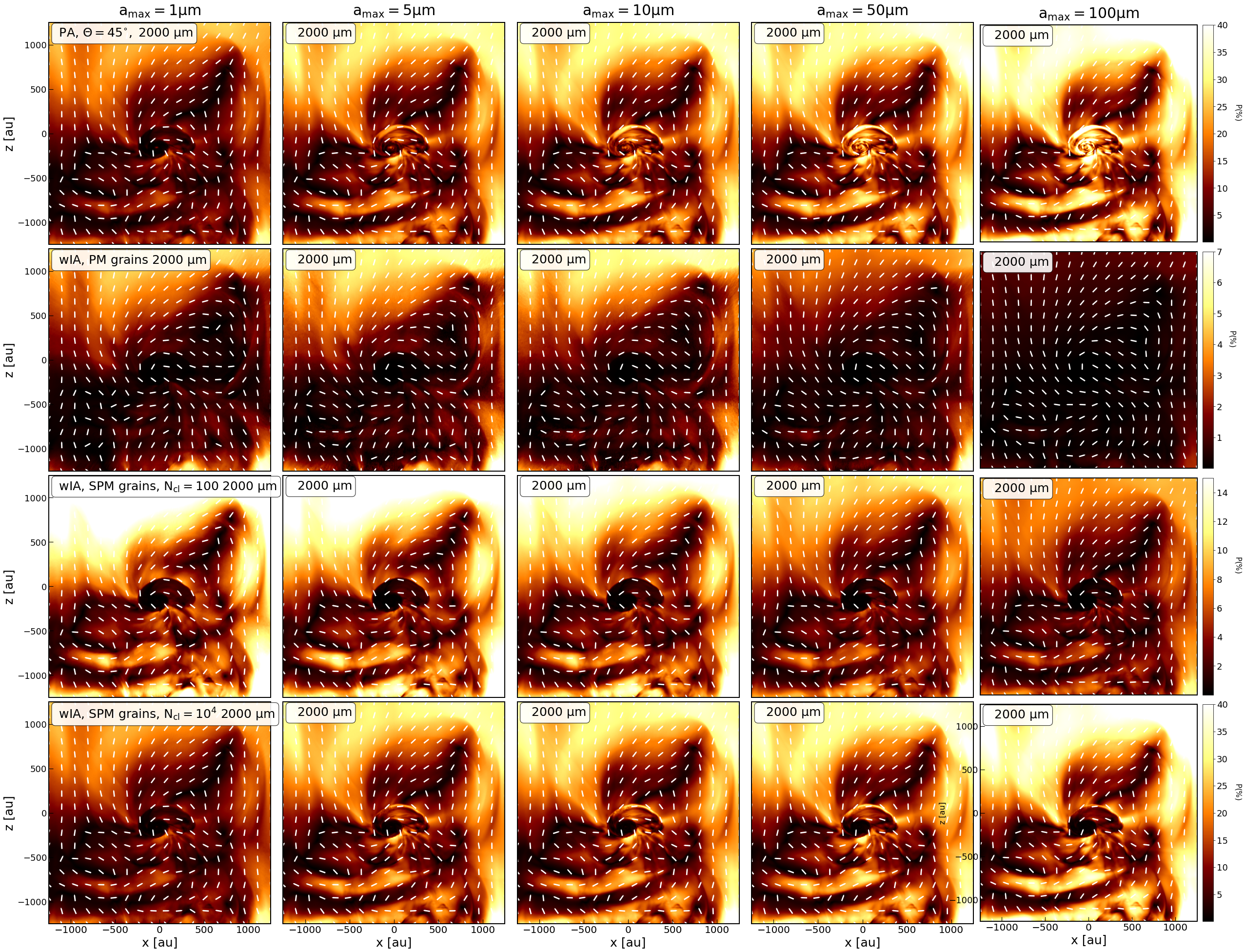}
    \caption{Inferred magnetic field obtained from dust polarization at 2 mm for model wIA with different grain magnetic properties and maximum grain size. Generally, both PM and SPM grains can trace well the spiral magnetic fields in the large envelope scale as model rIA (\ref{fig:pol_map_2mm_disk_rIA}). However, their obtained polarization degree is smaller than the results from model rIA due to the self-suppression of dust polarization emitting from grains with the right and wrong IA in the protostellar core.} 
     \label{fig:pol_map_2mm_wIA}
\end{figure*}

\section{Synthetic dust polarization obtained in the central region at other wavelengths}\label{sec:pol_multi_wavelength}
\subsection{Optical depth}\label{sec:optical_depth}
Figure \ref{fig:optical_depth} shows the optical depth in the inner 500 au region measured at 2mm, $870\mum$, $450\mum$, and $250\mum$, from left to right, respectively. The core is optically thin at $\lambda \geq 870\mum$ and the disk of radius 100 au starts to become optically thick at $450\mum$. 

\begin{figure*}
\centering
    \includegraphics[width=\textwidth,height=\textheight,keepaspectratio]{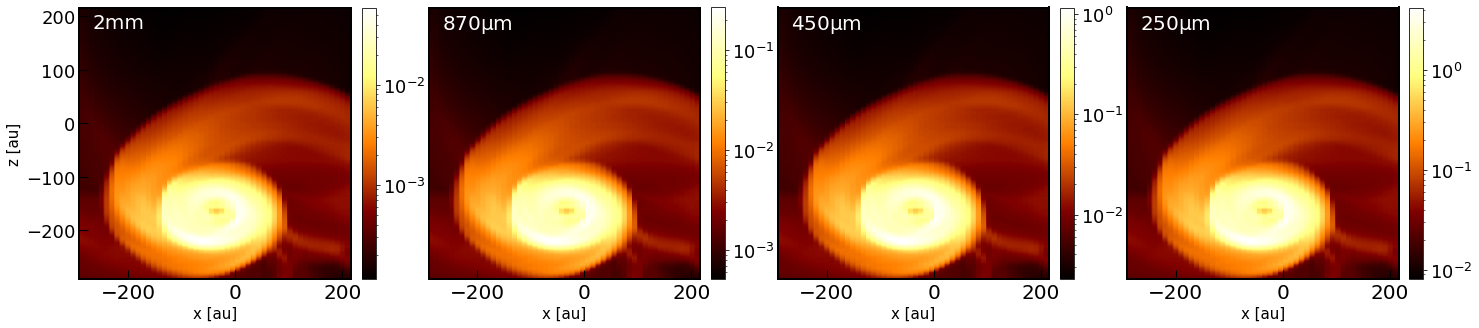}
    \caption{Optical depth in the central 500 au region at different wavelengths from 2mm to $250\mum$. The disk becomes optically thick at $450\mum$.} 
     \label{fig:optical_depth}
\end{figure*} 

\subsection{Optically thin wavelength $\lambda = 870\mum$}
\label{sec:870um}
Figures \ref{fig:pol_map_870um_rIA} and \ref{fig:pol_map_870um_wIA} show the inferred magnetic field map obtained at $870\mum$ from model rIA and wIA for different grain magnetic properties and maximum grain sizes. The map obtained from model PA is shown in the first row for comparison. Similar to the map inferred from dust polarization at optically thin wavelengths of 2mm, PM and SPM grains with low $N_{\rm cl}$ cannot clearly reveal the complex change of magnetic field within the disk structure due to the loss of grain alignment inside the disk. Only SPM grains with $N_{\rm cl} = 10^{4}$ can describe in detail the change of $\B$-fields in the innermost region of the disk as model PA. However, if a part of them has the wrong IA, the net polarization vector will be $\P \| \B$, thus, rotating $\P$ by $90^{\circ}$ will induce the wrong inferred magnetic field pattern inside the disk (\ref{fig:pol_map_870um_wIA}, fourth row)

\begin{figure*}
\centering
\includegraphics[width=\textwidth,height=\textheight,keepaspectratio]{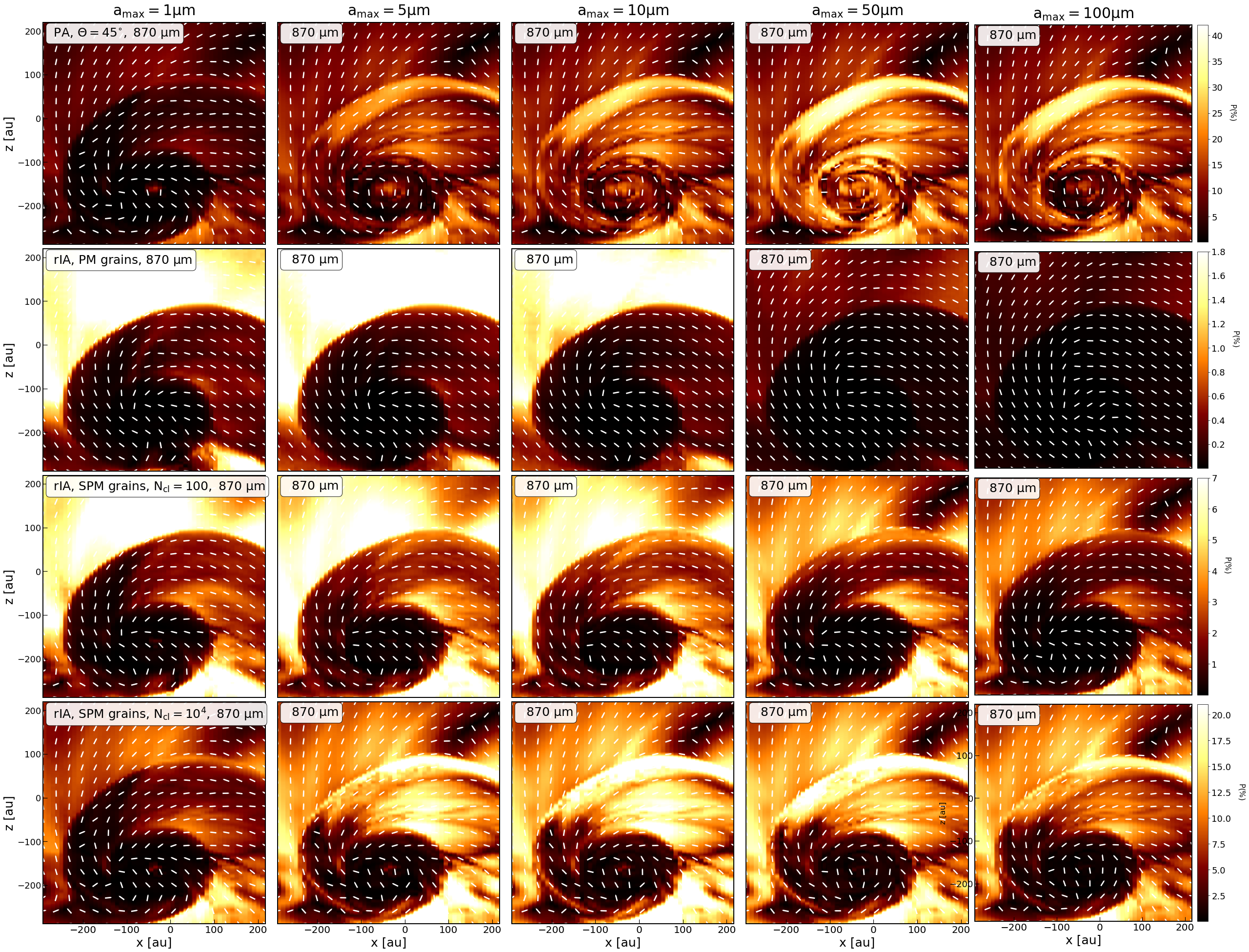}
    \caption{Effect of maximum grain size and grain magnetic properties on the inferred magnetic field map obtained in the inner 500 au region at $870\mum$. The upper row shows results for model PA, while the other rows are for model rIA. Similar to results obtained at 2mm (Figure \ref{fig:pol_map_2mm_disk_rIA}), only SPM grains with $N_{\rm cl} = 10^{4}$ can trace well the complex change of magnetic field from the inner part of the envelope to the disk driven by the accretion of gas toward the central region. However, their polarization degree is smaller than the results from model PA for all values of $a_{\rm max}$ due to the inefficient alignment of dust grains around the protostar. } 
     \label{fig:pol_map_870um_rIA}
\end{figure*} 

\begin{figure*}
\centering
    \includegraphics[width=\textwidth,height=\textheight,keepaspectratio]{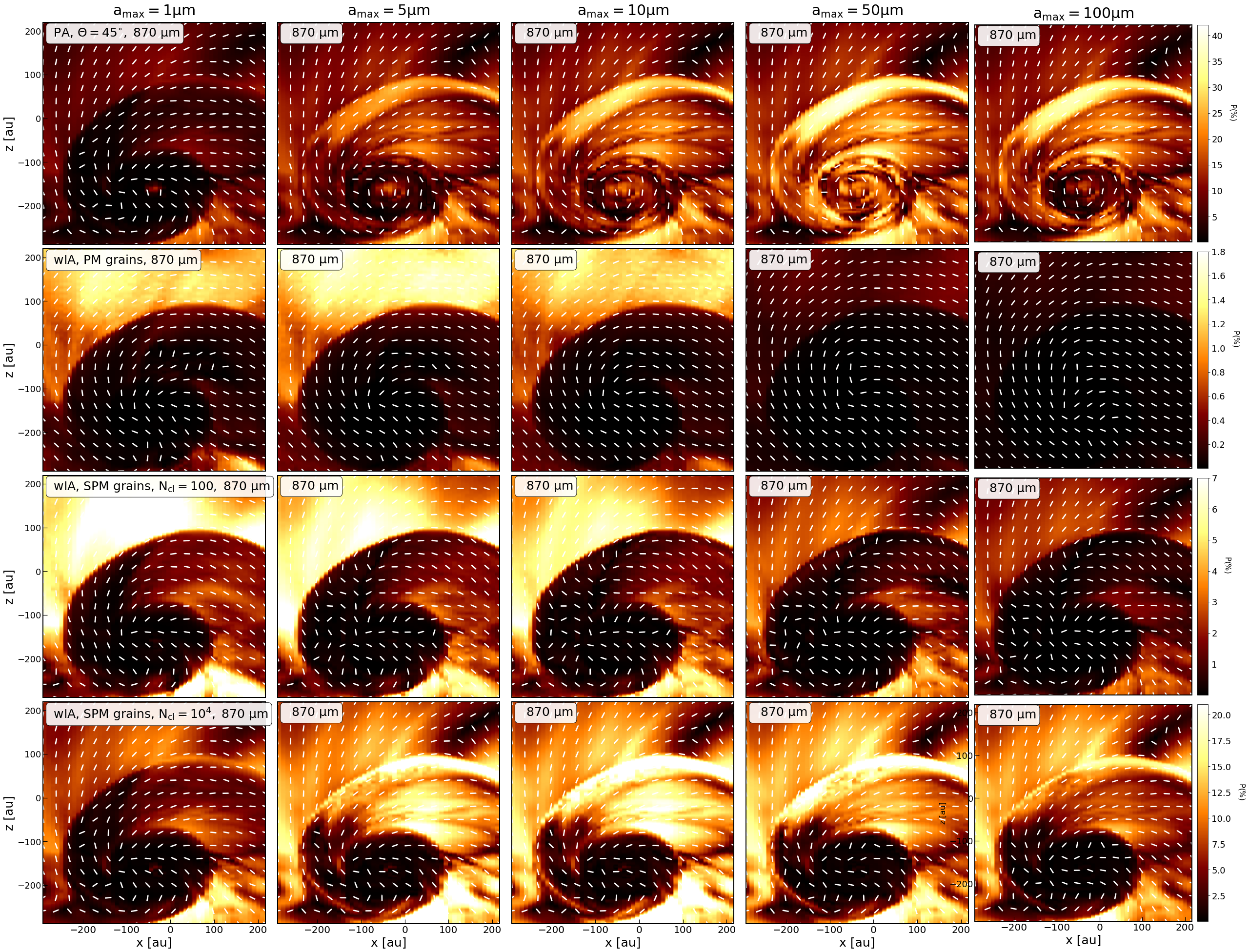}
    \caption{Similar as Figure \ref{fig:pol_map_870um_rIA} but for model wIA. The inferred magnetic fields at $870\mum$ in the inner 200 au region from SPM grains with $N_{\rm cl} = 10^{4}$ is the same as one detected at 2mm (Figure \ref{fig:pol_map_2mm_disk_wIA}), with $\B$ vectors are along the disk major axis resulted from the wrong interpretation of the polarization signal arising from wrong aligned dust grains. } 
    \label{fig:pol_map_870um_wIA}
\end{figure*}

\subsection{Optically thick wavelength $\lambda = 450\mum$}\label{sec:450um}

Figures \ref{fig:pol_map_450um_rIA} and \ref{fig:pol_map_450um_wIA} show similar results as Section \ref{sec:870um} but at optically thick wavelengths $450\mum$ (see Figure \ref{fig:optical_depth}). The first row shows the results for model PA, which reveals the $90^{\circ}$ flipping of inferred magnetic field in the disk caused by the change in the polarization mechanism from dichroic emission to dichroic extinction when grains grow to above $a_{\rm max} \geq 50\mum$. However, in both model rIA and wIA, one does not clearly see this flipping even for SPM grains with $N_{\rm cl} = 10^{4}$, which could be explained by the weak extinction of VLGs having low alignment degree with magnetic fields.
 
\begin{figure*}
\centering
\includegraphics[width=\textwidth,height=\textheight,keepaspectratio]{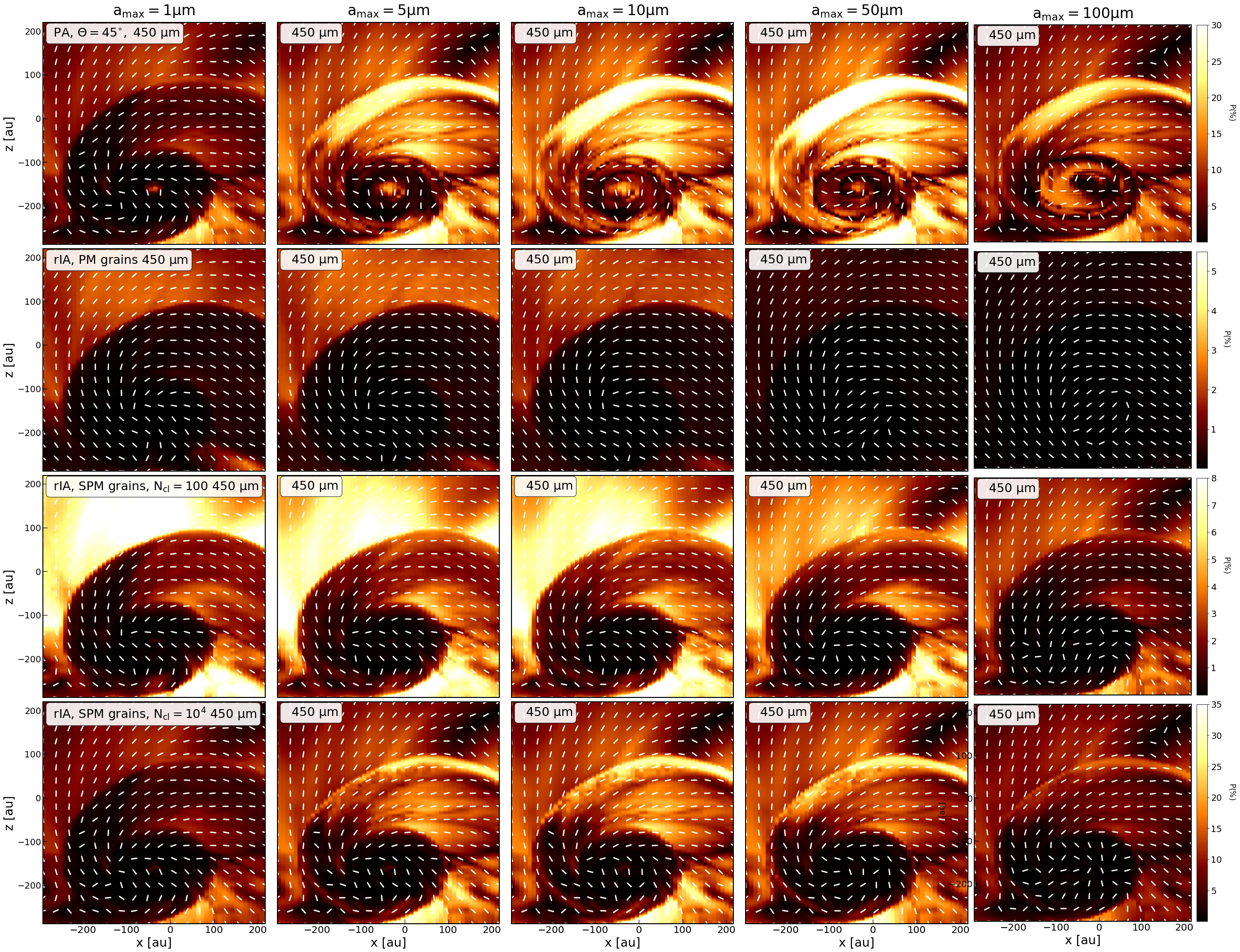}
  
    \caption{The inferred magnetic field map from dust polarization measured at $450\mum$. In model PA, the disk becomes optically thick at $450\mum$ (see Figure \ref{fig:optical_depth}), resulting in the $90^{\circ}$ flipping of $\B$ vectors in the disk from being parallel to perpendicular to the disk minor axis due to the change from dichroic emission to dichroic extinction when grains grow to above $a_{\rm max} \geq 50\mum$. However, for model rIA, one does not clearly see this transition due to the weak extinction of VLGs inside the disk. } 
     \label{fig:pol_map_450um_rIA}
\end{figure*} 

\begin{figure*}
\centering
    \includegraphics[width=\textwidth,height=\textheight,keepaspectratio]{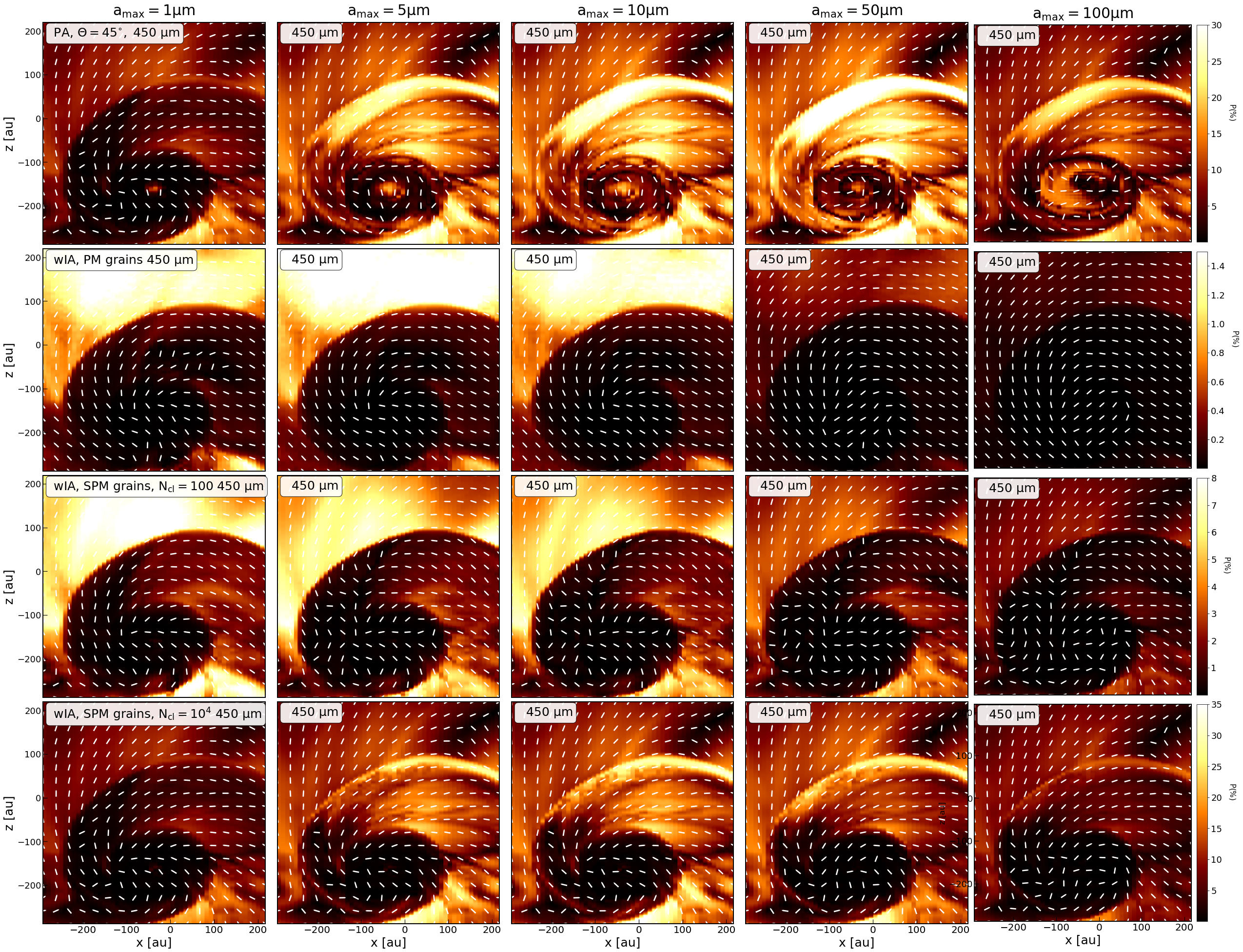}
    \caption{Similar results as Figure \ref{fig:pol_map_450um_rIA} but for model wIA. The inferred magnetic field maps from model wIA for both PM and SPM grains do not clearly show the $90^{\circ}$ flipping of $\B$ vectors for $a_{\rm max} \geq 50\mum$ due to the change in the polarization mechanism as model PA due to the inefficient alignment of VLGs with $\B$around the protostar.} 
    \label{fig:pol_map_450um_wIA}
\end{figure*} 

\section{Effect of grain growth on the polarization spectrum obtained in the inner region}\label{sec:p_lambda}
Figures \ref{fig:p_lambda_200-500au} show the variation of the mean polarization degree obtained in the inner region within $\sim 100- 500$ au and within the disk scale of $\sim 100$ au to the protostar with different maximum grain sizes from $a_{\rm max} = 1\mum$ to $a_{\rm max} = 100\mum$. The left and central panels are for SPM grains with high $N_{\rm cl} = 10^{4}$, and low $N_{\rm cl} = 100$, and the right panel is for PM grains. Similar to Figure \ref{fig:p_lambda_500au}, for SPM grains, $p(\%)$ clearly decreases toward millimeter wavelengths for $a_{\rm max} = 1\mum$ as a result of the weak emission of small grains at long wavelengths. The decline of $p$ with increasing $\lambda$ is shallower with grain growth, and $p$ then becomes to increase toward millimeter wavelengths if the maximum size reaches $a_{\rm max} \geq 50\mum$. However, if grains are PM grains, the polarization curve always has the long tail at millimeter wavelengths for all $a_{\rm max}$, which is caused by the misalignment of large aligned dust grains in the innermost region of the protostellar core (Figure \ref{fig:alar_amax}, upper panels).

\begin{figure*}
\centering
    \includegraphics[width=\textwidth,height=\textheight,keepaspectratio]{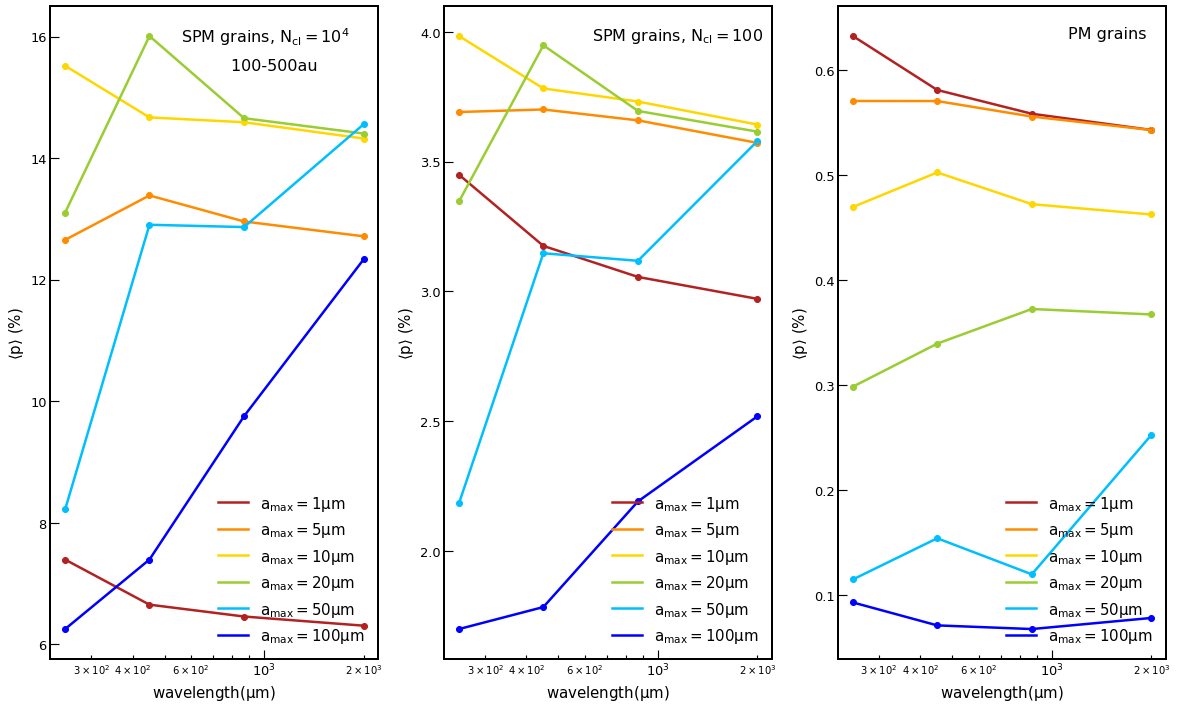}
    \includegraphics[width=\textwidth,height=\textheight,keepaspectratio]{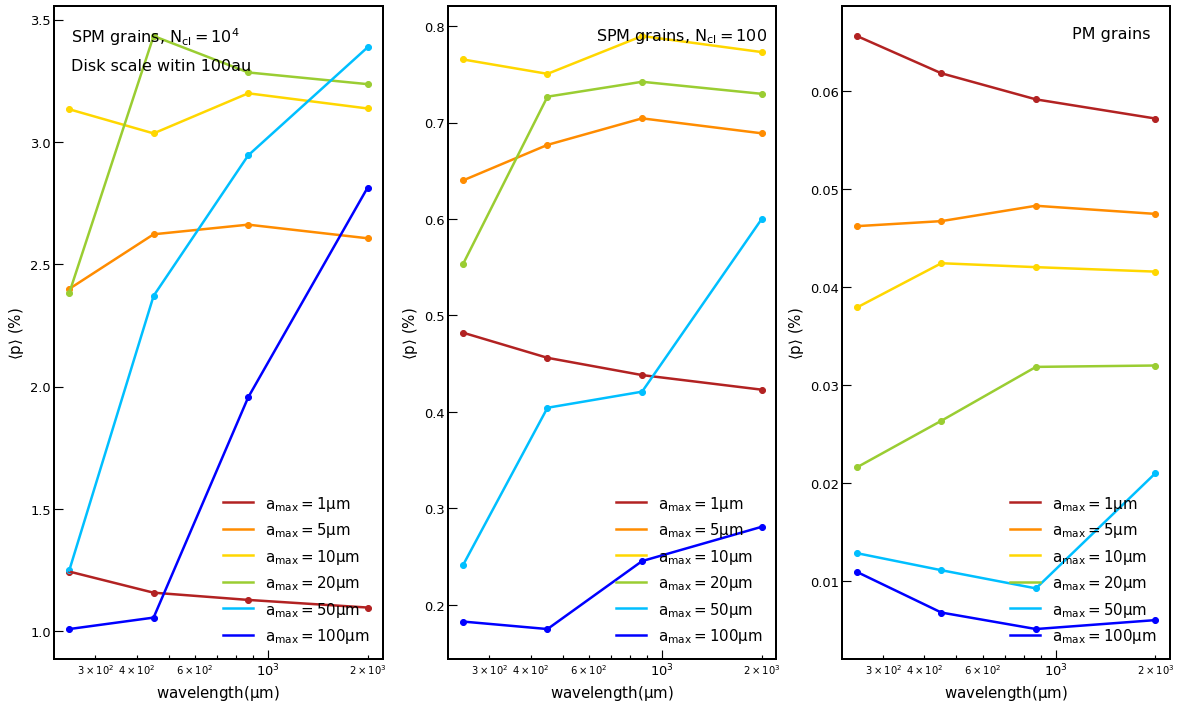}
    \caption{Effect of maximum grain size on the polarization spectrum from $250\mum$ to 2 mm obtained in the inner 100$-$500 au region (upper panel) and in the disk scale within 100 au (lower panel). The left panel shows results for SPM grains with $N_{\rm cl} = 10^{4}$, the central panel is for SPM grains with low $N_{\rm cl} = 100$, and the right panel is for PM grains. Similar to Figure \ref{fig:p_lambda_500au}, for all models of PM and SPM grains, one will obtain the clear decrease of $p$ toward millimeter wavelengths if the maximum grain size is small of $a_{\rm max} \sim 1\mum$. The reduction of $p$ with wavelengths is weaker as grains grow and $p$ changes to increase toward millimeter wavelengths if $a_{\rm max}$ exceeds $\geq 50\mum$. However, the increase of $p$ with $\lambda$ with $a_{\rm max}\geq 50\mum$ is less prominent for grains having low levels of iron inclusions because of the weak alignment of VLGs with magnetic fields in the protostellar core.} 
     \label{fig:p_lambda_200-500au}
\end{figure*} 
 
\section{Effects of inclination angle on the slope of $p-I$}\label{sec:alpha_Theta}
Figure \ref{fig:alpha_inclination_angle} shows the comparison between model PA (black line) and model rIA about the variation of $\alpha$ with the inclination angle $\Theta$ from 0 (the face-on direction along the North direction) to $180^{\circ}$ (face-on along the South direction), assuming $a_{\rm max} = 10\mum$, and $\lambda = 2$ mm. The left panel is for the envelope scale beyond 500 au and the center panel is for the region between $\sim 200-500$ au. For model PA (grains are perfectly aligned with $\B$), $\alpha_{\rm > 500}$ changes from positive value, i.e., $p$ increases with $I$ with $\alpha_{\rm > 500} \sim 0.2$, to the negative value of $\alpha_{\rm > 500} \sim -0.2$, when we change the observed direction from the face-on to the edge-on direction. The increase of $p$ with intensity in the envelope at $\Theta = 0^{\circ}$ or $180^{\circ}$ is caused by the change in direction of the hourglass shape magnetic field, from parallel to the LOS (along z$-$ direction) in $\sim 1000$ au, to be perpendicular to the observed LOS driven by the development of the spiral $\B$ field inside the accretion gas motion in the inner region (Figure \ref{fig:magnetic_field}, right panel).  In contrast, along the edge-on direction, one will see the large-scale hourglass-shaped $\B$-fields have increasing toroidal component (which is perpendicular to the LOS) toward the inner region, resulting in negative $p-I$. This tendency of $p-I$ with $\Theta$ almost matches in the case of SPM grains with high $N_{\rm cl} = 10^{4}$ in model rIA because these grains can have perfect alignment with $\B$ beyond $200$ au. However, for grains with lower $N_{\rm cl} = 100$, $\alpha_{\rm > 500}$ and $\alpha_{\rm 200-500}$ are smaller than model PA, i.e., stronger depolarization effect, induced by the imperfect alignment of grains with magnetic fields there.

\begin{figure*}
\centering
    \includegraphics[width=\textwidth,height=\textheight,keepaspectratio]{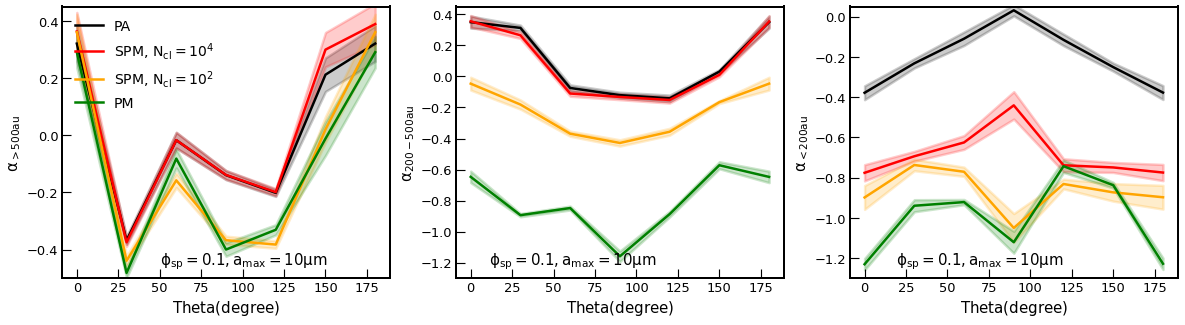}
    \caption{Variation of the slope of $p-I$ relation beyond 500 au, $\alpha_{\geq 500}$ (left panel), in the inner 200-500 au, $\alpha_{200-500}$ (center panel), and within the disk of 200 au $\alpha_{\leq 200}$ (right panel) obtained at 2mm as the function of the inclination angle $\Theta$ for different models of grain magnetic properties, assuming $a_{\rm max} = 10\mum$. The results from model PA are plotted in the black line for comparison. In model PA, $p$ can increase or decrease with intensity depending on the change in the projected geometry of magnetic fields on the POS with the observed direction. In model rIA, SPM grains with high $N_{\rm cl} = 10^{4}$ show a similar variation of $p$ with intensity beyond 200 au as model PA because they can have perfect alignment with $\B$by MRAT alignment. However, in the inner region, the slope of $p-I$ is much steeper compared with model PA due to the inefficient alignment of dust grains around the protostar. For grains with lower levels of iron inclusions, $p$ always decreases with intensity in the entire protostellar core, i.e., negative $\alpha$, and the slope of $p-I$ is also much steeper compared with model PA.} 
     \label{fig:alpha_inclination_angle}
\end{figure*}

Moving to the disk scale (right panel), for model PA, the presence of the poloidal $\B$ field component along the outflow direction induces the depolarization effect with the slope up to $\alpha_{\rm < 200} \sim -0.2$ when we observe the protostar along the face-on direction. Taking into account the realistic model of grain alignment (model rIA), $\alpha_{\rm <200}$ significantly decreases to the value below $\sim -0.4$ for all values of $\Theta$ and different grain types, implying the dominant effects of grain alignment state in controlling the slope of $p-I$ within 200 au. In particular, $\alpha_{\rm <200}$ for PM grains can vary from $\sim -0.8$ to $-1$ for $\Theta = [25-150^{\circ}]$ due to the misalignment of grains inside the disk. This value could reach $-1.2$ at $\Theta = 0^{\circ}$ and $\Theta = 180^{\circ}$ owing to the additional contribution from the parallel of $\B$-fields along the LOS. For SPM grains, $\alpha_{\rm <200}$ belongs to $[\sim -0.4, 0.6]$ for $\Theta \sim [25-150^{\circ}$ as a result of the imperfect alignment of large micron-sized grains with $\B$ around the protostar, and this value can reach to $-0.8$ when observing the disk along the face-on direction.

\section{Fitting the slope of $p-I$}\label{sec:fitting_slope}
Figure \ref{fig:fitting_amax} shows the fitting between the variation of $p-I$ found from synthetic observations with the power law $p\sim I^{\alpha}$ for different maximum grain sizes, assuming $\Theta = 45^{\circ}$. For each panel, the black line shows the result for model PA, the orbrick, red, orange, and green lines are for SPM grains with $N_{\rm cl} = 10^{4}$, $10^{3}$, $100$, $10$, respectively, and blue line is for PM grains. The legend of each line shows the value of $\alpha$ found for the slope of $p-I$ at $\geq 500$ au, between $200-500$ au, and within $\leq 200$ au, respectively. 

Figure \ref{fig:fitting_theta} shows similar results as Figure \ref{fig:fitting_amax} but for different inclination angles $\Theta$ from $0^{\circ}$ to $180^{\circ}$ to the North direction, assuming $a_{\rm max} = 10\mum$. The black line shows the result for model PA, the green and orange lines are for SPM grains with $N_{\rm cl} =10^{4}$ and $100$, and the red line is for PM grains. The meaning of the legend of each line is the same as Figure \ref{fig:fitting_amax}.

\begin{figure*}
\centering
   \includegraphics[width=\textwidth,height=\textheight,keepaspectratio]{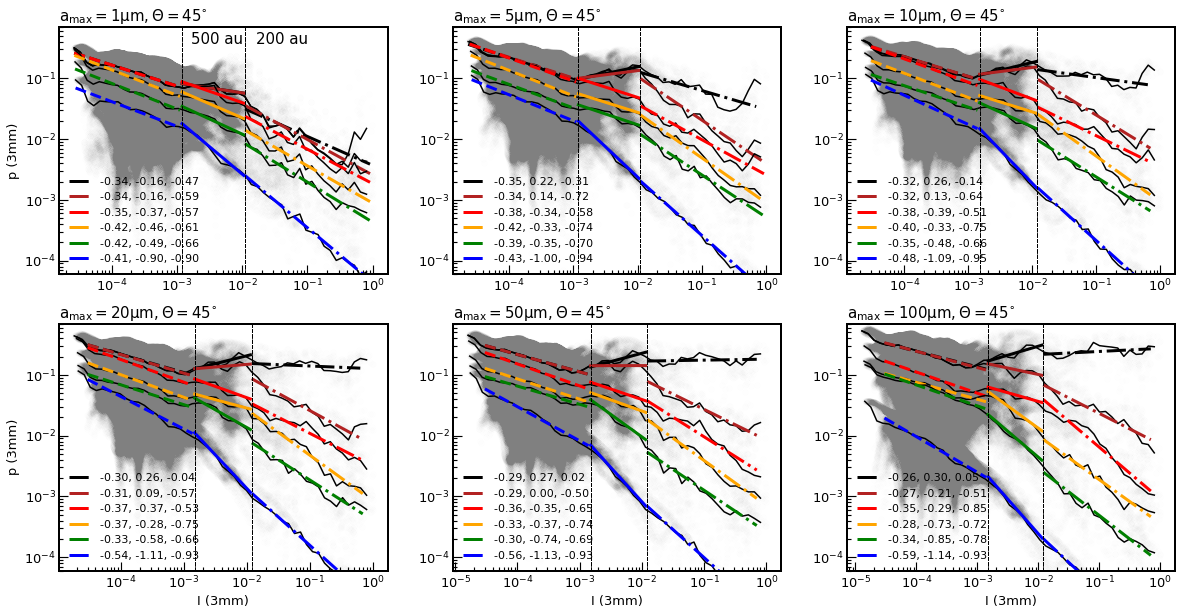}
   \caption{Fitting between the curve $p-I$ from simulations with the power law $p\sim I^{\alpha}$ for different grain magnetic properties and maximum grain sizes. The blue curve is for PM grains, and others are for SPM grains with $N_{\rm cl} = 10$ to $10^{4}$. The slope of $p-I$ in three areas: beyond 500, within 200-500au, and within 200au, is marked in the legend of each color, assuming $\Theta = 45^{\circ}$ and model rIA.} 
    \label{fig:fitting_amax}
\end{figure*}

\begin{figure*}
\centering
   \includegraphics[width=\textwidth,height=\textheight,keepaspectratio]{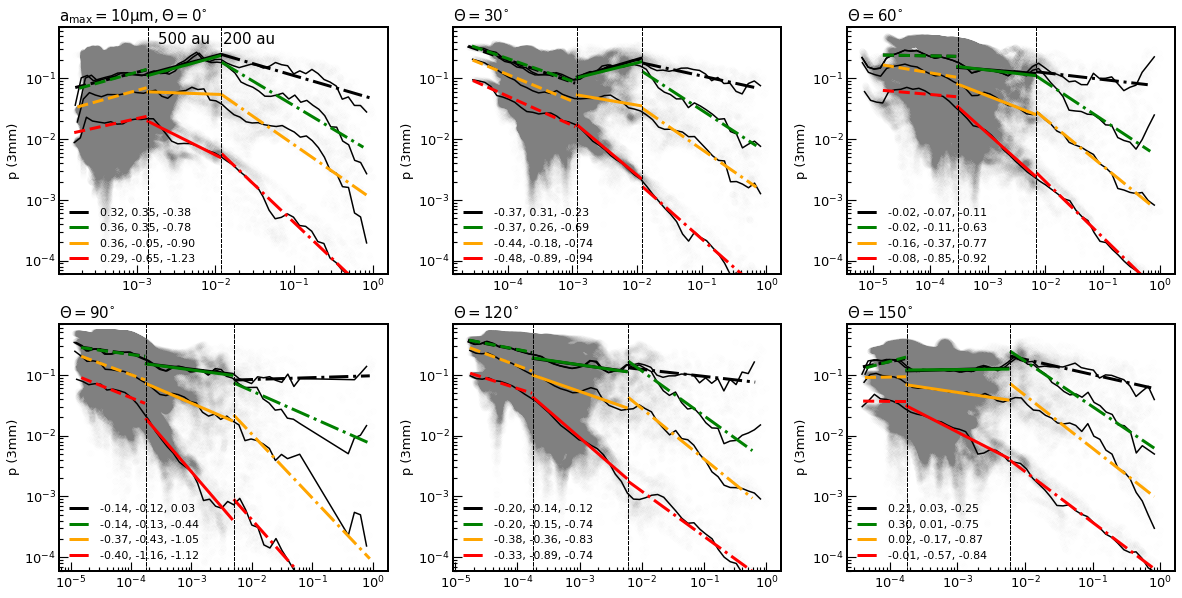}
   \caption{The slope of $p-I$ for different dust model (model PA, black curve) and model rIA (color curves) and different inclination angle $\Theta$ from $0^{\circ}$ to $150^{\circ}$, assuming $a_{\rm max} = 10\mu m$.} 
    \label{fig:fitting_theta}
\end{figure*}

\end{document}